\documentclass[aps, reprint, superscriptaddress, nofootinbib]{revtex4-1} 
\usepackage[dvipdfmx]{graphicx}
\usepackage{dcolumn}
\usepackage{amsmath}
\usepackage{amssymb}
\usepackage{color}
\usepackage{bm}
\usepackage{braket}
\usepackage{txfonts}

\begin{document}
\title{Time Glasses: Symmetry Broken Chaotic Phase with a Finite Gap}
\author{Taiki Haga}
\email[]{taiki.haga@omu.ac.jp}
\affiliation{Department of Physics and Electronics, Osaka Metropolitan University, Sakai-shi, Osaka 599-8531, Japan}
\date{\today}

\begin{abstract}
We introduce the \textit{time glass}, a non-periodic analogue of the discrete time crystal that arises in periodically driven dissipative quantum many-body systems.
This phase is defined by two key features: (i) spatial long-range order arising from the spontaneous breaking of an internal symmetry, and (ii) temporally chaotic oscillations of the order parameter, whose lifetime diverges with system size.
In other words, a time glass is a state of matter in which all components evolve in a synchronized yet chaotic manner.
To characterize the time glass phase, we focus on the spectral gap of the one-cycle (Floquet) Liouvillian, which determines the decay rate of the slowest relaxation mode.
Theoretical arguments and numerical studies of periodically driven dissipative Ising models show that, in the time glass phase, the Liouvillian gap remains finite in the thermodynamic limit, in contrast to time crystals where the gap closes exponentially with system size.
We further demonstrate that the Liouvillian gap converges to the decay rate of the order-parameter autocorrelation derived from the classical (mean-field) dynamics in the thermodynamic limit.
This result establishes a direct correspondence between microscopic spectral features and emergent macroscopic dynamics in driven dissipative quantum systems.
At first glance, the existence of a nonzero Liouvillian gap appears incompatible with the presence of indefinitely persistent chaotic oscillations.
We resolve this apparent paradox by showing that the quantum R\'enyi divergence between a localized coherent initial state and the highly delocalized steady state grows unboundedly with system size.
This divergence allows long-lived transients to persist even in the presence of a finite Liouvillian gap.
\end{abstract}

\maketitle

\section{Introduction}
Synchronization, the spontaneous adjustment of rhythms among many interacting units, is a pervasive phenomenon that spans the full hierarchy of physical systems.
When the coupling between individual units overcomes intrinsic disorder or noise, collective phases emerge in which a macroscopic observable oscillates with a well-defined phase and frequency.
In recent years, interest has shifted to quantum manifestations of synchronization, most prominently the discovery of discrete time crystals (DTCs) in periodically driven many-body systems \cite{Sacha-17, Else-20, Zaletel-23, Khemani-16, Else-16, Yao-17}.
A DTC displays rigid oscillations at an integer multiple of the drive period and persists indefinitely in the thermodynamic limit, thereby realizing a form of temporal long-range order.
Experimental realizations in trapped ions, nitrogen-vacancy centers, and superconducting qubits have confirmed the robustness of these oscillations against moderate imperfections and noise, establishing time-crystalline order as a genuine nonequilibrium phase of matter \cite{Zhang-17, Choi-17, Mi-22}.
The present work considers how the paradigm of synchronization can be extended beyond periodic order into intrinsically chaotic regimes, leading us to the concept of a \emph{time glass}, which is a non-periodic counterpart of the DTC.

Before introducing the time glass, we briefly recall key aspects of time-crystalline order, emphasizing the roles of dissipation and spatial long-range order.
The concept of DTCs was first established in isolated systems, where many-body localization induced by disorder plays a crucial role in preventing runaway heating~\cite{Khemani-16, Else-16, Yao-17}.
It is natural to extend the notion of time-crystalline order to open systems, where dissipation arises due to interactions between the system and its environment.
In such driven dissipative quantum systems, the time evolution is governed by a quantum master equation with a temporally periodic Hamiltonian and jump operators.
Coupling to the environment typically induces both damping and noise: while damping suppresses heating, noise can disrupt phase coherence in the system's oscillations.
A substantial body of research has explored the conditions under which dissipation stabilizes time-crystalline order in open systems \cite{Wang-18, Gong-18, Zhu-19, Gambetta-19, Lazarides-20, Riera-Campeny-20, Kesler-21, Cabot-24}.
It is particularly noteworthy that, in certain dissipative settings, time crystals can arise even in the absence of periodic driving.  
Such phases, termed continuous time crystals, exhibit spontaneous time-translational symmetry breaking under time-independent Liouvillian dynamics \cite{Iemini-18, Tucker-18, Buca-19-1, Buca-19-2, Lledo-19, Seibold-20, Booker-20, Prazeres-21, Piccitto-21, Carollo-22, Hajdusek-22, Kongkhambut-22, Krishna-23, Dutta-25}.

In the time crystal phase, the spatial long-range order associated with spontaneous breaking of an internal symmetry, often a discrete symmetry like $\mathbb{Z}_2$, is fundamental to establishing their unique dynamical order \cite{Khemani-16, Else-16, Yao-17}.
Although the underlying Hamiltonian or master equation respects this symmetry, many-body interactions drive the system to select one of several degenerate states.
This symmetry breaking synchronizes the individual components, leading to collective oscillatory behavior with a period different from that of the external drive.
Moreover, the broken symmetry offers protection against perturbations, which ensures that the time crystal phase remains stable even in the presence of noise or disorder.
It is also noteworthy that in isolated time crystals, where many-body localization prevents runaway heating, the long-range order is characterized by a spin glass order parameter rather than by conventional magnetization.

In recent years, the chaotic behavior of open quantum many-body systems has drawn increasing attention.
Approaches originally developed for isolated systems, such as random matrix theory and out-of-time-ordered correlations, have been extended to dissipative settings \cite{Yusipov-19, Sa-20, Hamazaki-20, Li-21, Prasad-22, Mahoney-24, Zhang-19, Touil-21, Zanardi-21, Schuster-23}.
Much of these studies focus on ``microscopic" chaos, which refers to the irregular dynamics of individual components like spins or particles.
By contrast, in the context of time crystals, where synchronization of microscopic degrees of freedom leads to collective oscillations, it is natural to consider ``macroscopic" chaos, wherein nearly all components move in a coordinated manner, and macroscopic observables exhibit chaotic oscillations.
This distinction parallels the difference between the random thermal motion of fluid molecules at microscopic scales and the coherent turbulent flow at macroscopic scales.
While numerous studies have identified signatures of macroscopic chaos in open quantum many-body systems within the semiclassical or mean-field approximation \cite{Wang-18, Gong-18, Zhu-19, Hartmann-17, Xu-22, Solanki-24-1, Greilich-24, Zhihao-24, Carollo-24, Solanki-24-2, Yang-25}, relatively few investigations have addressed how such behavior can be described directly from the full quantum master equation.
Recent work has begun to fill this gap \cite{Zhihao-24, Carollo-24, Solanki-24-2}, but the systematic study of macroscopic chaos from a quantum perspective remains an open challenge.

\begin{figure}
\centering
\includegraphics[width=8.6cm]{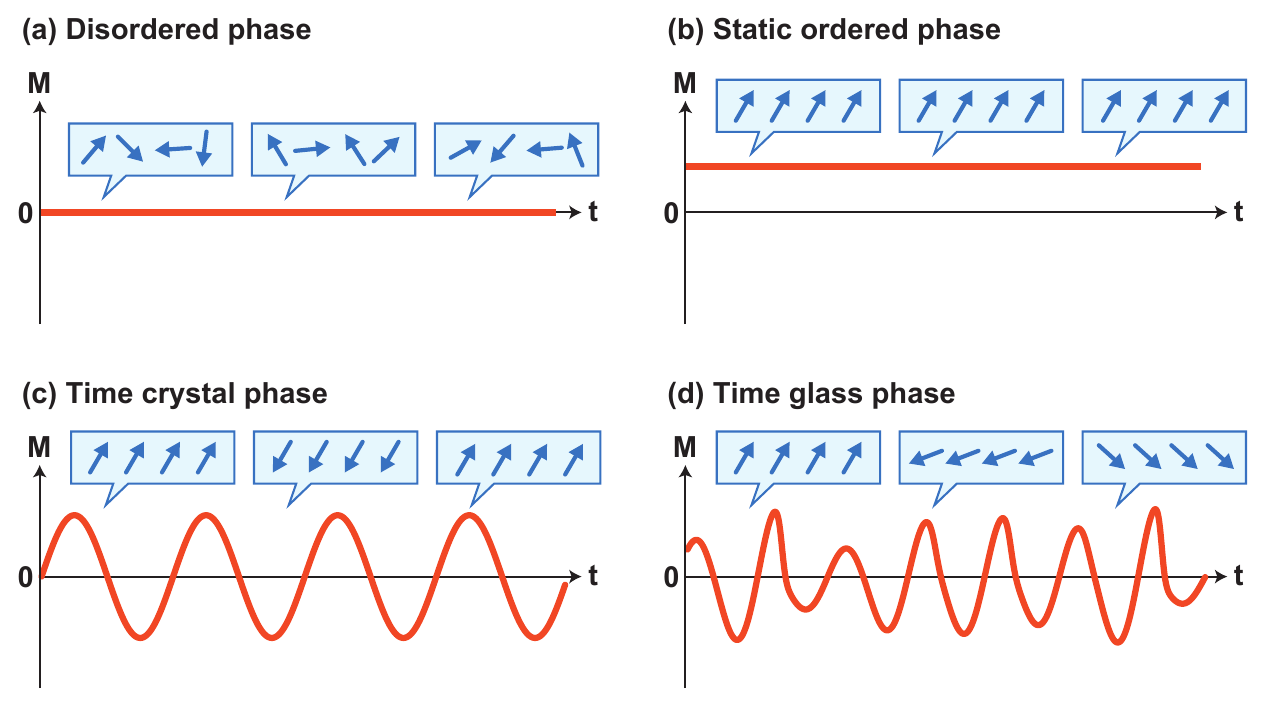}
\caption{Schematic illustrations of the time evolution of the order parameter $M$ in four phases of a spin system.
(a) Disordered phase: Spins point in random directions, resulting in a vanishing order parameter.
(b) Static ordered phase: Spins uniformly align along a specific direction, yielding a nonzero, time-independent order parameter.
(c) Time crystal phase: The system exhibits spatial long-range order, and the order parameter remains nonzero while undergoing periodic oscillations that persist indefinitely in the thermodynamic limit.
(d) Time glass phase: Although the system develops spatial long-range order similar to the time crystal, the order parameter displays irregular, chaotic oscillations that persist indefinitely in the thermodynamic limit.}
\label{fig_time_glass}
\end{figure}

In this work, we investigate a dynamical phase in driven dissipative systems, termed the \emph{time glass}, which serves as a minimal model for macroscopic chaos.
Figure \ref{fig_time_glass} illustrates four possible phases of a spin system: (a) disordered, (b) static ordered, (c) time crystal, and (d) time glass. 
Here, $M$ represents the order parameter of the system.
In the disordered phase, spins point in random directions, which makes the order parameter vanish [see Fig.~\ref{fig_time_glass}(a)].
In the static ordered phase, spins align uniformly in a specific direction, which results in a nonzero but time-independent order parameter [see Fig.~\ref{fig_time_glass}(b)].
Both of these phases are typically encountered in thermal equilibrium.
By contrast, a time crystal exhibits spatial long-range order, and its order parameter undergoes periodic oscillations [see Fig.~\ref{fig_time_glass}(c)].
The lifetime of these oscillations grows with system size, eventually persisting indefinitely in the thermodynamic limit.
In the proposed time glass phase, the system likewise exhibits spatial long-range order, but the order parameter shows chaotic oscillations [see Fig.~\ref{fig_time_glass}(d)].
Remarkably, the timescale of these chaotic oscillations also diverges as the system size increases, which leads to an indefinitely sustained chaotic dynamics in the thermodynamic limit.
The time glass encapsulates the key ingredients required for macroscopic chaos: (i) synchronization among the microscopic degrees of freedom and (ii) chaotic temporal dynamics of the synchronized collective motion.
Thus, this phase provides a concrete realization of macroscopic chaos within a driven dissipative many-body system.

The goal of this study is to characterize the time glass phase by examining the spectral properties of the time evolution map.
In dissipative systems, the eigenvalues of the time evolution map determine both the decay rates and the frequencies of the various modes.
We focus in particular on the spectral gap, which governs the decay rate of the slowest mode.
It is known that in the time crystal phase, non-decaying modes with finite frequencies emerge, causing the spectral gap to close in the thermodynamic limit.
However, our theoretical arguments and numerical results on periodically driven Ising models show that, in the time glass phase, the spectral gap remains finite in the thermodynamic limit.
Furthermore, we demonstrate that the spectral gap converges to the decay rate of the autocorrelation function of the order parameter for the chaotic dynamics emerged in the thermodynamic limit.
At first glance, maintaining a nonzero spectral gap seems incompatible with persistent chaotic oscillations, because one would expect all non-steady modes to decay quickly.
We resolve this apparent paradox by noting that the overlap between a localized initial state and the extended steady state is extremely small.

This paper is organized as follows.
In Sec.~\ref{sec:overview}, we summarize the main results of this work.
We outline the general setup of periodically driven dissipative systems and introduce the concept of the time glass.
Readers seeking only the core ideas may focus on this section.
In Sec.~\ref{sec:models}, we introduce the periodically driven dissipative Ising models analyzed throughout this paper, specifically a kicked collective spin model and a kicked spin chain model.
In Sec.~\ref{sec:classical_dynamics}, we discuss the classical dynamics of these models in the thermodynamic limit.
We show that the time evolution of the order parameter is described by a set of nonlinear equations, which exhibits chaos within certain parameter regimes.
In Sec.~\ref{sec:autocorrelation}, we numerically calculate the autocorrelation function of the order parameter, using it to identify both time crystal and time glass phases.
In Sec.~\ref{sec:spectral_gap}, we investigate the spectral gap of the time evolution map for these phases.
We demonstrate that, in the time glass phase, the gap remains finite as the system size grows and coincides with the decay rate of the autocorrelation function.
In Sec.~\ref{sec:relaxation_time}, we discuss the system size dependence of the relaxation time in the time glass phase.
We show that it increases logarithmically with the system size.
We explain how the divergence of the relaxation time is reconciled with the presence of a finite gap.
Finally, Sec.~\ref{sec:discussion} discusses the limitations of the present study and proposes open problems for future research, while Sec.~\ref{sec:conclusion} offers concluding remarks.

The technical and supporting details are provided in the Appendices: 
Appendix \ref{sec:quantum_classical_correspondence} presents a heuristic derivation that demonstrates the convergence of quantum autocorrelations to their classical counterparts in the semiclassical limit. 
Appendix \ref{sec:linear_stability_analysis} provides the linear stability analysis of the fixed point in the classical dynamics. 
Appendix \ref{sec:quantum_trajectory_method} reviews the quantum trajectory method used for calculating two-point correlation functions. 
Appendix \ref{sec:symmetric_sector_spin_chain} summarizes the matrix elements of the Liouvillian in the symmetric sector for the all-to-all coupled spin model. 
Appendix \ref{appendix:Liouvillian_gap_of_stochastic_system} investigates the Liouvillian gap of an analogous classical stochastic system, where we find that, for chaotic cases, the gap opens and provides the mixing rate in the noiseless limit. 
Appendix \ref{appendix:clarifying_the_discrepancy} discusses the reconciliation between our findings and those reported in previous studies.

\section{Overview of Time Glasses}
\label{sec:overview}

In this section, we provide a comprehensive overview of the time glass phase in driven dissipative quantum systems. 
We begin by presenting the general theoretical framework, after which we define the time glass rigorously from two perspectives: the correlation-based definition and the trajectory-based definition. 
We then establish the equivalence between these two definitions by invoking the quantum-classical correspondence of autocorrelations, as formalized in Eq.~\eqref{autocorrelation_quantum_classical_correspondence}. 
Next, we discuss the spectral features of the time glass phase, demonstrating that the spectral gap coincides with the decay rate of the order-parameter autocorrelation function derived from the classical chaotic dynamics, as encapsulated by our central result, Eq.~\eqref{gap_mixing_rate}. 
Finally, we address the critical mechanism that allows for indefinitely persistent chaotic dynamics despite the presence of a finite spectral gap.

\subsection{General setup}

For Markovian dissipative systems, the time evolution of the density matrix $\hat{\rho}$ is governed by the quantum master equation \cite{Lindblad-76, Gorini-76, Breuer, Rivas}:
\begin{equation}
\partial_t \hat{\rho} = - i [\hat{H}(t), \hat{\rho}] + \sum_{k} \left( \hat{L}_k(t) \hat{\rho} \hat{L}_k(t)^\dag - \frac{1}{2} \{ \hat{L}_k(t)^\dag \hat{L}_k(t), \hat{\rho} \} \right),
\label{master_equation_general}
\end{equation}
where $\hat{H}(t)$ is the time-dependent Hamiltonian and $\hat{L}_k(t)$ are the time-dependent jump operators.
We assume periodic driving with period $T$, so that $\hat{H}(t+T)=\hat{H}(t)$ and $\hat{L}_k(t+T)=\hat{L}_k(t)$ for all $t$.
The generator of this evolution is often referred to as the Liouvillian and denoted $\mathcal{L}_t$,
\begin{equation}
\partial_t \hat{\rho} = \mathcal{L}_t \hat{\rho}.
\end{equation}
The Liouvillian also satisfies $\mathcal{L}_{t+T}=\mathcal{L}_t$ for all $t$.

The corresponding Floquet map is
\begin{equation}
\mathcal{U} = \mathcal{T} \exp \left[ \int_0^T \mathcal{L}_t dt \right],
\end{equation}
where $\mathcal{T}$ denotes the time-ordering operator.
The stroboscopic time evolution of the density matrix is given by
\begin{equation}
\hat{\rho}(nT) = \mathcal{U} \hat{\rho}((n-1)T) = \mathcal{U}^n \hat{\rho}(0).
\end{equation}
It is noteworthy that, unlike in isolated systems, $\mathcal{U}$ generally cannot be expressed by a time-independent effective Liouvillian.
More precisely, for isolated systems with $\hat{L}_k(t)=0$, one can always write $\mathcal{U}=\exp(t \mathcal{L}_{\text{eff}})$ with $\mathcal{L}_{\text{eff}} = -i[\hat{H}_{\text{eff}}, \cdot]$, where $\hat{H}_{\text{eff}}$ is a time-independent effective Hamiltonian.
In the presence of dissipation, however, there is no guarantee that a time-independent $\mathcal{L}_{\text{eff}}$ of the form Eq.~\eqref{master_equation_general} exists \cite{Hartmann-17, Wolf-08, Schnell-20, Mizuta-21}.
This irreducibility to a static effective system is what makes driven dissipative quantum systems richer than their isolated counterparts.

We consider the eigenvalues $\lambda_\alpha$ and eigenmodes $\hat{\rho}_\alpha$ $(\alpha=0, 1, \ldots)$ of the Floquet map:
\begin{equation}
\mathcal{U} \hat{\rho}_\alpha = \lambda_\alpha \hat{\rho}_\alpha.
\end{equation}
By definition, there is always a stroboscopic steady state $\hat{\rho}_0$ associated with the eigenvalue $\lambda_0=1$, satisfying $\mathcal{U} \hat{\rho}_0 = \hat{\rho}_0$.
In the following, we denote the stroboscopic steady state $\hat{\rho}_0$ as $\hat{\rho}_{\text{ss}}$.
In isolated systems where $\hat{L}_k(t)=0$, $\mathcal{U}$ is unitary, and all eigenvalues lie on the unit circle, $|\lambda_\alpha|=1$.
However, once dissipation is introduced, every eigenvalue resides within or on the boundary of the unit disk, $|\lambda_\alpha| \leq 1$.
We label the eigenvalues and corresponding eigenmodes in descending order of their magnitudes, $|\lambda_0|=1 \geq |\lambda_1| \geq \cdots$.
For any non-steady eigenmode associated with an eigenvalue $\lambda_\alpha \neq 1$, the trace-preserving property of $\mathcal{U}$ implies $\mathrm{Tr}[\hat{\rho}_\alpha]=0$.
Thus, such modes are not physical density matrices.
The normalization of these eigenmodes is not fixed by physical constraints, but it is standard to choose a convenient Hilbert--Schmidt normalization, $\mathrm{Tr}[\hat{\rho}_\alpha^\dagger \hat{\rho}_\alpha]=1$.

Suppose an initial density matrix $\hat{\rho}(0)$ is expanded in terms of eigenmodes:
\begin{equation}
\hat{\rho}(0) = \hat{\rho}_{\text{ss}} + \sum_{\alpha \geq 1} c_\alpha \hat{\rho}_\alpha.
\end{equation}
The positivity of the physical density matrix $\hat{\rho}(0)$ is encoded in the coefficients $c_\alpha$.  
Determining whether a given set $\{c_\alpha\}$ yields a positive operator is generally a nontrivial constraint-satisfaction problem.
However, this issue does not affect the validity of the expansion itself: if $\mathcal{U}$ is diagonalizable, its eigenmodes $\{\hat{\rho}_\alpha\}$ form a complete basis of the operator space, and any physical density matrix admits a unique decomposition of the form above.
Under stroboscopic evolution, the state at time $nT$ is given by
\begin{equation}
\hat{\rho}(nT) = \hat{\rho}_{\text{ss}} + \sum_{\alpha \geq 1} c_\alpha (\lambda_\alpha)^n \hat{\rho}_\alpha.
\label{eigenmode_expansion}
\end{equation}
After a sufficiently long time ($n \gg 1$), the distance between $\hat{\rho}(nT)$ and the steady state $\hat{\rho}_{\text{ss}}$ behaves as
\begin{equation}
|| \hat{\rho}(nT) - \hat{\rho}_{\text{ss}} || \sim |\lambda_1|^n = e^{n \log |\lambda_1|},
\end{equation}
where $|| \ldots ||$ denotes an appropriate norm and $\lambda_1$ is the eigenvalue with the second largest magnitude.
We thus define the Liouvillian gap $\Delta$ as
\begin{equation}
\Delta := - \frac{1}{T} \log |\lambda_1|.
\label{Liouvillian_gap}
\end{equation}
This gap characterizes the asymptotic decay rate to the steady state.
It is known that the Liouvillian gap $\Delta$ plays a crucial role as an indicator of phase transitions in steady states of dissipative quantum systems \cite{Kessler-12, Honing-12, Horstmann-13, Casteels-16, Casteels-17, Fitzpatrick-17, Vicentini-18, Minganti-18, Imamoglu-18, Rota-18, Ferreira-19}.

\subsection{Correlation-based definition of time glasses}

We now introduce the definitions of the time glass and related phases. 
Although our primary focus is on periodically driven systems, the following discussion also applies to undriven scenarios with a time-independent Liouvillian.
Suppose the system possesses an internal symmetry, for instance a $\mathbb{Z}_2$ symmetry associated with a global spin flip.
This means there exists a unitary operator $V$ such that the Liouvillian satisfies $\mathcal{L}_t(V^\dag \hat{\rho} V)=V^\dag \mathcal{L}_t(\hat{\rho}) V$ for any state $\hat{\rho}$ and any time $t$.
Let $\hat{S}(x)$ be an observable related to this internal symmetry at position $x$; for a spin system on a lattice, $x$ might be a discrete site index.
For the case of the $\mathbb{Z}_2$ symmetry, $\hat{S}(x)$ transforms as $V^\dag \hat{S}(x) V = - \hat{S}(x)$.
Since the (unique) stroboscopic steady state is invariant with respect to this transformation, $V^\dag \hat{\rho}_{\text{ss}} V = \hat{\rho}_{\text{ss}}$, the expectation value of $\hat{S}(x)$ vanishes,
\begin{equation}
\langle \hat{S}(x) \rangle_{\text{ss}} := \text{Tr}\left[ \hat{S}(x) \hat{\rho}_{\text{ss}} \right] = 0.
\end{equation}

We write the stroboscopic evolution of $\hat{S}(x)$ in the Heisenberg picture as
\begin{equation}
\hat{S}(x, nT) = (\mathcal{U}^\dag)^n \hat{S}(x),
\end{equation}
where \(\mathcal{U}^\dag\) denotes the adjoint of \(\mathcal{U}\), defined via the relation \(\mathrm{Tr}[\hat{A} \, \mathcal{U}(\hat{B})] = \mathrm{Tr}[\mathcal{U}^\dag(\hat{A}) \, \hat{B}]\) for arbitrary operators \(\hat{A}\) and \(\hat{B}\).
We then define the spatio-temporal correlation function by
\begin{align}
C(x, x'; nT) &:= \langle \hat{S}(x, nT) \hat{S}(x', 0) \rangle_{\text{ss}} \nonumber \\
&= \text{Tr}\left[ \hat{S}(x) \mathcal{U}^n\left(\hat{S}(x') \hat{\rho}_{\text{ss}}\right) \right].
\end{align}
The order parameter $\hat{M}$ is given by
\begin{equation}
\hat{M} = \frac{1}{V} \int dx \hat{S}(x),
\end{equation}
where $V$ is the volume of the system.
Accordingly, the autocorrelation function of the order parameter is
\begin{equation}
C_M(nT) = \langle \hat{M}(nT) \hat{M}(0) \rangle_{\text{ss}} = \frac{1}{V^2} \int dx dx' C(x, x'; nT).
\label{C_general_def}
\end{equation}
We consider its behavior in the thermodynamic limit,
\begin{equation}
C_M^\infty(t) = \lim_{V \to \infty} C_M(t).
\end{equation}
In particular, the static order parameter can be given by the equal-time correlation $C_M^\infty(0)$.

\begin{figure}
\centering
\includegraphics[width=8.6cm]{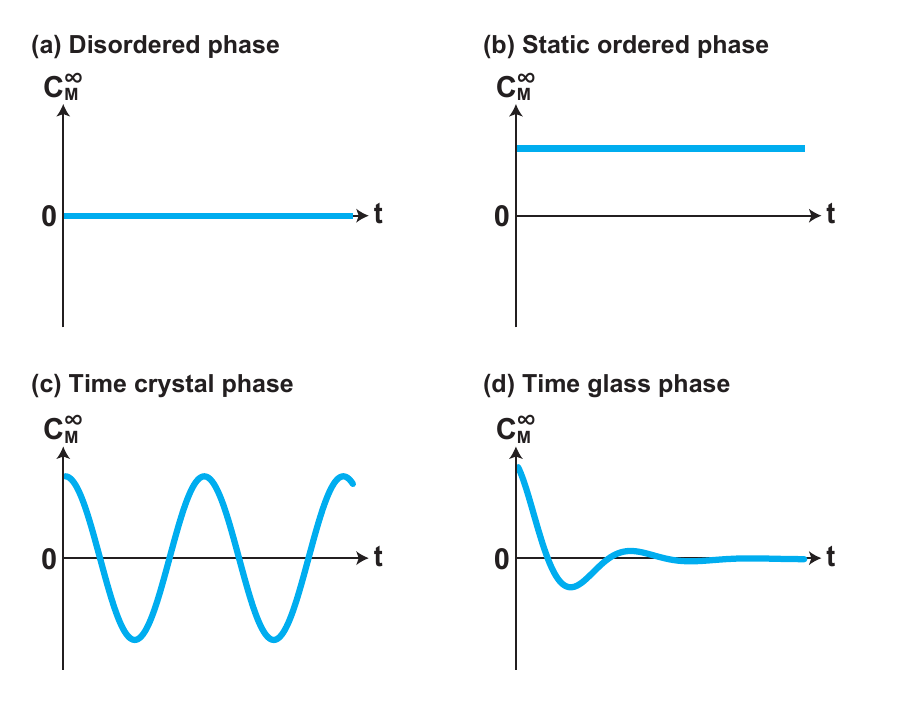}
\caption{Schematic illustrations of the autocorrelation $C_M^\infty(t)$ in the thermodynamic limit.
(a) Disordered phase: The autocorrelation is zero at all times. 
(b) Static ordered phase: The autocorrelation remains nonzero and constant.
(c) Time crystal: The autocorrelation exhibits persistent periodic oscillations.
(d) Time glass: Although $C_M^\infty(0)>0$, indicating a finite order parameter, the autocorrelation decays to zero for $t \to \infty$.}
\label{fig_autocorrelation_schematic}
\end{figure}

By using the autocorrelation function $C_M^\infty(t)$ in the thermodynamic limit, we can distinguish the four phases shown in Fig.~\ref{fig_time_glass} as follows:
\begin{enumerate}
\item \textit{Disordered phase:} The autocorrelation function is identically zero: $C_M^\infty(t)=0$ for all $t$ [see Fig.~\ref{fig_autocorrelation_schematic}(a)].
\item \textit{Static ordered phase:} A finite order parameter exists due to spontaneous symmetry breaking, $C_M^\infty(0)>0$.
The autocorrelation remains a nonzero constant for all $t$, i.e., $C_M^\infty(t) = \text{const.} > 0$ [see Fig.~\ref{fig_autocorrelation_schematic}(b)].
\item \textit{Time crystal:} A finite order parameter again arises due to spontaneous symmetry breaking, $C_M^\infty(0)>0$, but now the autocorrelation function exhibits persistent periodic oscillations [see Fig.~\ref{fig_autocorrelation_schematic}(c)].
\item \textit{Time glass:} Like the time crystal, a finite order parameter exists, $C_M^\infty(0)>0$.
However, $C_M^\infty(t)$ decays to zero for large $t$ due to chaotic dynamics of the order parameter [see Fig.~\ref{fig_autocorrelation_schematic}(d)].
\end{enumerate}
We emphasize that this classification scheme, relying only on the macroscopic order parameter's correlation function in the thermodynamic limit, is equally applicable for both quantum and classical many-body systems.

It is important to highlight that time glasses are fundamentally distinct from conventional spatio-temporal chaos, which does not exhibit spatially synchronized chaotic oscillations.
For example, the complex Ginzburg-Landau equation can give rise to defect turbulence \cite{Aranson-02}, wherein the $U(1)$ order parameter field exhibits chaotic dynamics accompanied by the proliferation of topological defects (vortices).
Within the autocorrelation framework discussed earlier, such states correspond to disordered phases characterized by vanishing autocorrelation, $C_M^\infty(t)=0$, as illustrated in Fig.~\ref{fig_autocorrelation_schematic}(a).
The same equation can also exhibit phase turbulence \cite{Aranson-02}, wherein the phase of the $U(1)$ order parameter field remains chaotic, but no topological defects are generated.
In this case, the global magnetization $M$ fluctuates around a nonzero mean value, and the autocorrelation remains constant over time, as shown in Fig.~\ref{fig_autocorrelation_schematic}(b).
Neither of these scenarios, defect turbulence or phase turbulence, qualifies as a time glass under our definition. 

A useful way to view the time glass is as a spin glass in which the roles of space and time are reversed.
In the spin glass phase, the spins remain frozen in time while they point in random directions, so no net magnetization develops.
This means the spin glass is effectively ordered in time (spins do not evolve) but disordered in space (random orientations).
By contrast, a time glass is disordered in time but exhibits spatial order.
To illustrate this analogy, let us consider the spatio-temporal correlation function $C(x, t)=\langle S(x, t) S(0, 0) \rangle$ in a spin glass, where the average over quenched disorder is taken after the thermal average.
One can integrate over a time interval $T$ and then let $T \to \infty$, defining
\begin{equation}
C_M^\infty(x) = \lim_{T \to \infty} \frac{1}{T}  \int_0^T dt C(x, t).
\end{equation}
In a static ordered phase, $C_M^\infty(x)$ remains a nonzero constant for all $x$.
In the spin glass phase, $C_M^\infty(0)$ is also nonzero because the temporal correlation $C(0, t)$ does not decay for $t \to \infty$.
However, $C_M^\infty(x)$ vanishes when $x \to \infty$ due to random spin configurations.
This behavior is the same as the criterion for a time glass with space $x$ and time $t$ interchanged, reinforcing that the term ``time glass" appropriately describes the phase illustrated in Fig.~\ref{fig_time_glass}(d).

\subsection{Trajectory-based definition of time glasses}

We remark that a time glass can also be characterized from the viewpoint of dynamics in the expectation values of observables.
Consider an initial density matrix chosen as a coherent pure state with minimal quantum uncertainty,  $\hat{\rho}(0)=\ket{\psi_0} \bra{\psi_0}$.
For instance, in a spin system, one may take a product state of identical spin coherent states:
\begin{equation}
\ket{\psi_0} = \bigotimes_{i=1}^N \ket{\theta, \phi}_i,
\label{initial_coherent_state}
\end{equation}
where $\ket{\theta, \phi}_i$ denotes the spin coherent state characterized by the polar angle $\theta$ and the azimuthal angle $\phi$ for the $i$th spin [see Eq.~\eqref{spin_coherent_state_def}].
We then examine the time evolution of the order parameter,
\begin{equation}
m(t) = \mathrm{Tr}[\hat{M} \hat{\rho}(t)].
\end{equation}
For any finite open quantum system, the state $\hat{\rho}(t)$ relaxes to its steady state $\hat{\rho}_{\mathrm{ss}}$ at long times, and correspondingly $m(t)$ approaches a constant.
However, the behavior can fundamentally change when the thermodynamic limit $V \to \infty$ is taken before the long-time limit.
We assume that, for each fixed $t$, the limit 
\begin{equation}
m_{\infty}(t) = \lim_{V \to \infty} m(t)
\end{equation}
exists.
In a time crystal phase, $m_{\infty}(t)$ exhibits persistent periodic oscillations that never decay.
In contrast, we define the time glass phase as the regime in which $m_{\infty}(t)$ shows indefinitely sustained chaotic temporal evolution.
This dynamical definition is fully consistent with the intuitive picture illustrated in Fig.~\ref{fig_time_glass}.

The appearance of macroscopic chaos at the level of individual trajectories is well established in fully connected (mean-field) spin models.
In particular, Ref.~\cite{Carollo-24} rigorously demonstrated that, when the initial state is a coherent pure state, the thermodynamic-limit trajectory $m_{\infty}(t)$ exactly follows the corresponding mean-field equations of motion.
More precisely, there exists a relaxation timescale \( \tau_{\mathrm{rel}}(N) \) that diverges with \( N \), such that for times \( t \ll \tau_{\mathrm{rel}}(N) \), the system collectively behaves like a classical spin.
If these mean-field equations are chaotic in a given parameter regime, the resulting dynamical phase should therefore be classified as a time glass.
It should be emphasized, however, that for spin systems defined on finite-dimensional lattices, where mean-field theory is no longer exact, the emergence of a time glass phase is far from guaranteed and becomes a highly nontrivial phenomenon.

We now clarify the connection between the two definitions of a time glass: one based on the two-time correlation function $C_M^\infty(t)$, and the other based on the single-time expectation value $m_{\infty}(t)$.
Suppose that $m_{\infty}(t)$ obeys a closed set of deterministic equations of motion and exhibits chaotic behavior.
When the resulting dynamics is \emph{ergodic}, we may define a ``classical" autocorrelation $C_{\mathrm{cl}}(t)$ from a single trajectory of $m_{\infty}(t)$ as
\begin{equation}
C_{\mathrm{cl}}(nT) = \lim_{K \to \infty} \frac{1}{K} \sum_{k=1}^K m_{\infty}((n+k)T) m_{\infty}(kT).
\label{classical_autocorrelation_general}
\end{equation}
Such autocorrelations typically decay exponentially,
\begin{equation}
|C_{\mathrm{cl}}(t)| \propto e^{-gt},
\end{equation}
where $g$ denotes the mixing rate of the classical dynamics.
We conjecture that, provided the emerging classical dynamics is ergodic, the classical autocorrelation $C_{\mathrm{cl}}(t)$ coincides with the thermodynamic-limit quantum autocorrelation,
\begin{equation}
C_M^\infty(t) = C_{\mathrm{cl}}(t).
\label{autocorrelation_quantum_classical_correspondence}
\end{equation}
This correspondence implies that, in the time glass phase, the quantum autocorrelation $C_M^\infty(t)$ also decays exponentially with the classical mixing rate $g$, giving rise to the picture summarized in Fig.~\ref{fig_autocorrelation_schematic}.

It is important to note that the quantum-classical correspondence of the autocorrelation given by Eq.~\eqref{autocorrelation_quantum_classical_correspondence} is not trivial.  
Consider an initial state \( \hat{\rho}(0) \) chosen as a coherent pure state given by Eq.~\eqref{initial_coherent_state}.
We rewrite the quantum autocorrelation function Eq.~\eqref{C_general_def} in terms of the time evolution from \( \hat{\rho}(0) \) as follows:
\begin{equation}
C_M(nT) = \lim_{K \to \infty} \frac{1}{K} \sum_{k = 1}^K \mathrm{Tr} \left[ \hat{M} \mathcal{U}^n \left( \hat{M} \mathcal{U}^k (\hat{\rho}(0)) \right) \right].
\label{C_M_def_time_average}
\end{equation}  
Since \( \hat{\rho}_{\mathrm{ss}} = \lim_{n \to \infty} \mathcal{U}^n (\hat{\rho}(0)) \), this expression is equivalent to Eq.~\eqref{C_general_def}.

Now consider taking the thermodynamic limit \( N \to \infty \) before the long time limit \( K \to \infty \) in Eq.~\eqref{C_M_def_time_average}.  
Let \( m_\infty(t) \) denote the classical trajectory of \( \hat{M} \) corresponding to the initial condition \( \hat{\rho}(0) \).
Then we have  
\begin{equation}
\lim_{N \to \infty} \mathrm{Tr} \left[ \hat{M} \mathcal{U}^n \left( \hat{M} \mathcal{U}^k (\hat{\rho}(0)) \right) \right] = m_\infty((n+k)T) m_\infty(kT).
\end{equation}  
The classical autocorrelation Eq.~\eqref{classical_autocorrelation_general} can be written as  
\begin{equation}
C_{\mathrm{cl}}(nT) = \lim_{K \to \infty} \lim_{N \to \infty} \frac{1}{K} \sum_{k = 1}^K \mathrm{Tr} \left[ \hat{M} \mathcal{U}^n \left( \hat{M} \mathcal{U}^k (\hat{\rho}(0)) \right) \right].
\label{C_cl_double_limits}
\end{equation}  
Therefore, the condition Eq.~\eqref{autocorrelation_quantum_classical_correspondence} is equivalent to the commutativity of the thermodynamic limit and the long-time limit in Eq.~\eqref{C_cl_double_limits}.

In general, however, these two limits do \emph{not} commute.
To illustrate a situation in which the correspondence Eq.~\eqref{autocorrelation_quantum_classical_correspondence} fails, consider a case where the emergent classical dynamics in the thermodynamic limit is \emph{nonergodic} and possesses two distinct attractors, $\mathcal{A}_1$ and $\mathcal{A}_2$.
We assume further that $\mathcal{A}_1$ has a larger basin of attractor than $\mathcal{A}_2$.
Then, the quantum steady state $\hat{\rho}_{\mathrm{ss}}$ carries a larger weight on $\mathcal{A}_1$.
As the system size $N$ increases, quantum fluctuations are suppressed and tunneling between the two attractors becomes increasingly rare, making $\hat{\rho}_{\mathrm{ss}}$ more strongly localized on $\mathcal{A}_1$.
Consequently, in the thermodynamic limit, the autocorrelation $C_M^\infty(t)$ reflects only the dynamics of $\mathcal{A}_1$.
In contrast, if the thermodynamic limit is taken \emph{before} the long-time limit and the initial coherent state lies in the basin of the secondary attractor $\mathcal{A}_2$, then the resulting classical trajectory will remain trapped in $\mathcal{A}_2$ forever.
In this situation, the classical autocorrelation $C_{\mathrm{cl}}(t)$ is determined by the properties of $\mathcal{A}_2$, and therefore $C_M^\infty(t) \neq C_{\mathrm{cl}}(t)$.
This example illustrates that the quantum-classical correspondence of autocorrelations relies critically on the ergodicity of the emergent classical dynamics.

In Appendix~\ref{sec:quantum_classical_correspondence}, we derive the quantum-classical correspondence of autocorrelations stated in Eq.~\eqref{autocorrelation_quantum_classical_correspondence} for fully connected (mean-field) spin models.
Our argument relies crucially on the theorem proved in Ref.~\cite{Carollo-24}, which ensures that quantum expectation values converge to the corresponding classical trajectories in the thermodynamic limit.
We further employ a physically reasonable assumption connecting the quantum steady state to the classical invariant distribution in the semiclassical regime.
Together, these ingredients establish the equivalence between the correlation-based and trajectory-based definitions of time glass phase.

\subsection{Spectral features}
\label{sec:overview_spectral features}

\begin{figure*}
\centering
\includegraphics[width=\textwidth]{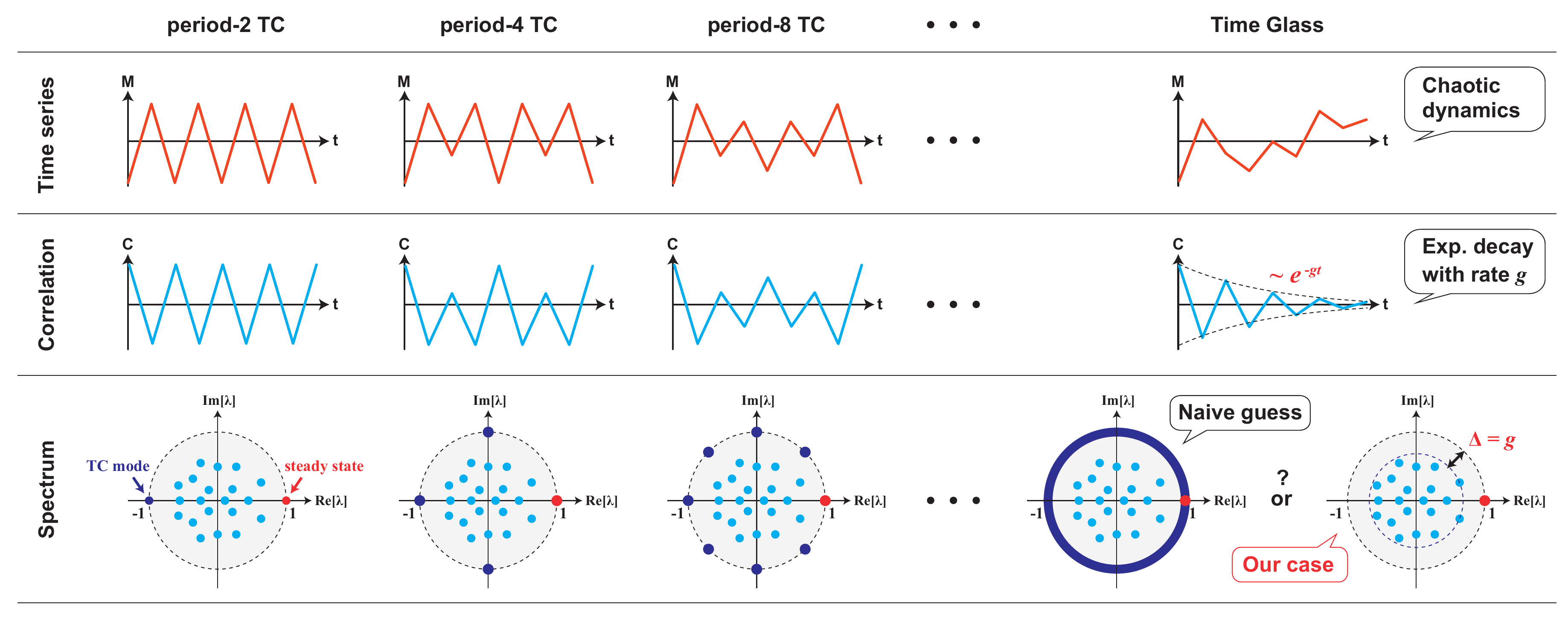}
\caption{Relationship between the dynamics of the order parameter and the spectrum of the time evolution map.
The first row shows the time evolution of the order parameter $M$ for a period-2 time crystal, a period-4 time crystal, a period-8 time crystal, and a time glass.
The second row shows the corresponding autocorrelations of $M$, and the third row shows the spectra of the time evolution map $\mathcal{U}$.
In the time crystal, the autocorrelation of $M$ shares the same period as its time-series oscillations.
By contrast, in the time glass, the autocorrelation decays exponentially with a rate $g$.
From a spectral perspective, a $p$-period time crystal arises when there are non-decaying eigenmodes with eigenvalues $\lambda_\alpha = e^{2 n \pi / p} \: (n = 0, 1, \ldots, p-1)$.
For a time glass, although one might naively expect infinitely many eigenvalues to cluster on the unit circle, our results show this is not the case.
Instead, a finite gap $\Delta$ appears, matching the decay rate $g$ of the classical autocorrelation.}
\label{fig_spectrum_schematic}
\end{figure*}

Before discussing the spectral characteristics of the time glass, let us first review those of time crystals.
Figure \ref{fig_spectrum_schematic} illustrates examples of period-2, period-4, and period-8 time crystals. 
In the top row, we show the time evolution of the order parameter $M(t)$ in the thermodynamic limit.
The second row shows the corresponding autocorrelation $C_M^\infty(t)$.
Note that the autocorrelation of $M(t)$ oscillates with the same period as the order parameter itself.

The spectral features of these time crystals are illustrated in the third row of Fig.~\ref{fig_spectrum_schematic}.
Each dot represents an eigenvalue $\lambda_\alpha$ of the time evolution map $\mathcal{U}$, which satisfies $|\lambda_\alpha| \leq 1$.
The red dot at $\lambda=1$ corresponds to the steady state.
For a period-$p$ time crystal, there are eigenvalues $\lambda_\alpha = e^{2 n \pi / p} \: (n = 0, 1, \ldots, p-1)$ on the unit circle \cite{Gong-18, Riera-Campeny-20}, which are indicated by thick blue dots.
The existence of such non-decaying eigenmodes means that the Liouvillian gap $\Delta$ vanishes in the thermodynamic limit.

Let us now consider the spectrum associated with the time glass.
In many dynamical systems, such as the logistic map, chaos emerges through a period-doubling route, where the system successively transitions to $2^n$-period states.
As more period-doubling occurs, additional eigenvalues appear on the unit circle.
Consequently, one might be tempted to assume that in the time glass, infinitely many eigenvalues would accumulate on the unit circle, as illustrated in Fig.~\ref{fig_spectrum_schematic} as ``Naive guess", causing the Liouvillian gap $\Delta$ to close in the thermodynamic limit.
Indeed, several recent studies adopt this perspective \cite{Carollo-24, Solanki-24-2}.

However, our results show that this intuition does not hold for time glasses whose emergent classical dynamics is \emph{ergodic}.
Instead, we find that the Liouvillian gap $\Delta$ remains nonzero in the thermodynamic limit and, moreover, converges to the classical mixing rate $g$ (see Fig.~\ref{fig_spectrum_schematic}):
\begin{equation}
\lim_{N \to \infty} \Delta = - \lim_{t \to \infty} \frac{1}{t} \log |C_{\mathrm{cl}}(t)|,
\label{gap_mixing_rate}
\end{equation}
where $C_{\mathrm{cl}}(t)$ is the classical autocorrelation defined by Eq.~\eqref{classical_autocorrelation_general}.
This correspondence is the central result of our work, establishing a direct connection between microscopic properties of the quantum master equation and the classical nonlinear behavior that emerges at the macroscopic level.

This result can be understood directly from the structure of the quantum autocorrelation function in Eq.~\eqref{C_general_def}.  
Suppose that \( \hat{M} \hat{\rho}_{\mathrm{ss}} \) is expanded in terms of the eigenmodes \( \hat{\rho}_\alpha \) of the Floquet map \( \mathcal{U} \) as  
\begin{equation}
\hat{M} \hat{\rho}_{\mathrm{ss}} = \sum_{\alpha>0} c_\alpha \hat{\rho}_\alpha.
\end{equation}
Note that the expansion does not contain the steady-state mode \(  \hat{\rho}_{\mathrm{ss}} \) because \( \mathrm{Tr}[\hat{M} \hat{\rho}_{\mathrm{ss}}] = 0 \) and \( \mathrm{Tr}[\hat{\rho}_{\mathrm{ss}}] \neq 0 \).
Substituting this into Eq.~\eqref{C_general_def}, we obtain  
\begin{equation}
C_M(nT) = \sum_{\alpha>0} c_\alpha (\lambda_\alpha)^n \mathrm{Tr}[\hat{M} \hat{\rho}_\alpha].
\end{equation}
Let \( \lambda_1 \) be the eigenvalue with the second largest modulus.  
For sufficiently large \( n \), the autocorrelation is dominated by this single mode:
\begin{equation}
|C_M(nT)| \simeq |c_1| |\lambda_1|^n |\mathrm{Tr}[\hat{M} \hat{\rho}_1]| \sim e^{-nT \Delta},
\end{equation}
which gives
\begin{equation}
\Delta = - \lim_{t \to \infty} \frac{1}{t} \log |C_M(t)|.
\end{equation}
Taking the thermodynamic limit, we obtain
\begin{equation}
\lim_{N \to \infty} \Delta = - \lim_{N \to \infty} \lim_{t \to \infty} \frac{1}{t} \log |C_M(t)|.
\label{gap_thermodynamic_limit}
\end{equation}
If the limits $N \to \infty$ and $t \to \infty$ in the right-hand side commute, \footnote{If the convergence of $\log |C_M(t)|/t$ to $\log |C_{\mathrm{cl}}(t)|/t$ is \emph{uniform} for all $t \in [0, \infty)$, then the limits $N \to \infty$ and $t \to \infty$ in Eq.~\eqref{gap_thermodynamic_limit} commute.} our main result Eq.~\eqref{gap_mixing_rate} follows from the quantum-classical correspondence of autocorrelations in Eq.~\eqref{autocorrelation_quantum_classical_correspondence}.

A distinctive feature of the time glass is that it retains a finite gap even in the presence of spontaneous symmetry breaking.
This behavior is highly unusual, as symmetry breaking in many-body systems is typically accompanied by gap closing.
For example, in the transverse-field Ising model, the energy gap between the ground state, given by the symmetric superposition of all-up and all-down states, and the first excited state, the corresponding antisymmetric superposition, vanishes exponentially with increasing system size in the symmetry-broken phase \cite{Sachdev}.
Similarly, in open quantum systems described by a quantum master equation, it is well established that the Liouvillian gap closes when the steady state exhibits symmetry breaking \cite{Kessler-12, Honing-12, Horstmann-13, Casteels-16, Casteels-17, Fitzpatrick-17, Vicentini-18, Minganti-18, Imamoglu-18, Rota-18, Ferreira-19}.
In contrast, the time glass phase uniquely combines symmetry breaking with a nonzero gap, marking a universal hallmark of this dynamical phase.

It is important to emphasize that the ergodicity of the emergent classical dynamics is essential for the existence of a finite gap in the time glass phase.
Consider instead the case in which the classical dynamics is nonergodic and possesses two distinct attractors, $\mathcal{A}_1$ and $\mathcal{A}_2$.
In such a situation, the quantum steady state $\hat{\rho}_{\mathrm{ss}}$ typically becomes a superposition of two distributions supported on $\mathcal{A}_1$ and $\mathcal{A}_2$, both with positive weight.
The first excited mode $\hat{\rho}_1$, by contrast, is given by the corresponding superposition with opposite signs.
The Liouvillian gap separating these two modes reflects the inverse of the tunneling time required for the system to transition between the two attractors.
As the system size increases, quantum fluctuations become increasingly suppressed, making such tunneling events exponentially rare.
Consequently, in this nonergodic case, the Liouvillian gap $\Delta$ is expected to vanish in the thermodynamic limit.

We here briefly remark on the distinction between time glasses and time quasicrystals \cite{Dumitrescu-18, Autti-18, Huang-18, Giergiel-18, Flicker-18, Giergiel-19, Zhao-19, Pizzi-19, He-25}.
A time quasicrystal is characterized by quasiperiodic (incommensurate) oscillations of macroscopic observables.
Its temporal signal can be written as a finite sum of harmonics \( M(t)=\sum_{j=1}^{d}A_j \cos(\omega_j t+\varphi_j) \), where the frequency ratios \( \omega_i / \omega_j \) are mutually irrational.
In systems that are driven periodically or quasiperiodically, a necessary condition for the spontaneous breaking of time-translational symmetry is that the response frequencies \( \omega_j \) differ from those of the external driving.
It is important to note that the Fourier spectrum of a time quasicrystal consists of discrete Bragg-like peaks, whereas time glasses exhibit a broad and continuous spectrum.
As a result, autocorrelations in time quasicrystals do not decay with time.
Moreover, in the case of time quasicrystals, the Floquet map \( \mathcal{U} \) possesses unit-modulus eigenvalues associated with the frequencies \( \{ \omega_j \} \) in the thermodynamic limit.
Consequently, the corresponding Liouvillian gap closes as the system size increases.

\subsection{Relaxation time}

In any finite dissipative system governed by a quantum master equation, the density matrix $\hat{\rho}(t)$ eventually settles into a time-independent steady state $\hat{\rho}_{\text{ss}}$ after a sufficiently long time.
We denote this relaxation time as $\tau_{\text{rel}}$.
In the time crystal phase, $\tau_{\text{rel}}$ corresponds to the duration for which periodic oscillations of $\hat{\rho}(t)$ persist.
Typically, $\tau_{\text{rel}}$ grows exponentially with the system size $N$, i.e., $\tau_{\text{rel}} \propto e^{cN}$ for some constant $c$ \cite{Gong-18}.
This exponential divergence of $\tau_{\text{rel}}$ in time crystals reflects the fact that the Liouvillian gap $\Delta$ vanishes exponentially with $N$, as $\Delta \propto e^{-cN}$.

Let us consider the case of time glasses, where $\tau_{\text{rel}}$ corresponds to the duration for which chaotic oscillations of $\hat{\rho}(t)$ persist.
In Sec.~\ref{sec:relaxation_time}, we show that $\tau_{\text{rel}}$ grows logarithmically with the system size: $\tau_{\text{rel}} \propto \log N$.
This logarithmic scaling precisely parallels the Ehrenfest time, which is the timescale up to which quantum expectation values reliably track the corresponding classical trajectories, and which grows logarithmically with the inverse of the effective Planck constant.
At first glance, the divergence of $\tau_{\text{rel}}$ in the thermodynamic limit may seem to conflict with the existence of a finite Liouvillian gap $\Delta$, because a nonzero gap implies that all non-steady eigenmodes decay within a time scale of $1/\Delta$.

This apparent paradox can be resolved by noting that while $\Delta$ governs the asymptotic convergence to the steady state after a sufficient long period, it does not necessarily determine how long the initial transient dynamics last.
More concretely, in Eq.~\eqref{eigenmode_expansion}, if the expansion coefficients $c_\alpha$ grow with $N$, then the relaxation time can diverge despite a nonzero gap \cite{Song-19, Mori-20, Haga-21, Bensa-21, Mori-23}.
The magnitude of $c_\alpha$ can be linked to the distance between the initial density matrix and the steady state.
We show that this distance diverges for certain initial states that exhibit classical properties, causing $\tau_{\text{rel}}$ to diverge in time glasses.

\section{Models}
\label{sec:models}

In this section, we introduce the models studied throughout the paper.
We consider a periodically kicked system described by the time-dependent Hamiltonian
\begin{equation}
\hat{H}(t) = \hat{H}_0 + \hat{H}_1 \sum_{n=-\infty}^{\infty} \delta(t - n),
\label{kicked_Hamiltonian_general}
\end{equation}
where $\hat{H}_0$ represents the static part of the Hamiltonian, and $\hat{H}_1$ corresponds to an instantaneous kick applied at integer times.
For simplicity, we set the driving period $T=1$.
Additionally, we assume that the jump operators $\hat{L}_k$ appearing in the quantum master equation \eqref{master_equation_general} are time independent.
Let us define the Liouvillian $\mathcal{L}_0$ for the static part of the master equation as
\begin{equation}
\mathcal{L}_0(\hat{\rho}) = - i [\hat{H}_0, \hat{\rho}] + \sum_{k} \left( \hat{L}_k \hat{\rho} \hat{L}_k^\dag - \frac{1}{2} \{ \hat{L}_k^\dag \hat{L}_k, \hat{\rho} \} \right).
\end{equation}
Then, the Floquet map is given by
\begin{equation}
\mathcal{U}(\hat{\rho}) = e^{-i \hat{H}_1} e^{\mathcal{L}_0}(\hat{\rho}) e^{i \hat{H}_1}.
\end{equation}

\subsection{Kicked collective spin}

We begin by introducing the periodically kicked collective spin model.
The collective spin operators $\hat{S}^x$, $\hat{S}^y$, and $\hat{S}^z$ are defined as
\begin{equation}
\hat{S}^\alpha = \frac{1}{2} \sum_{i=1}^N \hat{\sigma}_i^\alpha \quad (\alpha = x ,y, z),
\label{def_collective_spin_operator}
\end{equation}
where $\hat{\sigma}_i^\alpha$ are Pauli matrices for the $i$th spin-$1/2$ particle.
The system evolves under a time-dependent Hamiltonian with a static part and a periodic kick:
\begin{equation}
\hat{H}_0 = \omega_z \hat{S}^z, \quad \hat{H}_1 = \frac{\omega_{xx}}{S} (\hat{S}^x)^2,
\label{Hamiltonian_collective_spin}
\end{equation}
where $S=N/2$ is the total spin, and $\omega_z$ and $\omega_{xx}$ are tunable parameters.
For the dissipative dynamics, we use a single jump operator $\hat{L} = \sqrt{\kappa/S} \hat{S}^-$, where $\hat{S}^\pm = \hat{S}^x \pm i\hat{S}^y$ are the collective raising and lowering operators.
The parameter $\kappa$ quantifies the dissipation strength.
The resulting quantum master equation takes the form:
\begin{equation}
\partial_t \hat{\rho} = - i [\hat{H}(t), \hat{\rho}] + \frac{\kappa}{S} \left( \hat{S}^- \hat{\rho} \hat{S}^+ - \frac{1}{2} \{ \hat{S}^+ \hat{S}^-, \hat{\rho} \} \right).
\label{master_equation_collective_spin}
\end{equation}
Within the full Hilbert space of dimension $2^N$, we focus on the subspace that is symmetric under permutations of the spin-$1/2$ particles. 
This symmetric sector has dimension $2S+1$, where $S=N/2$.
We refer to this model as the kicked collective spin.
The dissipative term in Eq.~\eqref{master_equation_collective_spin} is commonly used to describe the collective decay of atoms coupled to optical cavity modes \cite{Walls-78, Drummond-78, Puri-79, Hannukainen-18}.

Let us briefly discuss the symmetries of this model.
First, both the Hamiltonian $\hat{H}(t)$ and the jump operator $\hat{L}$ are invariant under permutations of the $N$ spin-$1/2$ particles.
This implies that the system possesses a strong symmetry \cite{Buca-12} under particle exchange.
Second, the model also exhibits a weak $\mathbb{Z}_2$ symmetry \cite{Buca-12} under a $\pi$-rotation about the $z$-axis.
Specifically, we define the unitary operator for this rotation as
\begin{equation}
\hat{V} := \exp \left( i\pi \hat{S}^z \right).
\end{equation}
It is straightforward to verify that both $\hat{H}_0$ and $\hat{H}_1$ commute with $\hat{V}$, i.e., $[\hat{H}_0, \hat{V}] = [\hat{H}_1, \hat{V}] = 0$.
Moreover, for the Liouvillian $\mathcal{L}_t$ corresponding to Eq.~\eqref{master_equation_collective_spin}, the following relation holds for any density matrix $\hat{\rho}$:
\begin{equation}
\mathcal{L}_t(\hat{V}^\dag \hat{\rho} \hat{V})=\hat{V}^\dag \mathcal{L}_t(\hat{\rho}) \hat{V}.
\end{equation}
This implies that the transformation $\mathcal{V}(\hat{\rho}) := \hat{V}^\dag \hat{\rho} \hat{V}$ commutes with the Floquet map $\mathcal{U}$.
Since the eigenvalues of $\mathcal{V}$ are $\pm 1$, the eigenmodes $\hat{\rho}_\alpha$ of $\mathcal{U}$ satisfy either $\mathcal{V}(\hat{\rho}_\alpha)=\hat{\rho}_\alpha$ or $\mathcal{V}(\hat{\rho}_\alpha)=-\hat{\rho}_\alpha$.
In particular, the steady state $\hat{\rho}_{\text{ss}}$ must satisfy $\mathcal{V}(\hat{\rho}_{\text{ss}})=\hat{\rho}_{\text{ss}}$ because $\mathcal{V}$ preserves the trace and $\text{Tr}[\hat{\rho}_{\text{ss}}] \neq 0$.
As a consequence, the expectation values of $\hat{S}^x$ and $\hat{S}^y$ in the steady state must vanish:
\begin{equation}
\langle \hat{S}^x \rangle_{\text{ss}} = \langle \hat{S}^y \rangle_{\text{ss}} = 0.
\end{equation}

\subsection{Kicked spin chain}

As a second model, we consider a one-dimensional lattice of $N$ spin-$1/2$ particles. 
The Hamiltonian consists of a static part and a periodic kick, given respectively by
\begin{equation}
\hat{H}_0 = \frac{\omega_z}{2} \sum_{i=1}^N \hat{\sigma}_i^z, \quad \hat{H}_1 = \sum_{i,j=1}^N J_{ij} \hat{\sigma}_i^x \hat{\sigma}_j^x,
\label{Hamiltonian_spin_chain}
\end{equation}
where $\hat{\sigma}_i^\alpha$ are the Pauli matrices acting on the $i$th site of the lattice.
The coupling constants $J_{ij}$ represent long-range interactions that decay algebraically with distance, controlled by an exponent $\alpha$:
\begin{equation}
J_{ij} = 
\begin{cases}
\frac{J}{C_{N, \alpha} (r_{ij})^\alpha} & (i \neq j), \\
0 & (i=j),
\end{cases}
\label{spin_chain_interaction}
\end{equation}
where $r_{ij}$ denotes the shortest distance between sites $i$ and $j$ on a ring (periodic boundary conditions):
\begin{equation}
r_{ij} = \min (|i-j|, |i-j+N|, |i-j-N|).
\end{equation}
To ensure that the energy per spin remains finite in the thermodynamic limit, we introduce a normalization factor $C_{N, \alpha}$, defined as
\begin{equation}
C_{N, \alpha} = \sum_{i=2}^N \frac{1}{(r_{1i})^\alpha}.
\label{Kac_normalization}
\end{equation}

In this model, suppose that dissipation acts independently on each spin. 
Specifically, the jump operators are defined as $\hat{L}_i = \sqrt{\kappa} \hat{\sigma}_i^-$, where $\hat{\sigma}_i^\pm = \hat{\sigma}_i^x \pm i\hat{\sigma}_i^y$ are the spin raising and lowering operators for the $i$th spin.
The corresponding quantum master equation is given by
\begin{equation}
\partial_t \hat{\rho} = - i [\hat{H}(t), \hat{\rho}] +  \kappa \sum_{i=1}^N \left( \hat{\sigma}_i^- \hat{\rho} \hat{\sigma}_i^+ - \frac{1}{2} \{ \hat{\sigma}_i^+ \hat{\sigma}_i^-, \hat{\rho} \} \right).
\label{master_equation_spin_chain}
\end{equation}
As in the kicked collective spin model, this system exhibits a weak $\mathbb{Z}_2$ symmetry under a $\pi$-rotation about the $z$-axis.
However, in contrast to the collective spin case, the master equation \eqref{master_equation_spin_chain} is not symmetric under permutations of the $N$ spins, except in the special case of $\alpha=0$.

The interaction range in the model is controlled by the exponent $\alpha$.
In the limit of $\alpha \to \infty$, the interaction becomes short-ranged, effectively reducing to a nearest-neighbor coupling.
In this case, the normalization factor becomes $C_{N, \alpha}=2$, and the interaction Hamiltonian simplifies to
\begin{equation}
H_1 = J \sum_{i=1}^N \sigma_i^x \sigma_{i+1}^x.
\end{equation}
On the other hand, in the opposite limit $\alpha \to 0$, the interaction becomes fully long-ranged (all-to-all coupling), and the normalization factor approaches $C_{N, \alpha}=N-1 \simeq N$.
The interaction Hamiltonian then becomes
\begin{equation}
H_1 = \frac{J}{N} \sum_{i \neq j}^N \sigma_i^x \sigma_j^x = \frac{2J}{N} \sum_{i < j}^N \sigma_i^x \sigma_j^x.
\label{Hamiltonian_spin_chain_alpha_0}
\end{equation}
In this way, the exponent $\alpha$ effectively tunes the dimensionality of the system, interpolating between one-dimensional local interactions and infinite-range mean-field-like interactions.

\section{Classical dynamics}
\label{sec:classical_dynamics}

In this section, we examine the classical dynamics of the models introduced previously. 
In the thermodynamic limit, the expectation values of global spin observables evolve according to a closed set of nonlinear equations. 
Depending on the system parameters, these equations admit limit-cycle solutions or exhibit chaotic behavior, corresponding to the time crystal and time glass phases, respectively. 
Throughout this work, we use the term ``classical dynamics" to refer broadly to the deterministic dynamics of macroscopic observables, irrespective of whether the system admits a formal classical limit.

\subsection{Kicked collective spin}

\begin{figure}
\centering
\includegraphics[width=8.6cm]{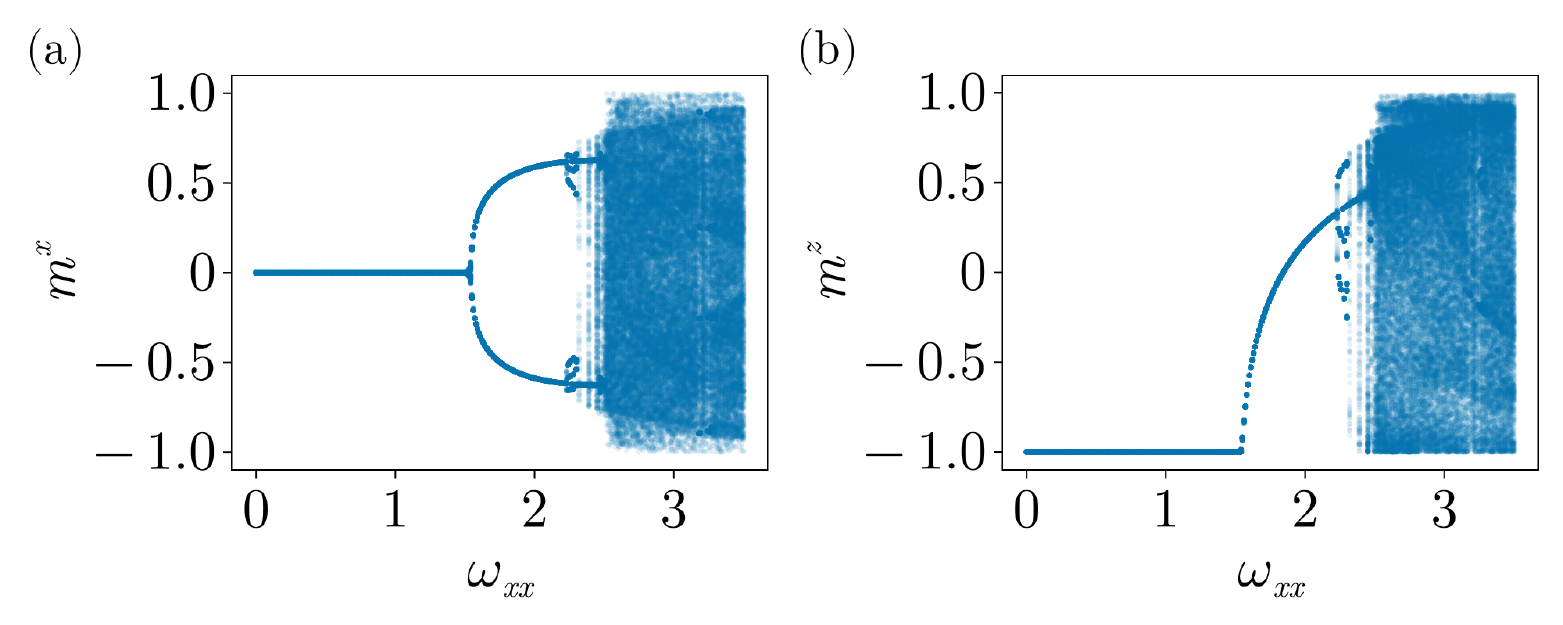}
\caption{Bifurcation diagrams for the classical dynamics of the kicked collective spin model.
Panels (a) and (b) show the long-time values of $m^x$ and $m^z$, respectively, plotted as functions of the kick strength $\omega_{xx}$, after discarding transient dynamics over the first $100$ cycles.
The parameters are fixed at $\omega_z=\pi/2$ and $\kappa=1$.
As $\omega_{xx}$ increases, the system undergoes a sequence of transitions from a stable fixed point, to a limit cycle, and eventually to chaotic behavior.}
\label{fig_collective_spin_bifurcation}
\end{figure}

We discuss the classical dynamics of the kicked collective spin model that emerges in the large-spin limit $S \to \infty$.
To this end, we introduce normalized spin variables defined by $m^\nu = \langle \hat{S}^\nu \rangle / S$, where $\nu=x, y, z$.
In the limit $S \to \infty$, quantum fluctuations of the spin operators become negligible, and factorization approximations such as $\langle S^x S^y \rangle \simeq \langle S^x \rangle \langle S^y \rangle$ become valid.
This allows us to derive a closed set of nonlinear equations governing the dynamics of $m^\nu$.

Under the static part of the master equation \eqref{master_equation_collective_spin}, the time evolution of $m^\nu$ is given by
\begin{equation}
\begin{split}
\frac{dm^x}{dt} &= - \omega_z m^y + \kappa m^x m^z, \\
\frac{dm^y}{dt} &= \omega_z m^x + \kappa m^y m^z, \\
\frac{dm^z}{dt} &= - \kappa [(m^x)^2 + (m^y)^2].
\end{split}
\label{collective_spin_classical_equation_1}
\end{equation}
These equations conserve the spin magnitude: $(m^x)^2+(m^y)^2+(m^z)^2=1$.
The terms involving $\omega_z$ describe coherent precession around the $z$-axis, while the dissipative terms proportional to $\kappa$ drive the spin components toward the negative $z$-direction.
Within this classical picture, the up-spin state $\boldsymbol{m}=(0, 0, 1)$ is an unstable fixed point, whereas the down-spin state $\boldsymbol{m}=(0, 0, -1)$ is a stable fixed point.

We next consider the dynamics induced by the periodic kick, represented by the unitary transformation $\hat{\rho} \to e^{-i \hat{H}_1} \hat{\rho} e^{i \hat{H}_1}$.
It is convenient to introduce a fictitious continuous time parameter $t$ and treat the kick as a unitary evolution over the interval $t \in [0, 1]$, governed by the master equation $\partial_t \hat{\rho} = -i [\hat{H}_1, \hat{\rho}]$. 
To compute the effect of a single kick, we integrate this equation from $t=0$ to $t=1$.
The corresponding classical equations of motion for the normalized spin components are given by
\begin{equation}
\begin{split}
\frac{dm^x}{dt} &= 0, \\
\frac{dm^y}{dt} &= -2\omega_{xx} m^x m^z, \\
\frac{dm^z}{dt} &= 2\omega_{xx} m^x m^y.
\end{split}
\label{collective_spin_classical_equation_2}
\end{equation}
These equations describe a rotation around the $x$-axis, where the angular velocity is proportional to $m^x$.
As a result, regions of the Bloch sphere with $m^x>0$ and $m^x<0$ rotate in opposite directions, inducing a twisting deformation about the $x$-axis.
This evolution also conserves the spin magnitude: $(m^x)^2+(m^y)^2+(m^z)^2=1$.
The complete one-cycle evolution of the classical spin is obtained by first integrating Eq.~\eqref{collective_spin_classical_equation_1} from $t=0$ to $t=1$, followed by integrating Eq.~\eqref{collective_spin_classical_equation_2} from $t=1$ to $t=2$.

Figure \ref{fig_collective_spin_bifurcation} shows bifurcation diagrams for the classical dynamics of the kicked collective spin model.
The parameters are fixed at $\omega_z=\pi/2$ and $\kappa=1$.
For each value of the kick strength $\omega_{xx}$, we compute the stroboscopic dynamics of $m^x$ and $m^z$ by integrating Eqs.~\eqref{collective_spin_classical_equation_1} and \eqref{collective_spin_classical_equation_2}, discarding the transient dynamics before plotting.
In the regime $0 \leq \omega_{xx} \lesssim 1.5$, the system relaxes to the down-spin state $\boldsymbol{m}=(0, 0, -1)$, which acts as a stable fixed point.
This behavior corresponds to a disordered phase.
A linear stability analysis, presented in Appendix \ref{sec:linear_stability_analysis}, shows that this fixed point becomes unstable at the critical kick strength $\omega_{xx}^c = (e^2+1)/2e \simeq 1.543$.
In the intermediate regime $1.5 \lesssim \omega_{xx} \lesssim 2.3$, the system undergoes spontaneous symmetry breaking, and a period-2 limit cycle emerges.
This corresponds to a time crystal phase.
As $\omega_{xx}$ increases further ($\omega_{xx} \gtrsim 2.3$), the system enters a chaotic regime, signaling the onset of a time glass phase.

\subsection{Kicked spin chain with all-to-all coupling}

\begin{figure}
\centering
\includegraphics[width=8.6cm]{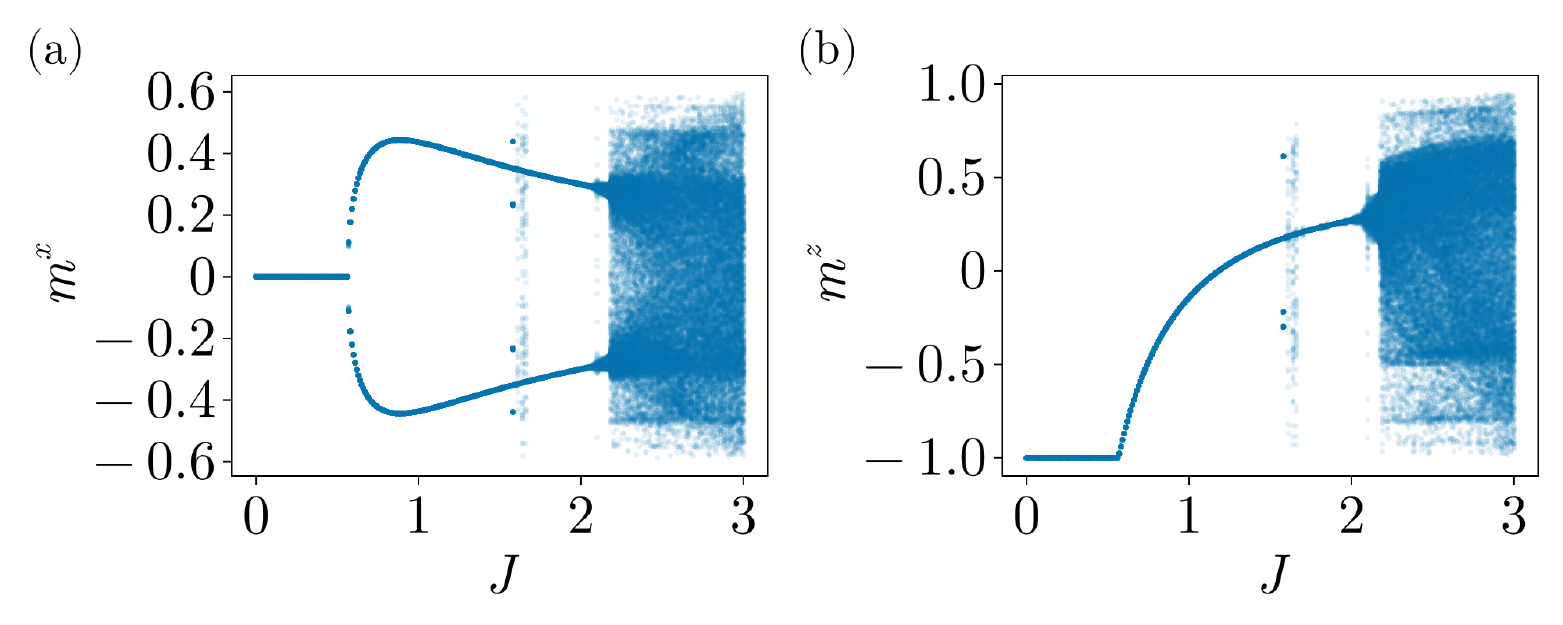}
\caption{Bifurcation diagrams for the classical dynamics of the kicked spin chain model with all-to-all coupling.
Panels (a) and (b) show the long-time values of $m^x$ and $m^z$, respectively, plotted as functions of the kick strength $J$, after discarding transient dynamics over the first $100$ cycles.
The parameters are fixed at $\omega_z=\pi/2$ and $\kappa=1$.
As $J$ increases, the system undergoes a sequence of transitions from a stable fixed point, to a limit cycle, and eventually to chaotic behavior.}
\label{fig_spin_chain_bifurcation}
\end{figure}

We now turn to the classical dynamics of the kicked spin chain in the case of all-to-all coupling, corresponding to the limit $\alpha=0$.
In this regime, the system becomes fully symmetric, and the total density matrix can be expressed as a tensor product of identical single-site density matrices:
\begin{equation}
\hat{\rho} = \otimes_{i=1}^N \hat{\chi}_i,
\label{density_matrix_factorization}
\end{equation}
where $\hat{\chi}_i$ denotes the reduced density matrix at site $i$.
The validity of the mean-field ansatz given by Eq.~\eqref{density_matrix_factorization} has been rigorously proved in Ref.~\cite{Carollo-24}.
The time evolution of each local density matrix is obtained by tracing out all degrees of freedom except for site $i$:
\begin{equation}
\partial_t \hat{\chi}_i = \mathrm{Tr}_{\neq i}[\partial_t \hat{\rho}],
\end{equation}
where $\mathrm{Tr}_{\neq i}$ denotes the partial trace over all sites except the $i$th.
Since all sites evolve identically, we omit the site index $i$ in $\hat{\chi}_i$ in the remainder of the analysis.

As in the kicked collective spin model, we divide the one-cycle evolution into two stages: the first half governed by the static Hamiltonian \(\hat{H}_0\), and the second half governed by the interaction Hamiltonian \(\hat{H}_1\). 
The master equation for the first half is given by
\begin{equation}
\partial_t \hat{\chi} = - \frac{i \omega_z}{2} [\hat{\sigma}^z, \hat{\chi}] + \kappa \left( \hat{\sigma}^- \hat{\chi} \hat{\sigma}^+ - \frac{1}{2} \{ \hat{\sigma}^+ \hat{\sigma}^-, \hat{\chi} \} \right),
\label{master_equation_spin_chain_mean_field_dissipation}
\end{equation}
while the second half is governed by
\begin{equation}
\partial_t \hat{\chi} = - i 2J m^x [\hat{\sigma}^x, \hat{\chi}],
\label{master_equation_spin_chain_mean_field_kick}
\end{equation}
where \(m^x = \mathrm{Tr}[\hat{\sigma}^x \hat{\chi}]\) denotes the expectation value of the local spin in the \(x\)-direction.
Expressing these equations in terms of the spin expectation values \(m^\nu\) (\(\nu = x, y, z\)), Eqs.~\eqref{master_equation_spin_chain_mean_field_dissipation} and \eqref{master_equation_spin_chain_mean_field_kick} become
\begin{equation}
\begin{split}
\frac{dm^x}{dt} &= - \omega_z m^y - \frac{\kappa}{2} m^x, \\
\frac{dm^y}{dt} &= \omega_z m^x - \frac{\kappa}{2} m^y, \\
\frac{dm^z}{dt} &= - \kappa (m^z + 1),
\end{split}
\label{spin_chain_mean_field_dissipation}
\end{equation}
and
\begin{equation}
\begin{split}
\frac{dm^x}{dt} &= 0, \\
\frac{dm^y}{dt} &= -4J m^x m^z, \\
\frac{dm^z}{dt} &= 4J m^x m^y,
\end{split}
\label{spin_chain_mean_field_kick}
\end{equation}
respectively.
These dynamics preserve the bound \((m^x)^2 + (m^y)^2 + (m^z)^2 \leq 1\), reflecting the positivity of the density matrix. 
The down-spin state \(\boldsymbol{m} = (0, 0, -1)\) is a stable fixed point. 
Notably, unlike in the kicked collective spin model, the up-spin state \(\boldsymbol{m} = (0, 0, 1)\) is not a fixed point in this system.

Figure \ref{fig_spin_chain_bifurcation} shows bifurcation diagrams for the classical dynamics of the kicked spin chain model.
The parameters are fixed at $\omega_z=\pi/2$ and $\kappa=1$.
For each value of the kick strength $J$, we compute the stroboscopic dynamics of $m^x$ and $m^z$ by integrating Eqs.~\eqref{spin_chain_mean_field_dissipation} and \eqref{spin_chain_mean_field_kick}, discarding the transient dynamics before plotting.
In the regime $0 \leq \omega_{xx} \lesssim 0.56$, the system relaxes to the down-spin state $\boldsymbol{m}=(0, 0, -1)$, which acts as a stable fixed point.
This behavior corresponds to a disordered phase.
A linear stability analysis, presented in Appendix \ref{sec:linear_stability_analysis}, shows that this fixed point becomes unstable at the critical kick strength $J_c = (e+1)/4e^{1/2} \simeq 0.564$.
In the intermediate regime $0.56 \lesssim J \lesssim 2.1$, the system undergoes spontaneous symmetry breaking, and a period-2 limit cycle emerges.
This corresponds to a time crystal phase.
As $J$ increases further ($J \gtrsim 2.1$), the system enters a chaotic regime, signaling the onset of a time glass phase.
It should be noted that for \( J \gtrsim 1.6 \), the system exhibits a long chaotic transient before the trajectory eventually settles into a limit cycle. 
This behavior appears as a faint cloud near \( J = 1.6 \) in Fig.~\ref{fig_spin_chain_bifurcation}, reflecting the transiently irregular dynamics.

\section{Autocorrelation diagnostics}
\label{sec:autocorrelation}

While both time crystals and time glasses exhibit spontaneous symmetry breaking, they can be clearly distinguished by examining the autocorrelation function of the order parameter, as illustrated in Fig.~\ref{fig_autocorrelation_schematic}. 
In terms of the Floquet map \(\mathcal{U}\), the Heisenberg representation of an observable \(\hat{A}\) at discrete time \(t = 0, 1, \ldots\) is given by \(\hat{A}(t) = (\mathcal{U}^\dag)^t(\hat{A})\).
We here focus on the real part of the autocorrelation function for the observable \(\hat{A}\),
\begin{equation}
C_A(t) = \text{Re} (\text{Tr}[\hat{A}(t) \hat{A} \hat{\rho}_{\mathrm{ss}}]) = \text{Re} (\text{Tr}[\hat{A} \, \mathcal{U}^t (\hat{A} \hat{\rho}_{\mathrm{ss}})]),
\label{Re_C_def}
\end{equation}
because its imaginary part is expected to vanish in the thermodynamic limit.
In this section, we demonstrate that the behavior of the autocorrelation function serves as a clear and consistent diagnostic for distinguishing between time crystal and time glass phases, in agreement with the classical dynamics described in Sec.~\ref{sec:classical_dynamics}.

\subsection{Kicked collective spin}

\begin{figure}
\centering
\includegraphics[width=8.6cm]{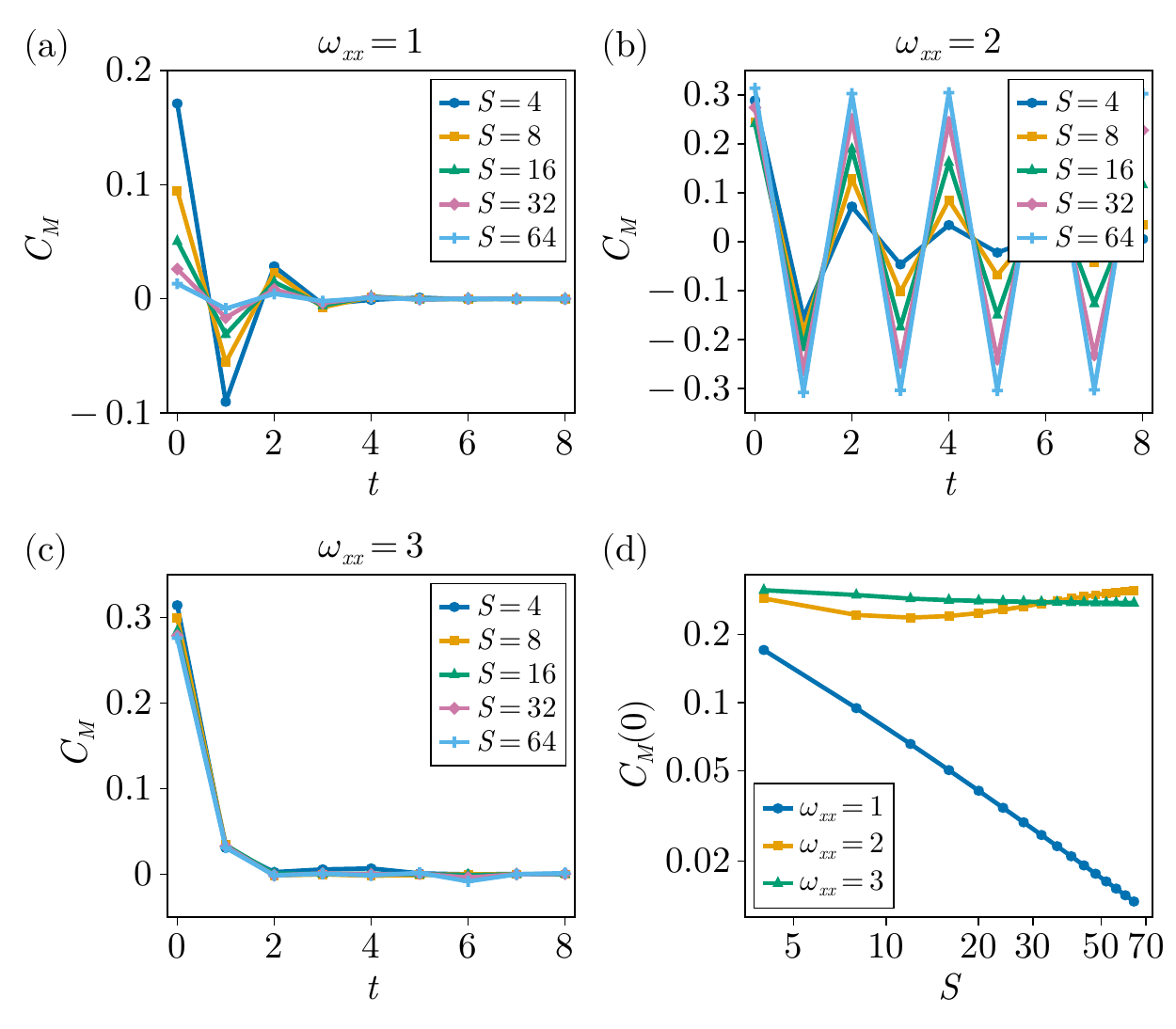}
\caption{Autocorrelation functions \( C_M(t) \) of the order parameter for the kicked collective spin model. 
Panels (a), (b), and (c) correspond to kick strengths \(\omega_{xx} = 1\), \(2\), and \(3\), representing the disordered phase, time crystal phase, and time glass phase, respectively.
The parameters are fixed at $\omega_z=\pi/2$ and $\kappa=1$.
Panel (d) shows \( C_M(0) \) as a function of the total spin \(S\), plotted on a double-logarithmic scale. 
The finite values of \( C_M(0) \) in the limit \( S \to \infty \) for \(\omega_{xx} = 2\) and \(3\) indicate the presence of spontaneous symmetry breaking in both the time crystal and time glass phases.}
\label{fig_collective_spin_autocorrelation}
\end{figure}

We investigate the behavior of the autocorrelation function for the kicked collective spin model. 
The order parameter is defined as \(\hat{M} = \hat{S}^x / S\), and its autocorrelation function \(C_M(t)\) is computed using Eq.~\eqref{Re_C_def} with \(\hat{A}\) replaced by \(\hat{M}\). 
Our primary interest lies in the large-spin limit of the autocorrelation, defined as
\begin{equation}
C_M^\infty(t) = \lim_{S \to \infty} C_M(t).
\end{equation}

Figure~\ref{fig_collective_spin_autocorrelation} shows \( C_M(t) \) for various values of \( S \).  
Panels (a), (b), and (c) correspond to kick strengths \( \omega_{xx} = 1 \), \( 2 \), and \( 3 \), respectively. 
According to the bifurcation diagram in Fig.~\ref{fig_collective_spin_bifurcation}, these values correspond to the disordered phase, time crystal phase, and time glass phase, respectively.
In the disordered phase (\( \omega_{xx} = 1 \)), the autocorrelation \( C_M(t) \) decays to zero for all \( t \), following \( C_M(t) \propto S^{-1} \).  
This decay is evident in Fig.~\ref{fig_collective_spin_autocorrelation}(d), which plots \( C_M(0) \) as a function of \( S \).  
As a result, we conclude that \( C_M^\infty(t) = 0 \) for all \( t \).
In the time crystal phase (\( \omega_{xx} = 2 \)), \( C_M(t) \) exhibits periodic oscillations whose decay timescale increases with \( S \).  
Thus, \( C_M^\infty(t) \) exhibits persistent oscillations in time.
In the time glass phase (\( \omega_{xx} = 3 \)), \( C_M(t) \) remains nonzero at small \( t \), but decays to zero as \( t \) increases.  
Figure~\ref{fig_collective_spin_autocorrelation}(d) confirms that \( C_M^\infty(0) > 0 \) for both the time crystal and time glass phases, indicating the presence of spontaneous symmetry breaking in both cases.

\subsection{Kicked spin chain}

\begin{figure}
\centering
\includegraphics[width=8.6cm]{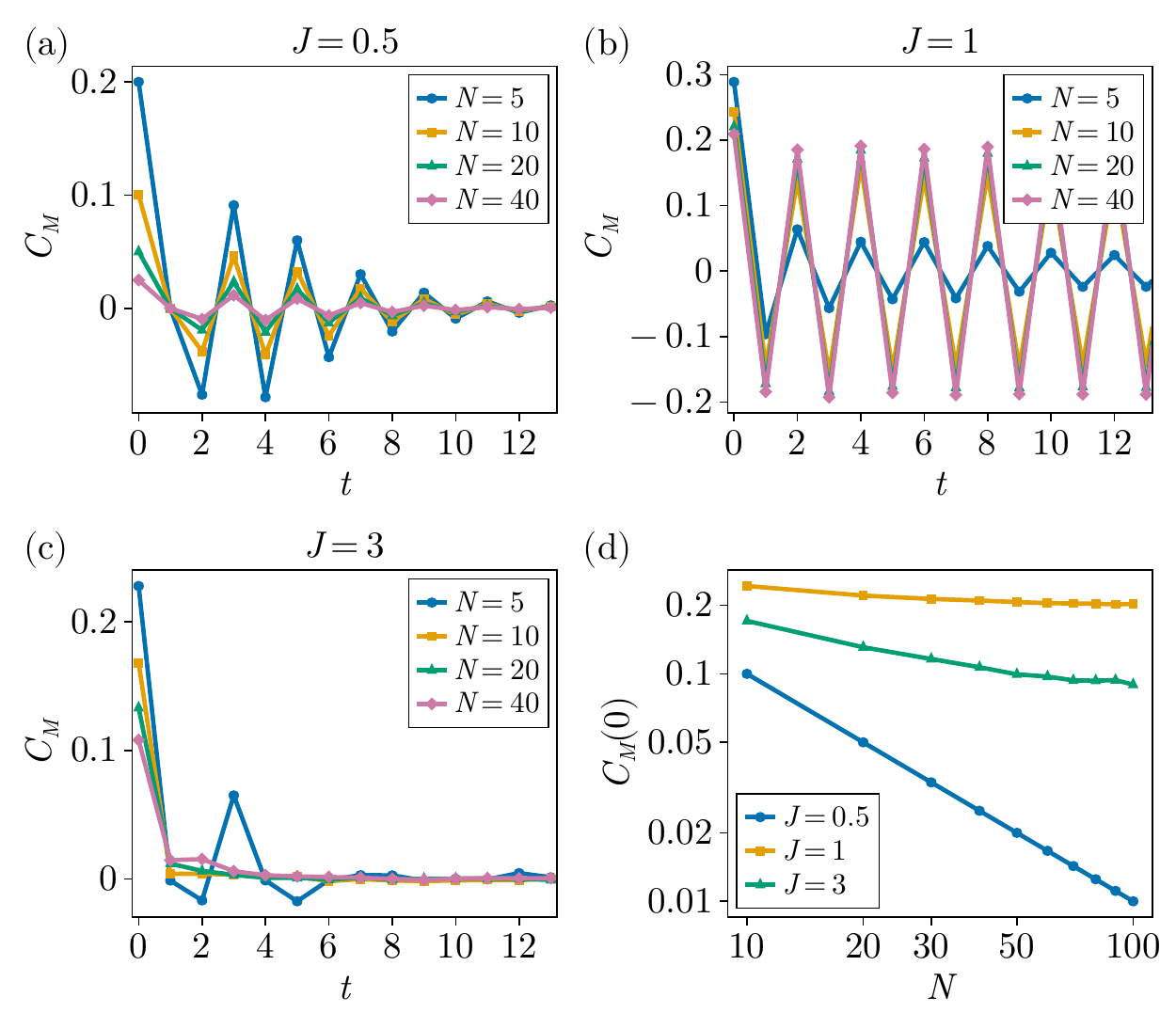}
\caption{Autocorrelation functions \( C_M(t) \) of the order parameter for the kicked spin chain model with all-to-all coupling. 
Panels (a), (b), and (c) correspond to kick strengths \(J = 0.5\), \(1\), and \(3\), representing the disordered phase, time crystal phase, and time glass phase, respectively.
The parameters are fixed at $\omega_z=\pi/2$ and $\kappa=1$.
The errors are comparable to the size of the symbols.
Panel (d) shows \( C_M(0) \) as a function of \(N\), plotted on a double-logarithmic scale. 
The finite values of \( C_M(0) \) in the limit \( N \to \infty \) for \(J = 1\) and \(3\) indicate the presence of spontaneous symmetry breaking in both the time crystal and time glass phases.}
\label{fig_spin_chain_autocorrelation}
\end{figure}

\begin{figure*}
\centering
\includegraphics[width=\textwidth]{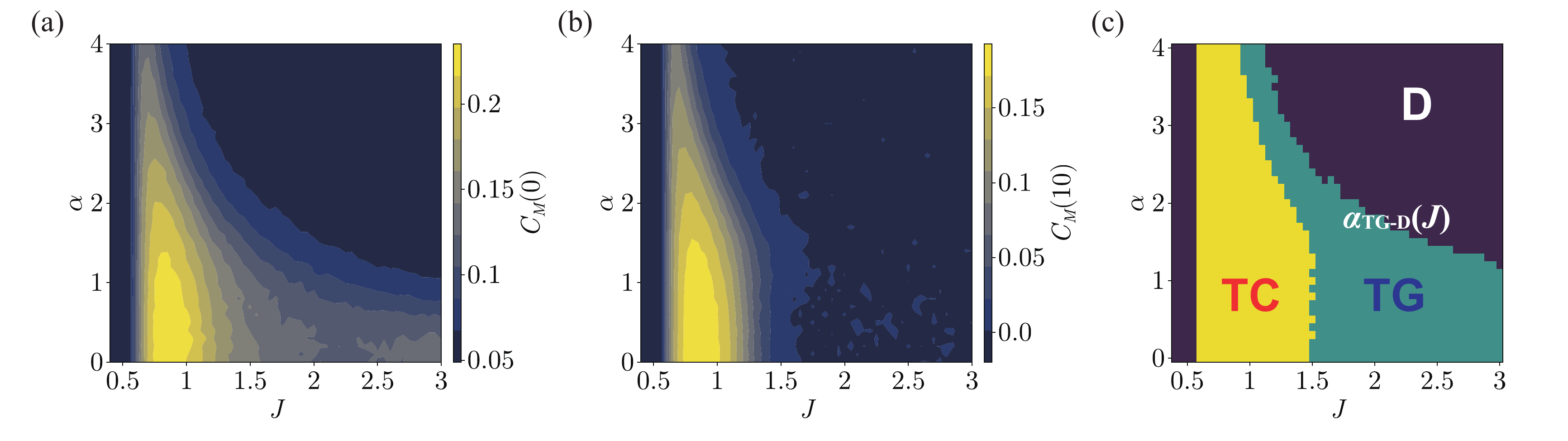}
\caption{Phase diagram of the kicked spin chain model as a function of the kick strength \( J \) and the interaction exponent \( \alpha \).  
The number of spins is fixed at \( N = 20 \), and the parameters are set to \( \omega_z = \pi/2 \) and \( \kappa = 1 \).  
(a) \( C_M(0) \) plotted as a function of \( J \) and \( \alpha \).  
(b) \( C_M(t_{\text{large}}) \) plotted as a function of \( J \) and \( \alpha \), where the lag time is set to \( t_{\text{large}} = 10 \).  
The disordered phase is identified by \( C_M(0) = C_M(t_{\text{large}}) = 0 \).  
The time crystal phase is identified by \( C_M(0) \neq 0 \) and \( C_M(t_{\text{large}}) \neq 0 \).  
The time glass phase is identified by \( C_M(0) \neq 0 \) and \( C_M(t_{\text{large}}) = 0 \).  
(c) Phase diagram for the disordered (D), time crystal (TC), and time glass (TG) phases.
The disordered phase (black) is defined by \( C_M(0) < 0.06 \).  
The time crystal phase (yellow) is defined by \( C_M(0) > 0.06 \) and \( C_M(t_{\text{large}}) > 0.01 \).  
The time glass phase (green) is defined by \( C_M(0) > 0.06 \) and \( C_M(t_{\text{large}}) < 0.01 \).
The phase boundary between the disordered phase and the time glass phase is indicated by \( \alpha_{\mathrm{TG-D}}(J) \).}
\label{fig_spin_chain_phase_diagram}
\end{figure*}

We next investigate the behavior of the autocorrelation function for the kicked spin chain model.  
In this case, direct integration of the quantum master equation~\eqref{master_equation_spin_chain} is computationally challenging due to the exponential growth of the Hilbert space dimension with system size.  
To overcome this difficulty, we employ the quantum trajectory method combined with a Gutzwiller-type approximation.

The quantum trajectory method represents the dissipative dynamics as an ensemble average over many stochastic realizations (``trajectories'') of pure-state wave functions \(\ket{\psi(t)}\) \cite{Dalay-14}, rather than evolving the full density matrix \(\hat{\rho}(t)\). 
The detailed procedure for computing the correlation function within the quantum trajectory framework is provided in Appendix~\ref{sec:quantum_trajectory_method}.  
To further reduce the computational cost, we employ a Gutzwiller-type approximation for the many-body pure state \(\ket{\psi(t)}\), assuming a factorized form:  
\begin{equation}
\ket{\psi(t)} \simeq \ket{\phi_1(t)} \otimes \ket{\phi_2(t)} \otimes \cdots \otimes \ket{\phi_N(t)},
\label{phi_product_approximation}
\end{equation}
where \(\ket{\phi_i(t)}\) denotes the local pure state at site \(i\).  
This approximation neglects entanglement between different sites. 
We expect that the inter-site entanglement is negligible for long-range interacting systems.
As a result, the computational complexity scales linearly with the number of particles \(N\), making it feasible to simulate large systems.

It is important to note that the tensor product ansatz \eqref{phi_product_approximation} assumed at the level of quantum trajectories differs from the mean-field approximation based on the tensor product form of the density matrix, Eq.~\eqref{density_matrix_factorization}, discussed in the previous section \cite{Dalay-14}.  
In what follows, we refer to the former as the \textit{Quantum Trajectory Gutzwiller Approximation} (QTGA), and the latter as the \textit{Density Matrix Gutzwiller Approximation} (DMGA).  
In QTGA, while quantum entanglement between sites is neglected, it can still capture inter-site correlations arising from classical fluctuations.  
In contrast, DMGA ignores all forms of inter-site correlations.  
The difference between these two approaches becomes particularly evident when applied to finite systems in the time crystal phase [see Fig.~\ref{fig_spin_chain_autocorrelation}(b)].  
Under QTGA, the expectation value of the spin exhibits damped oscillations that eventually relax to a steady value.  
The timescale of this decay grows with the system size \(N\).  
On the other hand, DMGA describes a single-site system evolving under a mean field, and thus no system-size dependence appears.
In this case, the spin expectation value continues to oscillate indefinitely.
Because QTGA partially incorporates the effects of fluctuations, it provides a useful framework for testing whether the behavior observed in DMGA is robust against such fluctuations.

For the kicked spin chain, the autocorrelation function \(C_M(t)\) of the order parameter is defined by Eq.~\eqref{Re_C_def} with \(\hat{A}\) replaced by \(\hat{M}= \sum_i \hat{\sigma}_i^x / N\). 
Our primary interest lies in the thermodynamic limit of the autocorrelation, defined as
\begin{equation}
C_M^\infty(t) = \lim_{N \to \infty} C_M(t).
\end{equation}
First, let us consider the case of all-to-all coupling (\( \alpha = 0\)).
Figure \ref{fig_spin_chain_autocorrelation} shows \(C_M(t)\) for various values of $N$.
In the calculation of \(C_M(t)\), we take the average over $1000$ trajectories.
Panels (a), (b), and (c) correspond to kick strengths \( J = 0.5 \), \( 1 \), and \( 3 \), respectively. 
According to the bifurcation diagram in Fig.~\ref{fig_spin_chain_bifurcation}, these values correspond to the disordered phase, time crystal phase, and time glass phase, respectively.
In the disordered phase (\( J = 0.5 \)), the autocorrelation \( C_M(t) \) decays to zero for all \( t \), following \( C_M(t) \propto N^{-1} \) [see Fig.~\ref{fig_spin_chain_autocorrelation}(d)]. 
As a result, we conclude that \( C_M^\infty(t) = 0 \) for all \( t \).
In the time crystal phase (\( J = 1 \)), \( C_M(t) \) exhibits periodic oscillations whose decay timescale increases with \( N \).  
Thus, \( C_M^\infty(t) \) exhibits persistent oscillations in time.
In the time glass phase (\( J = 3 \)), \( C_M(t) \) remains nonzero at small \( t \), but decays to zero as \( t \) increases.  
Figure~\ref{fig_spin_chain_autocorrelation}(d) confirms that \( C_M^\infty(0) > 0 \) for both the time crystal and time glass phases, indicating the presence of spontaneous symmetry breaking in both cases.

Beyond the mean-field case of all-to-all coupling (\( \alpha = 0\)), we consider the phase structure for general values of the interaction exponent \( \alpha \).  
As illustrated in Fig.~\ref{fig_autocorrelation_schematic}, the disordered, time crystal, and time glass phases can be identified using the following criteria:
\begin{itemize}
\item \textit{Disordered phase:} \( C_M^\infty(0) = C_M^\infty(\infty) = 0 \).
\item \textit{Time crystal phase:} \( C_M^\infty(0) \neq 0 \) and \( C_M^\infty(\infty) \neq 0 \).
\item \textit{Time glass phase:} \( C_M^\infty(0) \neq 0 \) and \( C_M^\infty(\infty) = 0 \).
\end{itemize}
In practice, we approximate \( C_M^\infty(\infty) \) by evaluating the autocorrelation function at a sufficiently large lag time.  
In our simulations, we set this lag time to \( t_{\text{large}} = 10 \).

Figures \ref{fig_spin_chain_phase_diagram}(a) and (b) show \( C_M(0) \) and \( C_M(t_{\text{large}}) \), respectively, as functions of the kick strength $J$ and the interaction exponent \( \alpha \).
The number of spins is $N=20$.
In the calculation of \(C_M(t)\), we take the average over $1000$ trajectories.
Following the classification criteria described earlier, the phase boundaries between the disordered phase (D), the time crystal phase (TC), and the time glass phase (TG) are shown in Fig.~\ref{fig_spin_chain_phase_diagram}(c).  
As the interaction exponent \( \alpha \) increases, the range of the interaction becomes shorter, making long-range order increasingly unstable.
In the limit \( \alpha \to \infty \), where only nearest-neighbor interactions remain, the system is expected to always be in the disordered phase.

First, let us consider the case of all-to-all coupling (\( \alpha = 0 \)).  
We observe that the transition between the disordered phase and the time crystal phase occurs near \( J \simeq 0.5 \) [see Fig.~\ref{fig_spin_chain_phase_diagram}(c)].  
This transition point coincides with the bifurcation point of the classical dynamics, as shown in Fig.~\ref{fig_spin_chain_bifurcation}.  
In addition, the transition between the time crystal and time glass phases occurs near \( J \simeq 1.5 \) [see Fig.~\ref{fig_spin_chain_phase_diagram}(c)].  
However, this transition point is smaller than the bifurcation point observed in the classical dynamics, around \( J \simeq 2.1 \), where global chaos onsets, as shown in Fig.~\ref{fig_spin_chain_bifurcation}.  
Instead, the value \( J \simeq 1.5 \) is close to the point where a long chaotic transient emerges, which appears as a faint cloud in Fig.~\ref{fig_spin_chain_bifurcation}.  
An important open question is how, in the thermodynamic limit, the transition point between the time crystal and time glass phases connects to the bifurcation structure of the classical dynamics.

In the all-to-all coupling case (\( \alpha = 0 \)), the system is always in the time glass phase for sufficiently large \( J \).  
However, when \( \alpha > 0 \) takes on a moderately large value, the system undergoes a phase transition from the time glass phase to the disordered phase at large \( J \).  
If we denote this phase boundary by \( \alpha_{\mathrm{TG-D}}(J) \), it remains unclear whether \( \lim_{J \to \infty} \alpha_{\mathrm{TG-D}}(J) \) is zero or finite.  
Since the normalization factor \eqref{Kac_normalization} diverges in the limit \( N \to \infty \) for \( 0 < \alpha \leq 1 \), models with \(\alpha\) in this range are expected to be qualitatively equivalent to the all-to-all coupling case.  
This suggests that \( \lim_{J \to \infty} \alpha_{\mathrm{TG-D}}(J) = 1 \) is a likely possibility.  
Additionally, note that the location of the phase transition between the disordered and time crystal phases near \( J = 0.5 \) is almost independent of \( \alpha \), which can be attributed to the normalization factor \eqref{Kac_normalization} properly scaling the interaction strength.

The fact that the time glass phase persists for \( \alpha > 1 \) suggests that this phase is stable in finite-dimensional systems with short-range interactions.  
The stability of the time glass phase under short-range interactions is also observed in a driven dissipative Dicke system (see Ref.~\cite{Zhu-19}, where the term ``irregular dynamics'' corresponds to the time glass in our terminology).
It is known that in classical one-dimensional Ising models with long-range interactions, long-range order at finite temperature does not exist for \( \alpha \geq 2 \).
On the other hand, Fig.~\ref{fig_spin_chain_phase_diagram} indicates that time crystal and time glass phases may still persist even for \( \alpha \geq 2 \), though further analysis is needed to confirm this.  
Finally, we note that this calculation neglects quantum correlations (entanglement) between different sites.  
For \( \alpha > 1 \), it is not ruled out that such quantum correlations could destabilize the time crystal or time glass phases.

\section{Spectral gap}
\label{sec:spectral_gap}

\begin{figure*}
\centering
\includegraphics[width=\textwidth]{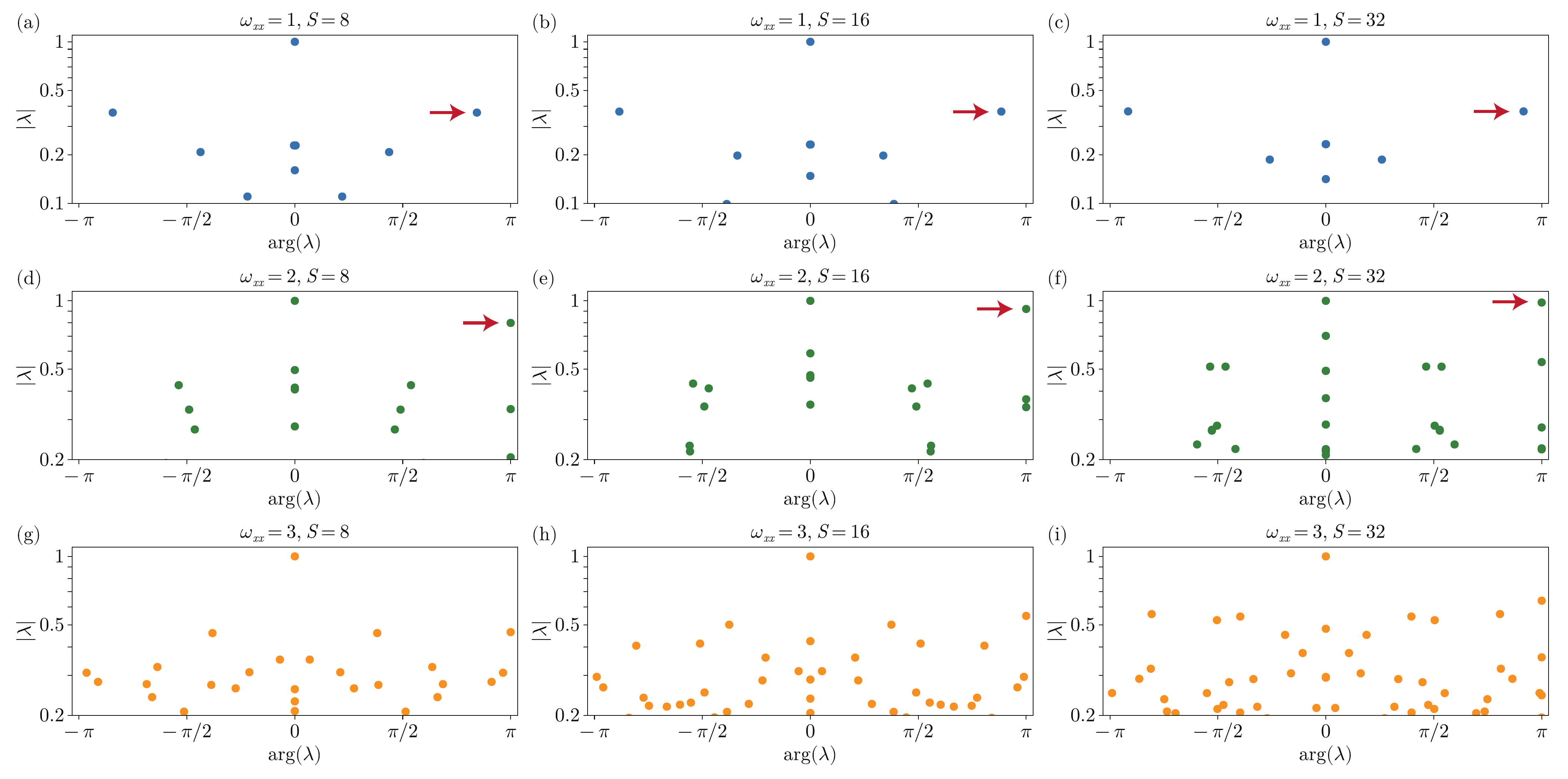}
\caption{Liouvillian spectra \( \lambda_\alpha \) for the kicked collective spin model.  
The horizontal axis represents the argument \( \text{arg}(\lambda_\alpha) \), and the vertical axis represents the modulus \( |\lambda_\alpha| \), plotted on a logarithmic scale.
Panels (a)-(c) show the spectra for kick strength \(\omega_{xx} = 1\) with \(S = 8\), \(16\), and \(32\), respectively.  
Panels (d)-(f) show the spectra for kick strength \(\omega_{xx} = 2\) with \(S = 8\), \(16\), and \(32\), respectively.  
Panels (g)-(i) show the spectra for kick strength \(\omega_{xx} = 3\) with \(S = 8\), \(16\), and \(32\), respectively.  
In all cases, the other parameters are fixed at \( \omega_z = \pi/2 \) and \( \kappa = 1 \).}
\label{fig_collective_spin_spectrum}
\end{figure*}

In this section, we investigate the behavior of the eigenvalues \(\lambda_\alpha\) of the Floquet map \(\mathcal{U}\).  
In particular, we focus on whether the Liouvillian gap \(\Delta\), defined by Eq.~\eqref{Liouvillian_gap}, closes or remains open in the thermodynamic limit \(N \to \infty\).  
Before proceeding to detailed analysis, we summarize the behavior of the Liouvillian gap \(\Delta\) in the disordered, time crystal, and time glass phases as follows:
\begin{itemize}
\item \textit{Disordered phase.}
The Liouvillian gap remains open: \(\Delta_{\infty} = \lim_{N \to \infty} \Delta > 0\).
In this case, the macroscopic dynamics in the thermodynamic limit are governed by an effective equation of motion with a stable fixed point.  
The value of \(\Delta_{\infty}\) corresponds to the eigenvalue of the Jacobian matrix at the fixed point.

\item \textit{Time crystal phase.}
The Liouvillian gap closes: \(\Delta_{\infty} = \lim_{N \to \infty} \Delta = 0\).
When the system exhibits oscillations with period \(p\), eigenvalues \(\lambda = e^{i 2n\pi/p} \: (n=0, 1,\ldots, p-1) \)  emerge in the thermodynamic limit.  
Moreover, the Liouvillian gap decays exponentially with the system size: \(\Delta \propto e^{-cN}\), where \( c \) is a constant.

\item \textit{Time glass phase.}
The Liouvillian gap remains open: \(\Delta_{\infty} = \lim_{N \to \infty} \Delta > 0\).
In this case, \(\Delta_{\infty}\) coincides with the decay rate of the autocorrelation function of the macroscopic observable.
\end{itemize}

The fact that the Liouvillian gap remains open in the time glass phase may seem counterintuitive.  
Based on the intuitive picture that chaos emerges from a series of period-doubling bifurcations in time crystals, one might expect that in the time glass phase, infinitely many eigenvalues accumulate along the unit circle, leading to a closing of the Liouvillian gap in the thermodynamic limit, as illustrated in Fig.~\ref{fig_spectrum_schematic} as ``Naive guess".  
However, as we demonstrate in the following analysis, this expectation is incorrect for our models.

\subsection{System-size dependence of the gap in kicked collective spin}

We first consider the Liouvillian spectrum for the kicked collective spin.
Figure~\ref{fig_collective_spin_spectrum} shows the eigenvalues \(\{ \lambda_\alpha \}\) of the Liouvillian for \(\omega_{xx} = 1\) (stable fixed point), \(\omega_{xx} = 2\) (time crystal), and \(\omega_{xx} = 3\) (time glass).  
Note that there is always a steady-state eigenvalue \(\lambda_0 = 1\).

In the case of \(\omega_{xx} = 1\), where a stable fixed point exists in the classical limit, the Liouvillian gap \(\Delta\) converges to a nonzero finite value as \(S\) increases.  
This is evident from the observation that the eigenvalue \(\lambda_1\), indicated by the red arrows, changes very little with increasing \(S\) [see Figs.~\ref{fig_collective_spin_spectrum}(a)-(c)].  
Moreover, each eigenvalue \(\lambda_\alpha\) (\(\alpha > 1\)) other than \(\lambda_1\) also appears to converge to a specific value in the limit \(S \to \infty\).  
The eigenvalue \(\lambda_1\) converges to the eigenvalue of the Jacobian matrix of the classical fixed point, which is calculated in Appendix~\ref{sec:linear_stability_analysis}.  
In fact, the eigenvalue of the Jacobian is \(-e^{-\kappa} \simeq -0.368\), and Figs.~\ref{fig_collective_spin_spectrum}(a)-(c) show that \(\lambda_1\) approaches this value as \(S\) increases.  

For \(\omega_{xx} = 2\), corresponding to the time crystal phase, it is notable that \(\lambda_1 \to -1\) as \(S \to \infty\), as indicated by the red arrows in Figs.~\ref{fig_collective_spin_spectrum}(d)-(f).
This behavior implies that the system exhibits period-2 oscillations.  
In the case of \(\omega_{xx} = 3\), corresponding to the time glass phase, the results suggest that the Liouvillian gap \(\Delta\) remains open as \(S \to \infty\) [see Figs.~\ref{fig_collective_spin_spectrum}(g)-(i)].  
It should be noted that for \(\omega_{xx} = 2\) and \(\omega_{xx} = 3\), there is no clear tendency for eigenvalues \(\lambda_\alpha\) (\(\alpha > 1\)) other than \(\lambda_1\) to converge to specific values up to \(S = 32\).

\begin{figure}
\centering
\includegraphics[width=8.6cm]{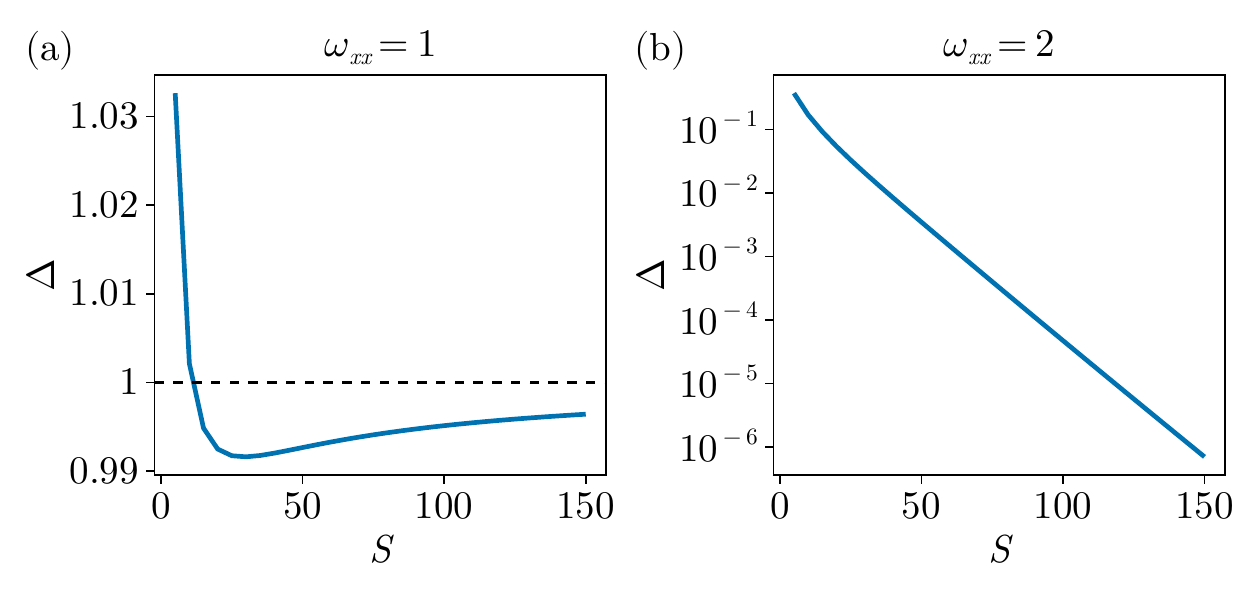}
\caption{Liouvillian gap \(\Delta\) as a function of \(S\) for the kicked collective spin model.  
(a) The case of \(\omega_{xx} = 1\), where a stable fixed point exists.  
The dashed line represents \(\kappa \), which is the negative logarithm of the eigenvalue of the Jacobian matrix near the fixed point in the classical dynamics.
(b) The case of \(\omega_{xx} = 2\), corresponding to the time crystal phase.  
The Liouvillian gap approaches zero exponentially as a function of \(S\).  
The other parameters are fixed at \(\omega_z = \pi/2\) and \(\kappa = 1\).}
\label{fig_collective_spin_gap_TC}
\end{figure}

Figure~\ref{fig_collective_spin_gap_TC} shows the Liouvillian gap \(\Delta\) as a function of \(S\) for the cases where a stable fixed point exists (\(\omega_{xx} = 1\)) and for the time crystal phase (\(\omega_{xx} = 2\)).  
In the case of \(\omega_{xx} = 1\), it is observed that \(\Delta\) converges to \(\kappa = 1\) in the limit \(S \to \infty\).  
As discussed earlier, this value corresponds to the negative logarithm of the eigenvalue of the Jacobian matrix near the classical fixed point.
On the other hand, for \(\omega_{xx} = 2\), \(\Delta\) approaches zero exponentially as a function of \(S\).  
Noting that the persistence time of the periodic oscillations is given by \(1/\Delta\), this behavior implies that the lifetime of the time crystal increases exponentially with the system size.
This result is consistent with previously known properties of time crystals \cite{Gong-18, Riera-Campeny-20}.

\begin{figure}
\centering
\includegraphics[width=8.6cm]{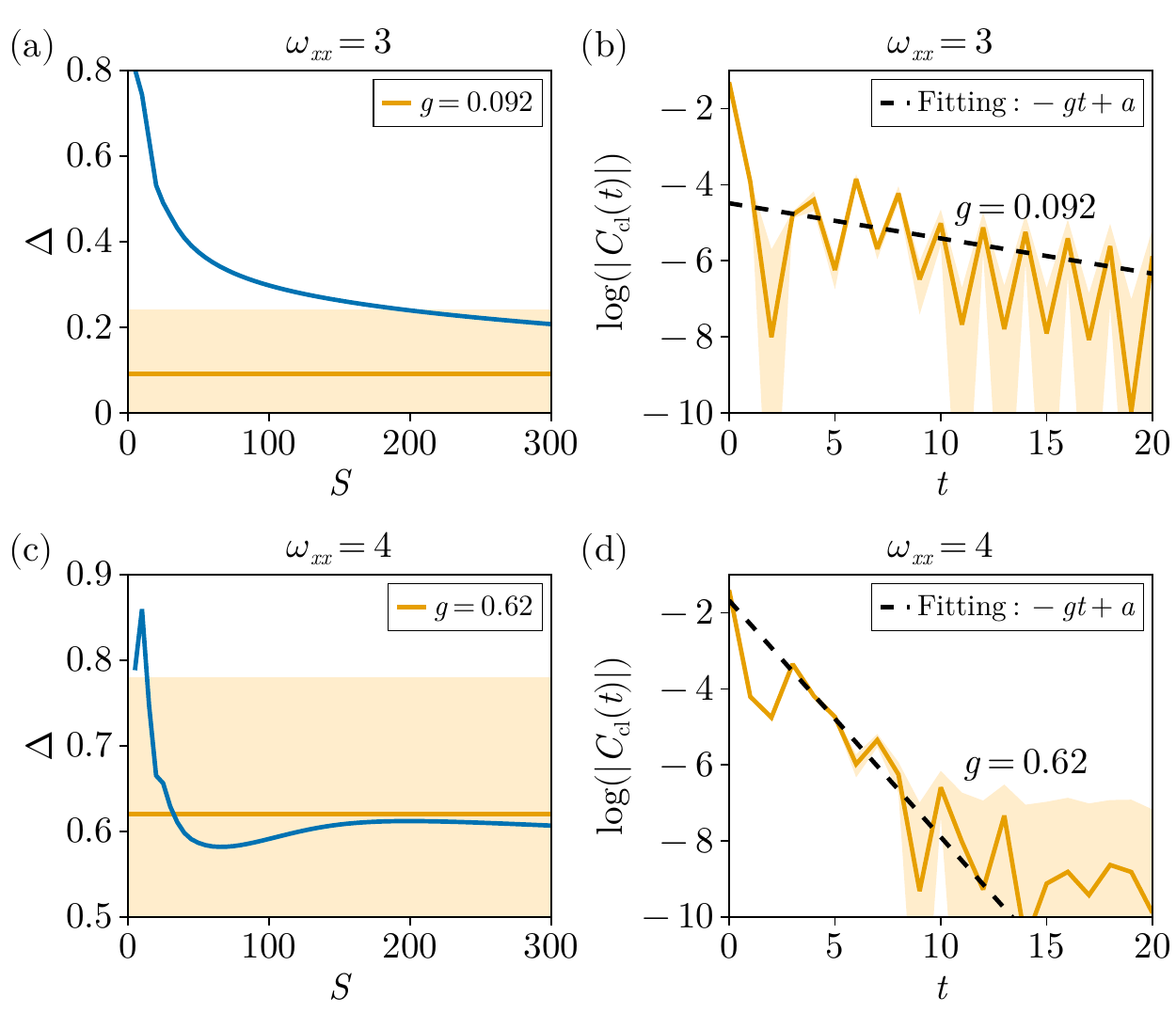}
\caption{Liouvillian gap \(\Delta\) and classical autocorrelation function for the kicked collective spin model in the time glass phase.
The values of the kick strength are \( \omega_{xx} = 3 \) for panels (a) and (b), and \( \omega_{xx} = 4 \) for panels (c) and (d).
The other parameters are fixed at \( \omega_z = \pi/2 \) and \( \kappa = 1 \).  
Panels (a) and (c) show the Liouvillian gap \( \Delta \) as a function of \( S \) for \( \omega_{xx} = 3 \) and \( 4 \), respectively.
Panels (b) and (d) show the logarithm of the autocorrelation function \( C_{\mathrm{cl}}(t) \) for the classical dynamics of \( m^x \). 
The shaded region indicates the standard deviation with respect to randomly sampled initial conditions.
The dashed lines represent fits to the form \( -g t + a \) over the range \([3, 10]\).
The fitted values of the decay rate are \( g = 0.092 \) for \( \omega_{xx} = 3 \) and \( g = 0.62 \) for \( \omega_{xx} = 4 \).
These values, together with their standard errors (shown as shaded bands), are plotted as horizontal lines in panels (a) and (c).
These results suggest that \( \Delta \) approaches \( g \) as \( S \to \infty \).}
\label{fig_collective_spin_gap_TG}
\end{figure}

Figures~\ref{fig_collective_spin_gap_TG}(a) and (c) show the Liouvillian gap \( \Delta \) as a function of \( S \) in the time glass phase for \( \omega_{xx} = 3 \) and \( 4 \), respectively.  
For \( \omega_{xx} = 4 \) [see Fig.~\ref{fig_collective_spin_gap_TG}(c)], it is evident that \( \Delta \) does not vanish in the classical limit \( S \to \infty \).
Here, we focus on the autocorrelation function \( C_{\mathrm{cl}}(t) \) \( (t=0,1,\ldots) \) of \( m^x \) in the classical dynamics, which is defined by
\begin{equation}
C_{\mathrm{cl}}(t) = \lim_{K \to \infty} \frac{1}{K} \sum_{k=1}^K m^x(t+k) m^x(k),
\label{autocorrelation_classical_dynamics_mx}
\end{equation}
where \( m^x(t) \) \( (t=0,1,\ldots) \) is the solution of Eqs.~\eqref{collective_spin_classical_equation_1} and \eqref{collective_spin_classical_equation_2}.
We assume that, in the time glass phase, the classical dynamics is ergodic and mixing, i.e., the autocorrelation decays exponentially as \( |C_{\mathrm{cl}}(t)| \propto e^{-g t} \) with a mixing rate \( g \).

We compare \( \Delta_{\infty} = \lim_{N \to \infty} \Delta \) with the mixing rate \( g \) of the classical dynamics.  
Figures~\ref{fig_collective_spin_gap_TG}(b) and (d) show \( \log(|C_{\mathrm{cl}}(t)|) \) as a function of \( t \) for  \( \omega_{xx} = 3 \) and \( 4 \), respectively.
The dashed lines indicate fits to the form \( -g t + a \) over the range \([3, 10]\), yielding \( g = 0.092 \) for \( \omega_{xx} = 3 \) and \( g = 0.62 \) for \( \omega_{xx} = 4 \).
The fitted values of \(g\), together with their standard errors (shown as shaded bands), are indicated by horizontal lines in Figs.~\ref{fig_collective_spin_gap_TG}(a) and (c). 
Although the fitting carries a large uncertainty due to the oscillatory behavior of \( \log(|C_{\mathrm{cl}}(t)|) \), the overall trend shows that \( \Delta \) approaches \( g \) as \( S \) increases.
For the weakly mixing case \( \omega_{xx} = 3 \), the small value of \( g \) makes it difficult to conclude whether \( \Delta \) remains finite as \( S \to \infty \).
In contrast, for the strongly mixing case \( \omega_{xx} = 4 \), \( \Delta \) converges clearly toward \( g \).
These observations support the conjecture that \( \Delta_{\infty} \) coincides with the decay rate of the autocorrelation function of macroscopically chaotic observables.

\begin{figure}
\centering
\includegraphics[width=8.6cm]{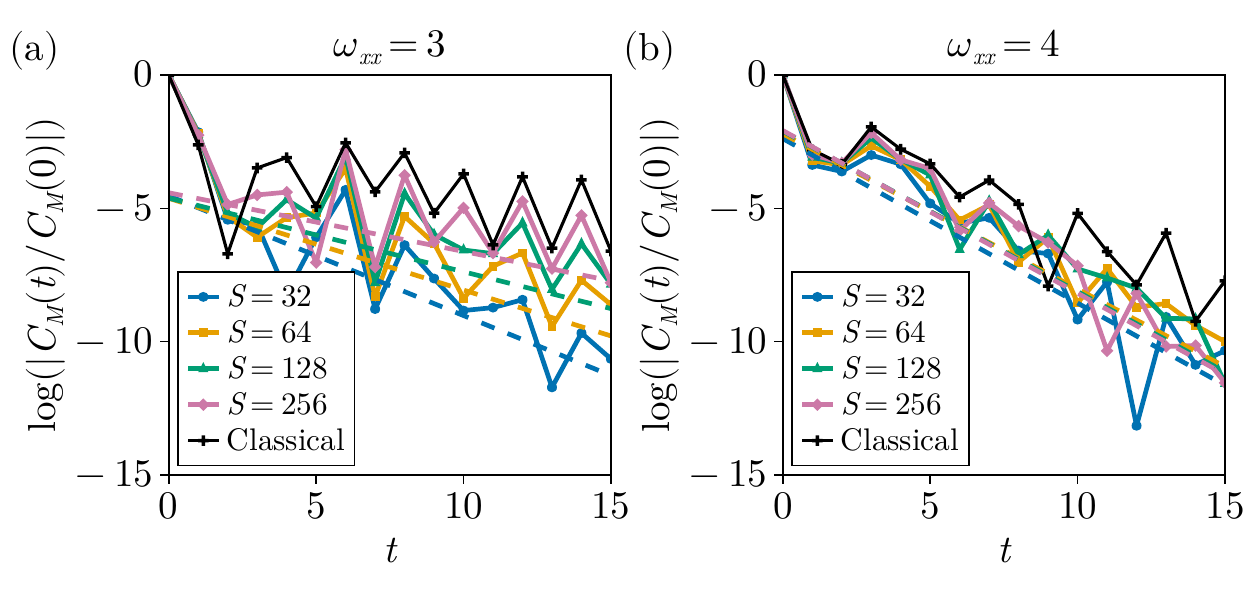}
\caption{Logarithm of the normalized autocorrelations \( C_M(t) \) of the order parameter for the kicked collective spin model with \(S = 32\), \(64\), \(128\), and \(256\).
The black line shows the classical autocorrelation \( C_{\mathrm{cl}}(t) \).
For small \(t\), \( C_M(t) \) approaches \( C_{\mathrm{cl}}(t) \) as \(S\) increases.
The dashed lines represent \(-\Delta t + a\), where $\Delta$ is the Liouvillian gap for each \(S\), and the offset \(a\) is chosen so that each line passes through the corresponding value of \( C_M(t) \) at \( t = 2\).
One can observe that the long-time decay of \( C_M(t) \) is accurately captured by $\Delta$.}
\label{fig_collective_spin_autocorrelation_correspondence}
\end{figure}

To further support the correspondence between the Liouvillian gap \(\Delta\) and the classical mixing rate \(g\), we examine how the quantum autocorrelation \( C_M(t) \) depends on the system size \(S\).
Figure \ref{fig_collective_spin_autocorrelation_correspondence} displays \( C_M(t) \) for several values of \(S\), together with the classical autocorrelation \( C_{\mathrm{cl}}(t) \) shown as a black line.
For short times, \( C_M(t) \) approaches \( C_{\mathrm{cl}}(t) \) as \(S\) increases, particularly clearly for \( \omega_{xx} = 4 \).
This behavior supports the quantum-classical correspondence of autocorrelations described in Eq.~\eqref{autocorrelation_quantum_classical_correspondence}.
The convergence becomes slower at larger times, which is expected in chaotic systems since the timescale over which \( C_M(t) \) tracks \( C_{\mathrm{cl}}(t) \) grows only logarithmically with \(S\) (see Sec.~\ref{sec:relaxation_time} and Appendix \ref{sec:quantum_classical_correspondence}).
The dashed lines in Fig.~\ref{fig_collective_spin_autocorrelation_correspondence} have slopes fixed at $-\Delta$ for each \(S\) (i.e., they are not fits to the data).
Their agreement with the long-time decay of \( C_M(t) \) demonstrates that
\begin{equation}
|C_M(t)| \propto e^{-t \Delta},
\label{quantum_correlation_decay}
\end{equation}
as discussed in Sec.~\ref{sec:overview_spectral features}.
Combining the convergence of \( C_M(t) \) toward \( C_{\mathrm{cl}}(t) \) with Eq.~\eqref{quantum_correlation_decay} naturally leads to the conclusion that \(\Delta\) converges to the classical mixing rate, \(\lim_{S \to \infty} \Delta = g\).
It is note worthy that the quantum autocorrelation remains below the classical counterpart at long times, 
\begin{equation}
|C_M(t)| \leq |C_{\mathrm{cl}}(t)|,
\end{equation}
which implies that the quantum system is more mixing than the classical one.
This bound provides additional support that \(\Delta\) does not vanish in the limit \(S \to \infty\).

We discuss the relationship between our results and several previous studies \cite{Carollo-24, Solanki-24-2}.
In Ref.~\cite{Carollo-24}, the authors investigated a periodically driven collective spin system and found that the corresponding classical (mean-field) dynamics exhibit chaotic behavior in certain parameter regimes.  
Remarkably, they concluded that when the classical dynamics are chaotic, the Liouvillian gap \( \Delta \) of the Floquet map vanishes as \( \Delta \propto N^{-1} \).
At first glance, this finding appears to contradict our results, which predict a nonzero gap in the thermodynamic limit.
In Appendix \ref{appendix:clarifying_the_discrepancy}, we provide a detailed analysis of the behavior of \( \Delta \) for the model studied in Ref.~\cite{Carollo-24}.
For the parameter set considered in that work, we find that the convergence of \( \Delta \) with increasing system size is very slow, making it difficult to determine whether its thermodynamic-limit value \( \Delta_\infty \) is finite or vanishing.
Such slow convergence can easily give rise to an apparently closing gap within accessible numerical ranges.
Therefore, the available data are not necessarily inconsistent with the existence of a small but finite gap \( \Delta_\infty \) in the thermodynamic limit.
In the same appendix, we also investigate another parameter set of the same model in the time glass regime and find that \( \Delta \) converges well to a finite value \( \Delta_\infty \), which closely matches the mixing rate of the corresponding classical dynamics, consistent with our results presented in this paper.

\subsection{System-size dependence of the gap in kicked spin chain with all-to-all coupling}

Next, we investigate the behavior of the Liouvillian gap for the kicked spin chain with all-to-all coupling (\( \alpha = 0 \)).  
In this case, the system is invariant under permutations of the \( N \) spins.  
Therefore, by restricting the analysis to the permutation-invariant subspace (symmetric sector), the dimension of the Hilbert space can be significantly reduced (see Appendix \ref{sec:symmetric_sector_spin_chain} for details).  
It should be noted that it is not obvious whether the eigenvalue \( \lambda_1 \), which has the second largest absolute value and determines the gap \( \Delta \), belongs to the symmetric sector.  
(A slower relaxing mode might exist in the asymmetric sector.)  
However, in the following, we proceed under the assumption that \( \lambda_1 \) belongs to the symmetric sector.  
By comparing with results from exact diagonalization including all sectors for small \( N \), we have confirmed that this assumption holds in the parameter range where the time glass phase appears.

\begin{figure}
\centering
\includegraphics[width=8.6cm]{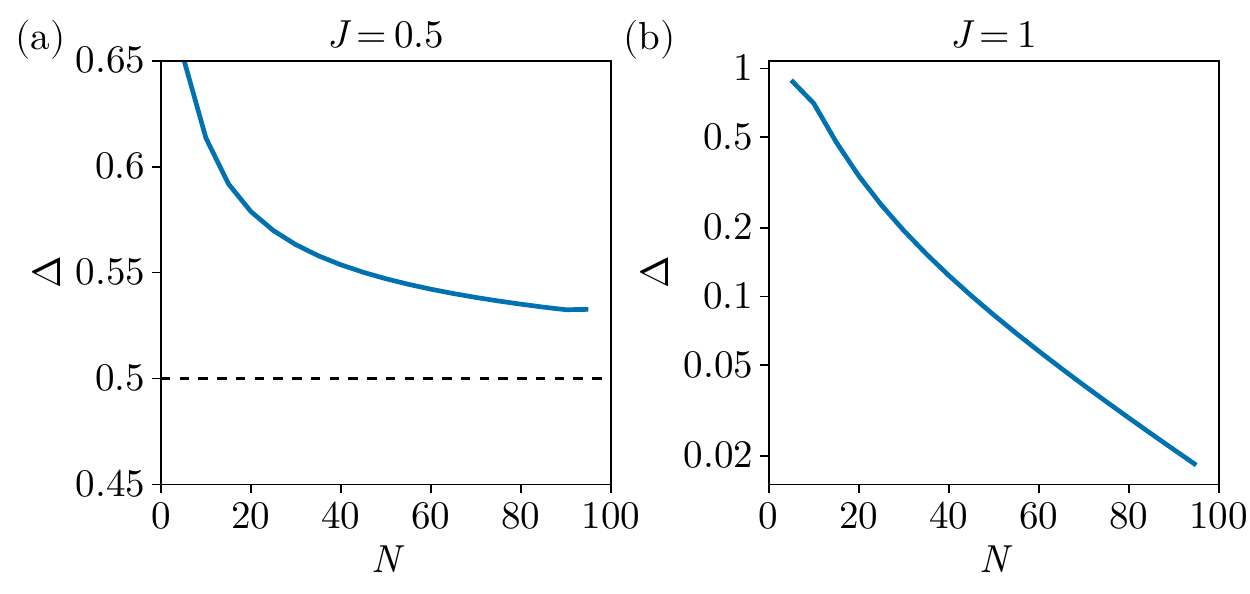}
\caption{Liouvillian gap \(\Delta\) as a function of \(N\) for the kicked spin chain model with all-to-all coupling.  
(a) The case of \(J = 0.5\), where a stable fixed point exists.  
The dashed line represents \(\kappa/2 \), which is the negative logarithm of the eigenvalue of the Jacobian matrix near the fixed point in the classical dynamics.
(b) The case of \(J = 1\), corresponding to the time crystal phase.  
The Liouvillian gap approaches zero exponentially as a function of \(N\).  
The other parameters are fixed at \(\omega_z = \pi/2\) and \(\kappa = 1\).}
\label{fig_spin_chain_gap_TC}
\end{figure}

Figure~\ref{fig_spin_chain_gap_TC} shows the Liouvillian gap \(\Delta\) as a function of \(N\) for the cases where a stable fixed point exists (\(J = 0.5\)) and for the time crystal phase (\(J = 1\)).  
In the case of \(J = 0.5\), it is observed that \(\Delta\) converges to \(\kappa/2\) in the limit \(N \to \infty\).  
This value corresponds to the negative logarithm of the eigenvalue of the Jacobian matrix near the classical fixed point (see Appendix \ref{sec:linear_stability_analysis}).
On the other hand, for \(J = 1\), \(\Delta\) approaches zero exponentially as a function of \(N\), which is consistent with the typical behavior observed in time crystals.

\begin{figure}
\centering
\includegraphics[width=8.6cm]{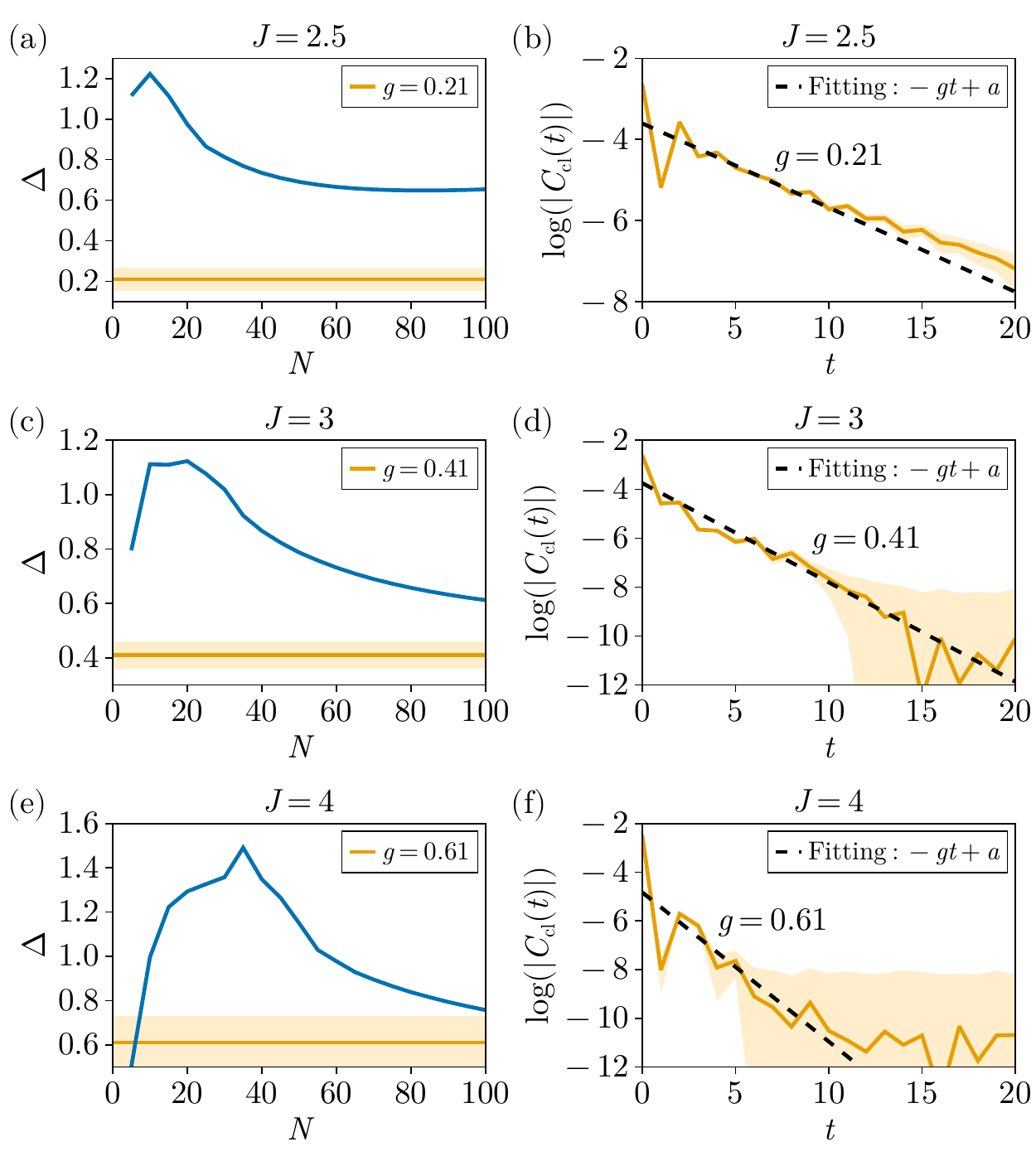}
\caption{Liouvillian gap \(\Delta\) and classical autocorrelation function for the kicked spin chain model with all-to-all couping in the time glass phase.
The values of the kick strength are \( J = 2.5 \) for panels (a) and (b), \( J = 3 \) for panels (c) and (d), and \( J = 4 \) for panels (e) and (f).
The other parameters are fixed at \( \omega_z = \pi/2 \) and \( \kappa = 1 \).  
Panels (a), (c), and (e) show the Liouvillian gap \( \Delta \) as a function of \( N \) for \( J = 2.5 \), \( 3 \), and \( 4 \), respectively.  
Panels (b), (d), and (f) show the logarithm of the autocorrelation function \( C_{\mathrm{cl}}(t) \) for the classical dynamics of \( m^x \).
The shaded region indicates the standard deviation with respect to randomly sampled initial conditions.
The dashed lines represent fits to the form \( -g t + a \) over the range \([0, 10]\).
The fitted values of the decay rate are \( g = 0.21 \) for \( J = 2.5 \), \( g = 0.41 \) for \( J = 3 \), and \( g = 0.61 \) for \( J = 4 \).
These values, together with their standard errors (shown as shaded bands), are plotted as horizontal lines in panels (a), (c), and (e).
For \( J = 3 \) and \( J = 4 \), one can observe that \( \Delta \) approaches \( g \) as \( N \to \infty \).}
\label{fig_spin_chain_gap_TG}
\end{figure}

Figures~\ref{fig_spin_chain_gap_TG}(a), (c), and (e) show the Liouvillian gap \( \Delta \) as a function of \( N \) in the time glass phase for \( J = 2.5 \), \(3\), and \(4\), respectively.
Figures~\ref{fig_spin_chain_gap_TG}(b), (d), and (f) display the logarithm of the classical autocorrelation function \( C_{\mathrm{cl}}(t) \), which is calculated by Eq.~\eqref{autocorrelation_classical_dynamics_mx} with the solution of Eqs.~\eqref{spin_chain_mean_field_dissipation} and \eqref{spin_chain_mean_field_kick}.
The dashed lines represent linear fits, and the slopes give the decay rates \( g \).  
The horizontal lines in Figs.~\ref{fig_spin_chain_gap_TG}(a), (c), and (e) indicate the values of \( g \) determined in this way.
The results for \( J = 3 \) and \( J = 4 \) appear consistent with the assumption that \( \Delta_{\infty} \) coincides with \( g \).  
However, for \( J = 2.5 \), the data suggest that \( \Delta_{\infty} \) may take a value clearly larger than \( g \).

The discrepancy between \( \Delta \) and \( g \) observed in the case of \( J = 2.5 \) requires further investigation.  
One possible explanation is the emergence of a ``periodic orbit window" near this parameter value.  
In many chaotic dynamical systems, chaos can disappear within a very narrow range of parameters, giving rise to periodic orbits.  
Near such windows, the autocorrelation function is expected to decay slowly or not decay at all, exhibiting persistent oscillations.  
In the corresponding quantum system, however, narrow periodic windows may be washed out by quantum fluctuations, causing the system to behave similarly to the surrounding chaotic regime.
In such cases, the quantum autocorrelation decays faster than the classical counterpart.
Figures~\ref{fig_spin_chain_gap_TG}(a) and (b) might correspond to such a situation.

\subsection{Kick-strength dependence of the gap}

\begin{figure}
\centering
\includegraphics[width=8.6cm]{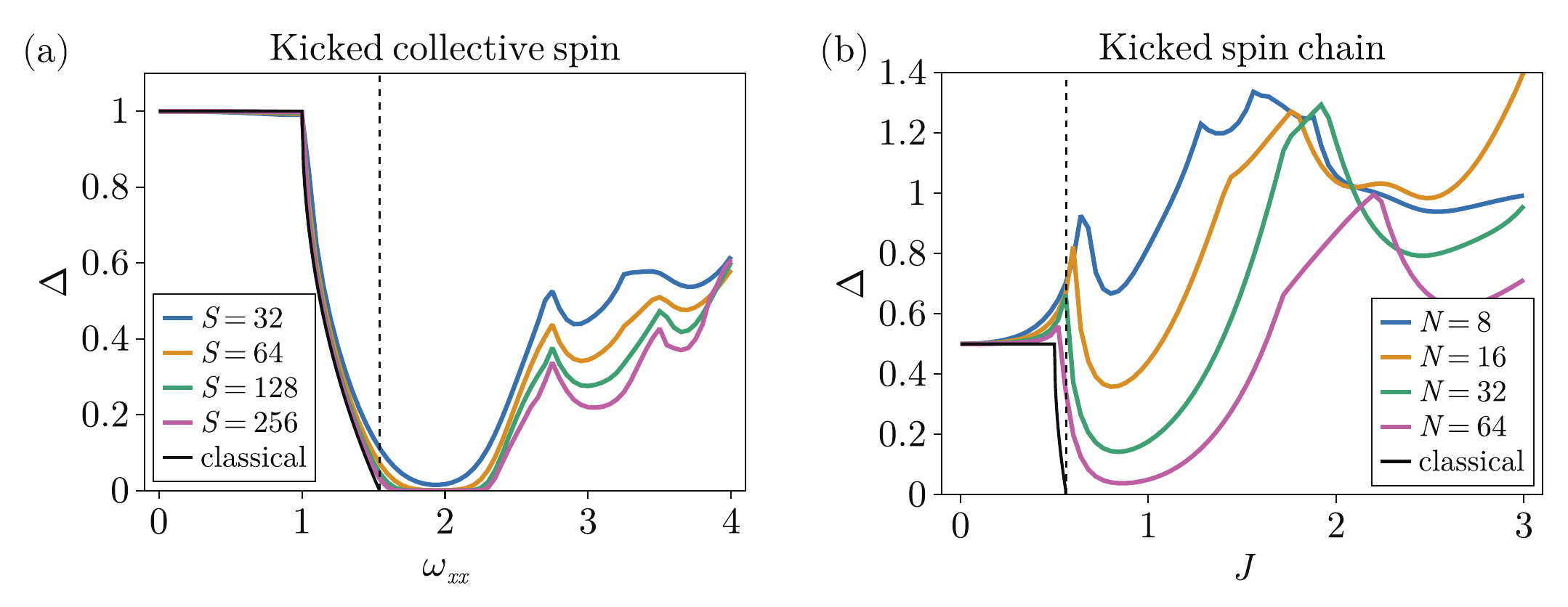}
\caption{Liouvillian gap \(\Delta\) as a function of the kick strength.
(a) \(\Delta\) for the kicked collective spin as a function of \(\omega_{xx}\) with \(S = 32\), \(64\), \(128\), and \(256\).
(b) \(\Delta\) for the kicked spin chain as a function of \(J\) with \(N = 8\), \(16\), \(32\), and \(64\).
The other parameters are fixed at \( \omega_z = \pi/2 \) and \( \kappa = 1 \). 
The solid black curves indicate the gap associated with the stable fixed point in the corresponding classical dynamics.
The dashed vertical lines mark the critical kick strength at which the fixed point loses stability and a limit cycle appears.
The gap opens at the transition from the time crystal phase to the time glass phase.}
\label{fig_gap_parameter_dependence}
\end{figure}

Let us consider how the Liouvillian gap \( \Delta \) depends on the kick strength.  
Figure~\ref{fig_gap_parameter_dependence}(a) shows \( \Delta \) for the kicked collective spin as a function of the kick strength \( \omega_{xx} \).  
For \( \omega_{xx} < \omega_{xx}^c \simeq 1.543 \), the corresponding classical dynamics possess a stable fixed point.  
The eigenvalues of the Jacobian matrix at this fixed point are given by Eq.~\eqref{collective_spin_jacobian_eigenvalue_fixed_point} in Appendix~\ref{sec:linear_stability_analysis}, under the condition \( \omega_z = \pi/2 \).
The associated gap \( \Delta_{\mathrm{cl}} \) can be defined as the negative logarithm of the largest eigenvalue,  
\begin{equation}
\Delta_{\mathrm{cl}} = 
\begin{cases}
\kappa & (0 \leq \omega_{xx} < 1), \\
\kappa - \log (\omega_{xx} + \sqrt{\omega_{xx}^2-1} ) & (1 \leq \omega_{xx} < \omega_{xx}^c).
\end{cases}
\end{equation}
This classical prediction is shown as the solid black curve in Fig.~\ref{fig_gap_parameter_dependence}(a).  
The critical kick strength \( \omega_{xx}^c \), at which \( \Delta_{\mathrm{cl}} \) vanishes and a limit cycle emerges, is indicated by the dashed vertical line.  
For \( \omega_{xx} < \omega_{xx}^c \), the Liouvillian gap \( \Delta \) converges to the classical counterpart \( \Delta_{\mathrm{cl}} \) as the spin size \( S \) increases.

For \( \omega_{xx}^c < \omega_{xx} \lesssim 2.3 \), the classical dynamics exhibit a limit cycle, as shown in Fig.~\ref{fig_collective_spin_bifurcation}.
In this regime, the system realizes the time crystal phase, where the Liouvillian gap \( \Delta \) vanishes in the limit \( S \to \infty \).
Around \( \omega_{xx} \simeq 2.3 \), the classical dynamics undergo a transition to a chaotic regime.  
This transition corresponds to the emergence of the time glass phase.  
The onset of this phase is accompanied by the opening of the Liouvillian gap, as shown in Fig.~\ref{fig_gap_parameter_dependence}(a).

Figure \ref{fig_gap_parameter_dependence}(a) strongly suggests that the opening of the Liouvillian gap $\Delta$ across the transition from the time crystal to the time glass phase is \emph{continuous}. 
This behavior is supported by the correspondence with classical nonlinear dynamics. 
In typical chaotic dynamical systems, such as the logistic map, the transition from a limit cycle to chaos is characterized by a continuous change in the maximal Lyapunov exponent, which crosses from negative to positive values. 
Since the mixing rate $g$, which governs the exponential decay of correlations, is directly related to the positive Lyapunov exponents, the mixing rate also continuously changes from zero to positive values upon entering the chaotic regime. 
This fundamental continuity in the classical chaotic metric implies that the quantum Liouvillian gap $\Delta$ must likewise exhibit a continuous opening at the transition to the time glass phase.

Next, let us consider the kicked spin chain with all-to-all coupling.
Figure~\ref{fig_gap_parameter_dependence}(b) shows the Liouvillian gap \( \Delta \) as a function of the kick strength \( J \).
For \( J < J_c \simeq 0.564 \), the corresponding classical dynamics possess a stable fixed point.
The eigenvalues of the Jacobian matrix at this fixed point are given by Eq.~\eqref{spin_chain_jacobian_eigenvalue_fixed_point} in Appendix~\ref{sec:linear_stability_analysis}.
The associated classical gap \( \Delta_{\mathrm{cl}} \), under the condition \( \omega_z = \pi/2 \), is given by
\begin{equation}
\Delta_{\mathrm{cl}} = 
\begin{cases}
\kappa/2 & (0 \leq J < 1/2), \\
\kappa/2 - \log (2J + \sqrt{4J^2-1} ) & (1/2 \leq J < J_c).
\end{cases}
\end{equation}
This classical prediction is shown as the solid black curve in Fig.~\ref{fig_gap_parameter_dependence}(b).
The critical kick strength \( J_c \), at which \( \Delta_{\mathrm{cl}} \) vanishes and a limit cycle emerges, is indicated by the dashed vertical line.  
For \( J < J_c \), the Liouvillian gap \( \Delta \) converges to the classical counterpart \( \Delta_{\mathrm{cl}} \) as the spin number \( N \) increases.
The behaviors in the time crystal and time glass phases are qualitatively similar to those observed in the kicked collective spin.

\subsection{Relationship to the Ruelle-Pollicott resonances}

We discuss the relationship between our main result, Eq.~\eqref{gap_mixing_rate}, and the Ruelle-Pollicott resonances in classical dynamical systems \cite{Ruelle-86, Pollicott-85, Hasegawa-92, Weber-00, Klus-16}.
Below, we briefly review the concept of the Ruelle-Pollicott resonances.
Consider a dynamical system defined by the time evolution  
\begin{equation}  
x_{t+1} = f(x_t),  
\end{equation}  
where \( x \) represents a state and \( f \) is a map.
The time evolution of a probability distribution \( p_t(x) \) is given by  
\begin{equation}  
p_{t+1}(x) = \int \delta(x - f(y)) p_t(y) dy =: \mathcal{P} p_t(x).
\label{def_Perron_Frobenius_operator}
\end{equation}
The linear operator \( \mathcal{P} \) is referred to as the Perron-Frobenius operator.
We consider the eigenvalues \( e^{-i\mu_\alpha} \) and eigenfunctions \( \phi_\alpha(x) \) of the Perron-Frobenius operator,  
\begin{equation}  
\mathcal{P} \phi_\alpha(x) = e^{-i\mu_\alpha} \phi_\alpha(x),  
\end{equation}  
with \( \text{Im} (\mu_\alpha) \leq 0 \).
Some eigenvalues lie on the unit circle (\( \text{Im} (\mu_\alpha) = 0 \)), while the others reside inside the unit circle (\( \text{Im} (\mu_\alpha) < 0 \)).
The latter eigenvalues are known as the Ruelle-Pollicott resonances.  

The important fact is that the Ruelle-Pollicott resonances determine the decay rates of correlation functions, i.e., the mixing rates.
The correlation function \( C_{AB}(t) \) for observables \( A(x) \) and \( B(x) \) is defined as  
\begin{equation}  
C_{AB}(t) = \langle A(x_{t+\tau}) B(x_{\tau}) \rangle_{\tau} - \langle A(x_{\tau}) \rangle_{\tau} \langle B(x_{\tau}) \rangle_{\tau},  
\end{equation}  
where \( \langle X \rangle_{\tau} = \lim_{T \to \infty} \frac{1}{T} \sum_{\tau=1}^T X_\tau \).
The Fourier transform of the correlation function is defined as  
\begin{equation}  
\hat{C}_{AB}(\omega) = \sum_{t=-\infty}^\infty C_{AB}(t) e^{i\omega t}.  
\end{equation}
It is important to note that the simple poles of \( \hat{C}_{AB}(\omega) \) correspond to the decay rates of \( C_{AB}(t) \).
Roughly speaking, it can be shown that the simple poles of \( \hat{C}_{AB}(\omega) \) coincide with the Ruelle-Pollicott resonances \( \mu_\alpha \) \cite{Ruelle-86, Pollicott-85}.
More concretely, the correlation function can be written as
\begin{equation}
C_{AB}(t) \sim \sum_{\alpha} c_\alpha e^{-i\mu_\alpha t},
\end{equation}
with appropriate constants \( c_\alpha \) that depend on \( A(x) \) and \( B(x) \).
In particular, the long time behavior of \( C_{AB}(t) \) is characterized by the Ruelle-Pollicott resonance \( e^{-i\mu_\alpha} \) with the largest modulus.

Since the Perron-Frobenius operator \( \mathcal{P} \) is an infinite-dimensional matrix defined on an appropriate functional space, the calculation of the Ruelle-Pollicott resonances requires truncating the functional space using a set of basis functions \cite{Weber-00, Klus-16}.
Let \( (\psi_1(x), \dots, \psi_k(x)) \) be a set of orthonormal basis functions.
The corresponding matrix elements of \( \mathcal{P} \) are given by  
\begin{equation}  
P_{ij} := \int dx \, \psi_i^*(x) \, \mathcal{P} \, \psi_j(x).  
\end{equation}
The diagonalization of the \( k \times k \) matrix \( P \) yields \( k \) eigenvalues.
In the limit \( k \to \infty \), new eigenvalues with small modulus continue to appear indefinitely.
However, the eigenvalues with large modulus converge to fixed values, which correspond to the Ruelle-Pollicott resonances.  
For example, in Ref.~\cite{Weber-00}, the eigenvalues and eigenfunctions of the Perron-Frobenius operator for the kicked top model are investigated.
The kicked top model describes the dynamics of a periodically kicked classical spin.
In this case, since the state space of the model is the sphere, the spherical harmonics are employed as the orthonormal basis functions for the functional space.  

The concept of Ruelle-Pollicott resonances has been extended to quantum systems to analyze thermalization in isolated settings \cite{Garcia-Mata-03, Nonnenmacher-03, Prosen-02, Prosen-04, Garcia-Mata-18, Znidaric-24, Mori-24, Yoshimura-24, Jacoby-25}.
In this context, the quantum evolution map $\mathcal{U}$ for the density matrix serves as the quantum analogue of the Perron-Frobenius operator $\mathcal{P}$. 
For isolated systems, \( \mathcal{U} \) is unitary, and all eigenvalues lie on the unit circle, implying the absence of decay modes.
To extract quantum Ruelle-Pollicott resonances, one introduces a dissipative evolution map \( \mathcal{U}(\gamma) \), which reduces to \( \mathcal{U} \) as the dissipation strength \( \gamma \to 0 \).
Since \( \mathcal{U}(\gamma) \) is non-unitary, it has eigenvalues with modulus less than unity.
By taking the limit \( \gamma \to 0 \) \emph{after} the thermodynamic limit, the decaying modes survive and define the quantum Ruelle-Pollicott resonances, which characterize the decorrelation associated with thermalization.

We restate our main finding in terms of Ruelle-Pollicott resonances.
In the classical limit (large spin limit $S \to \infty$), the quantum evolution map $\mathcal{U}$ for the time glass phase is conjectured to converge to the Perron-Frobenius operator $\mathcal{P}$ of the corresponding classical chaotic dynamics. 
This convergence implies that the spectral features of $\mathcal{U}$ directly reveal the Ruelle-Pollicott resonances of the classical system.
Specifically, the Liouvillian gap $\Delta$ corresponds to the dominant Ruelle-Pollicott resonance, providing the physical justification for the equivalence between $\Delta$ and the mixing rate.

Our work differs from prior studies on quantum Ruelle-Pollicott resonances in two fundamental aspects. 
First, previous works focused on microscopic chaos (irregularity of individual components), whereas we analyze macroscopic chaos, where chaotic behavior arises from the synchronized motion of many degrees of freedom. 
We conjecture that the eigenvalues of $\mathcal{U}$ converge, in the thermodynamic limit, to the Ruelle-Pollicott resonances of the \emph{collective dynamics} of macroscopic observables. 
Second, unlike earlier studies where dissipation was an artificial tool eventually removed ($\gamma \to 0$), dissipation is essential for stabilizing the time glass phase. 
Our results reveal a direct connection between the spectral gap of a quantum system with finite dissipation and the Ruelle-Pollicott resonances of a classical dissipative system.

\subsection{Liouvillian gap of classical stochastic systems in the noiseless limit}
\label{sec:Liouvillian_gap_of_stochastic_system}

It is known that the dynamics of a dissipative quantum system can be approximated by a classical stochastic system under suitable conditions.  
More concretely, in the semiclassical limit, the time evolution of the Wigner function is governed by the Fokker-Planck equation at leading order in \( \hbar \) \cite{Cabot-24, Dutta-25, Walls-78, Drummond-78, Drummond-80, Vogel-88, Dubois-21}.
Thus, it is instructive to consider the Liouvillian gap in classical stochastic systems that exhibit chaotic behavior without noise.
The behavior of the gap in such stochastic systems, particularly as the noise strength approaches zero, can provide valuable insight into the scaling of the Liouvillian gap in dissipative quantum systems in the large system-size limit.

We consider a dynamical system subject to noise:
\begin{equation}
x_{t+1} = f(x_t) + \sigma \xi_t,
\label{stochastic_dynamical_system_general}
\end{equation}
where \( \xi_t \) is an independent random variable drawn from a normal distribution with mean zero and unit variance.
The parameter \( \sigma \) denotes the standard deviation of the noise.
The conditional probability of \( x_{t+1} \) given \( x_t \) is written as
\begin{equation}
k(x_{t+1}|x_t) = \frac{1}{\sqrt{2\pi \sigma^2}} \exp \left(- \frac{[x_{t+1} - f(x_t)]^2}{2\sigma^2} \right),
\label{stochastic_dynamics_propagator}
\end{equation}
and the time evolution of the probability distribution \( p_t(x) \) is expressed as
\begin{equation}
p_{t+1}(x) = \int k(x|y) p_t(y) dy =: \mathcal{L} p_t (x).
\label{stochastic_dynamics_Liouvillian}
\end{equation}
We refer to the operator \( \mathcal{L} \), which governs the evolution of the probability distribution, as the Liouvillian.
The eigenvalues and spectral gap of \( \mathcal{L} \) can be defined analogously to those introduced in Sec.~\ref{sec:overview}.

\begin{figure*}
\centering
\includegraphics[width=\textwidth]{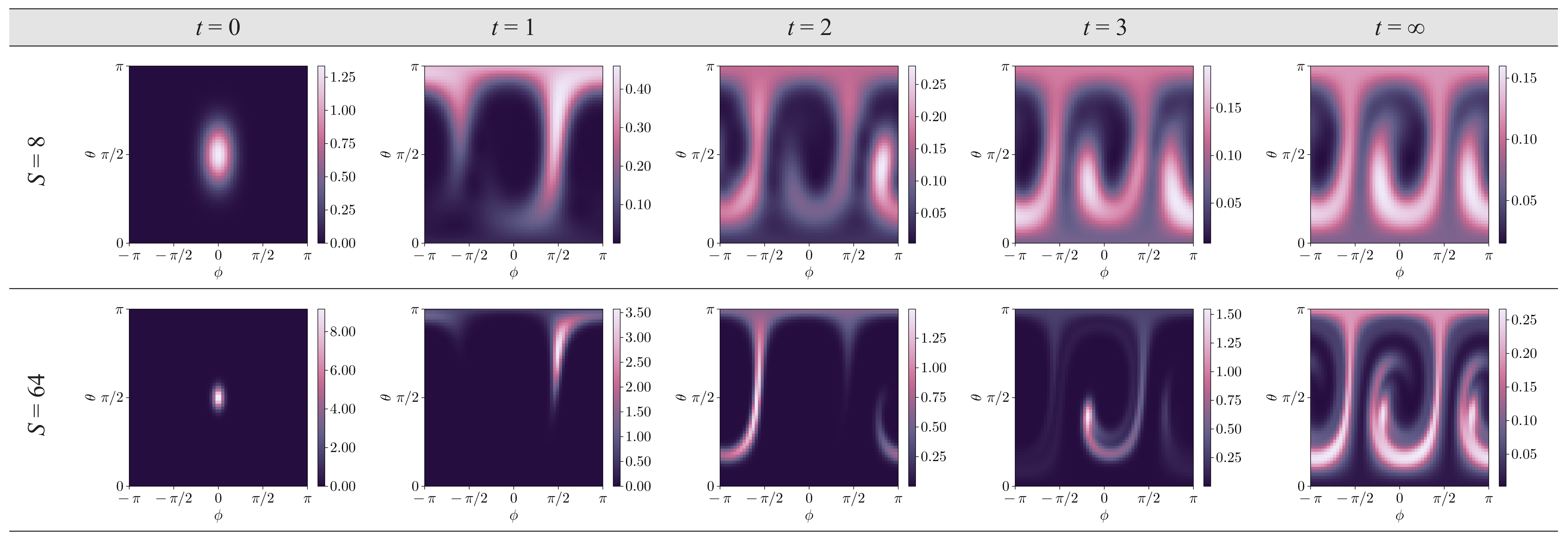}
\caption{Time evolution of the Husimi distribution \(Q(\theta, \phi)\) for the kicked collective spin model in the time glass phase.  
The upper row corresponds to \( S = 8 \), and the lower row to \( S = 64 \).  
The parameters are fixed at \( \omega_{xx} = 3 \), \( \omega_z = \pi/2 \), and \( \kappa = 1 \).  
The initial state is the spin coherent state with polar angle \( \theta = \pi/2 \) and azimuthal angle \( \phi = 0 \).  
Snapshots are shown, from left to right, at times \( t = 0,\,1,\,2,\,3,\,\infty \).
For \( S = 8 \), the Husimi distribution is already close to the extended steady state by \( t = 3 \), whereas for \( S = 64 \), it remains strongly localized at the same time.  
This contrast illustrates that the relaxation time diverges as \( S \to \infty \).}
\label{fig_collective_spin_husimi_evolution}
\end{figure*}

Assume that the deterministic dynamics \( (\sigma = 0) \) are chaotic.  
The natural question is how the Liouvillian gap behaves as the noise strength \( \sigma \) approaches zero.
In the limit \( \sigma \to 0 \), the stochastic propagator \( \mathcal{L} \) reduces to the Perron-Frobenius operator defined in Eq.~\eqref{def_Perron_Frobenius_operator}.  
Hence, it is plausible that the eigenvalues of \( \mathcal{L} \) coincide with the Ruelle-Pollicott resonances of the underlying deterministic map \cite{Nonnenmacher-03}.
Under this assumption, the Liouvillian gap in the noiseless limit should converge to the mixing rate of the deterministic dynamics, which governs the long-time decay of correlation functions.
Appendix~\ref{appendix:Liouvillian_gap_of_stochastic_system} confirms this conjecture numerically for the simple map \( f(x) = \sin x \).
Since a dissipative quantum system can often be mapped onto an effective classical stochastic problem, the behavior of the Liouvillian gap in the noiseless limit offers further support to our conclusions about the spectral gap in driven dissipative quantum systems.

\section{Relaxation time}
\label{sec:relaxation_time}

In this section, we discuss the relaxation time in the time glass phase.  
For finite open quantum systems, the density matrix \(\hat{\rho}(t)\) relaxes to a time-independent steady state \(\hat{\rho}_{\mathrm{ss}}\) within a finite timescale \(\tau_{\text{rel}}\).  
In the time glass phase, \(\tau_{\text{rel}}\) characterizes the duration over which chaotic oscillations persist in a finite system.  
As the system size increases, \(\tau_{\text{rel}}\) diverges.  
At first glance, this divergence appears to contradict the existence of a nonzero Liouvillian gap \(\Delta\), since it is often assumed that the relaxation time \(\tau_{\text{rel}}\) is given by  
\begin{equation}
\tau_{\mathrm{rel}} \sim \frac{1}{\Delta}.  
\label{relaxation_time_gap}
\end{equation}  
In the following, we explain how this apparent paradox is resolved by focusing on the kicked collective spin model.

As the initial state of the kicked collective spin model, we consider a coherent state defined as  
\begin{equation}
\ket{\theta, \phi} = \exp(-i \phi \hat{S}^z) \exp(-i \theta \hat{S}^y) \ket{S, S},
\label{spin_coherent_state_def}
\end{equation}  
where \(\ket{S, S}\) denotes the spin-up highest weight state.
The coherent state \(\ket{\theta, \phi}\) represents the quantum mechanical counterpart of a classical spin pointing in the direction specified by the polar angle \(\theta\) and the azimuthal angle \(\phi\) \cite{Robert_Combescure}.
The initial density matrix is given by 
\begin{equation}
\hat{\rho}_{\mathrm{ini}} = \ket{\theta, \phi} \bra{\theta, \phi}.
\label{rho_ini_coherent}
\end{equation}
Although the subsequent time evolved state \(\rho(t)\) generally deviates from a coherent state, in the semiclassical regime where \(S\) is sufficiently large, it is expected to remain close to a coherent state up to a certain time.  
The timescale \(\tau_{\mathrm{rel}}\) at which a crossover occurs from this localized classical-like state to a delocalized quantum state defines the relaxation time.

To observe how the localized wave packet spreads over time, we focus on the Husimi function defined as  
\begin{equation}
Q(\theta, \phi) = \frac{2S+1}{4\pi} \bra{\theta, \phi} \rho \ket{\theta, \phi}.
\label{Husimi_function_def}
\end{equation}
Unlike the Wigner function, the Husimi function always takes non-negative values.  
The Husimi function can be interpreted as a smoothed version of the Wigner function, obtained by convolving it with an appropriate Gaussian function.  
The function \(Q(\theta, \phi)\) is normalized according to  
\begin{equation}
\int_0^{2\pi} d\phi \int_0^{\pi} d\theta \, \sin \theta \, Q(\theta, \phi) = 1.  
\end{equation}

Figure~\ref{fig_collective_spin_husimi_evolution} shows the time evolution of the Husimi function for the kicked collective spin model in the time glass phase.  
The initial state is the coherent state with polar angle \(\theta = \pi/2\) and azimuthal angle \(\phi = 0\).  
For the case of \(S = 8\) (upper row), by the third cycle \(t = 3\), the distribution nearly coincides with the steady state distribution shown in the rightmost panel (\(t = \infty\)).  
Thus, the relaxation time in this case can be estimated as \(\tau_{\mathrm{rel}} \approx 3\).  
In contrast, for \(S = 64\) (lower row), the wave packet remains sharply localized at \(t=3\).  
In this case, the relaxation time is expected to be larger than \(3\).  
As a result, we can observe that the relaxation time \(\tau_{\mathrm{rel}}\) diverges in the classical limit \(S \to \infty\).

Let us consider the \( S \)-dependence of the relaxation time \( \tau_{\text{rel}} \).  
The width of the Wigner (or Husimi) function of a spin coherent state in the \( (\theta, \phi) \) space typically scales as \( S^{-1/2} \) \cite{Dutta-25}.
In the semiclassical regime with \( S \gg 1 \), the time evolution of the Wigner function is approximately described by the classical Liouville equation, up to the time when fine structures smaller than the length scale determined by the uncertainty principle develop.
Let \( h_{\mathrm{KS}} \) denote the Kolmogorov-Sinai entropy of the corresponding classical dynamics, which is given by the sum of positive Lyapunov exponents.  
Under chaotic dynamics, the width of the Wigner function increases exponentially as \( e^{h_{\mathrm{KS}} t} \).  
The relaxation time \( \tau_{\text{rel}} \) corresponds to the time at which the width of the Wigner function becomes \( O(1) \), at which point the wavepacket extends over the entire phase space.  
Therefore, we obtain  
\begin{equation}
\tau_{\text{rel}} \sim \frac{\log S}{h_{\mathrm{KS}}}.
\label{tau_rel_S}
\end{equation}
It is important to note that Eq.~\eqref{tau_rel_S} corresponds to the Ehrenfest time, the timescale on which the quantum evolution closely follows the corresponding classical dynamics \cite{Robert_Combescure}.
In the present case, the factor \( S^{-1} \) plays the role of the Planck constant \( \hbar \).

\begin{figure}
\centering
\includegraphics[width=8.6cm]{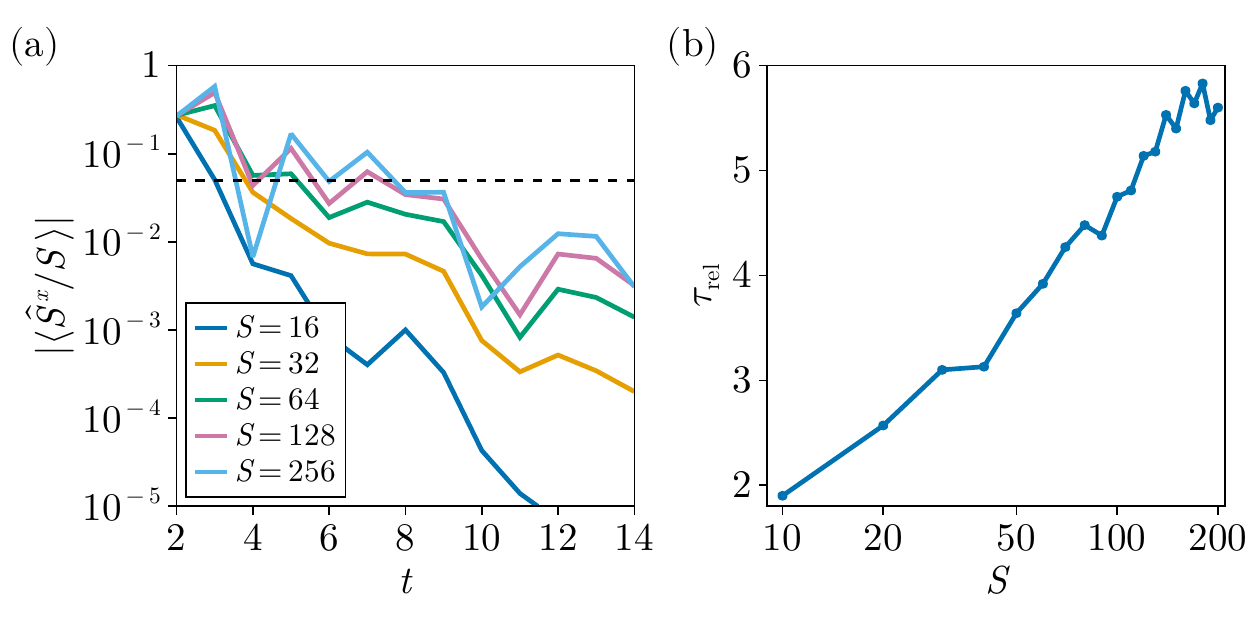}
\caption{Relaxation time of the kicked collective spin model in the time glass phase.
The parameters are fixed at \( \omega_{xx} = 3 \), \( \omega_z = \pi/2 \), and \( \kappa = 1 \). 
(a) Time evolution of \( |\langle \hat{S}^x / S \rangle| \) for \( S = 16,\, 32,\, 64,\, 128,\, 256 \).  
The initial state is the coherent state with polar angle \( \theta = \pi/2 \) and azimuthal angle \( \phi = 0 \).  
The dashed line indicates the value 0.05.
The relaxation time \( \tau_{\text{rel}} \) is defined as the largest time at which \( |\langle \hat{S}^x / S \rangle| \) exceeds 0.05.
(b) \( \tau_{\text{rel}} \) as a function of \( S \).  
For each value of \( S \), the average of \( \tau_{\text{rel}} \) is calculated over 100 initial coherent states corresponding to uniformly random samples of \( (\theta, \phi) \).  
The horizontal axis is plotted on a logarithmic scale, and the trend \( \tau_{\text{rel}} \propto \log S \) is observed.}
\label{fig_collective_spin_relaxation_time}
\end{figure}

To confirm the logarithmic dependence of the relaxation time given by Eq.~\eqref{tau_rel_S}, we investigate the decay of \( \langle \hat{S}^x \rangle \).  
Figure~\ref{fig_collective_spin_relaxation_time}(a) shows the time evolution of \( |\langle \hat{S}^x / S \rangle| \) for different values of \( S \).  
We define the relaxation time \( \tau_{\text{rel}} \) as the largest time at which \( |\langle \hat{S}^x / S \rangle| > 0.05 \).  
As \( S \) increases, the decay of \( |\langle \hat{S}^x / S \rangle| \) becomes slower and \( \tau_{\text{rel}} \) correspondingly increases.  
Figure~\ref{fig_collective_spin_relaxation_time}(b) displays \( \tau_{\text{rel}} \) as a function of \( S \).  
We here take the average of \( \tau_{\text{rel}} \) over 100 initial coherent states, each corresponding to a uniformly random sample of \( (\theta, \phi) \).  
The results clearly exhibit the logarithmic dependence predicted by Eq.~\eqref{tau_rel_S}.

We now turn to the paradox of the divergent relaxation time in the time glass phase despite the presence of a finite Liouvillian gap.  
The discrepancy between the relaxation time and the Liouvillian gap in open quantum systems has attracted significant attention in recent years \cite{Song-19, Mori-20, Haga-21, Bensa-21, Mori-23}.
An important point is that, while the Liouvillian gap determines the asymptotic relaxation rate at long times, it does not govern the duration of the initial transient dynamics.
In the eigenmode expansion given by Eq.~\eqref{eigenmode_expansion}, many eigenmodes with various decay rates contribute to the transient behavior at early times.
If the coefficients \( c_{\alpha} \) in Eq.~\eqref{eigenmode_expansion} corresponding to slowly decaying modes have large magnitudes, the transient dynamics can persist for arbitrarily long times, even when the spectrum has a finite gap.
In general, when the initial state \( \hat{\rho}_{\mathrm{ini}} \) is far from the steady state \( \hat{\rho}_{\mathrm{ss}} \), the expansion coefficients \( c_{\alpha} \) tend to have larger magnitudes.
Therefore, the distance between \( \hat{\rho}_{\mathrm{ini}} \) and \( \hat{\rho}_{\mathrm{ss}} \) must be taken into account when evaluating the relaxation time.

Following Ref.~\cite{Mori-23}, we introduce a refined bound relating the relaxation time \( \tau_{\text{rel}} \) and the Liouvillian gap \(\Delta\),
\begin{equation}
\tau_{\text{rel}} \lesssim \frac{S_2(\hat{\rho}_{\mathrm{ini}}|\hat{\rho}_{\mathrm{ss}})}{\Delta},
\label{relaxation_time_gap_Renyi}
\end{equation}
where 
\begin{equation}
S_2(\hat{\rho}|\hat{\sigma}) := \log \text{Tr}[\hat{\rho}^2 \hat{\sigma}^{-1}]
\end{equation}
is the quantum R\'enyi 2-divergence.
This quantity characterizes the distance between the initial state and the steady state.
While Ref.~\cite{Mori-23} derives a more rigorous bound on \( \tau_{\text{rel}} \) using the symmetrized Liouvillian gap, the simpler estimate in Eq.~\eqref{relaxation_time_gap_Renyi} based on the ordinary Liouvillian gap is sufficient for our present purpose.

The inequality \eqref{relaxation_time_gap_Renyi} can be derived using arguments similar to those presented in Ref.~\cite{Mori-23}.
To estimate the relaxation time, we consider the deviation of an observable \( \hat{O} \) from its steady-state expectation value,
\begin{align}
\Delta O(t) := |\text{Tr}[\hat{O} (\hat{\rho}(t) - \hat{\rho}_{\mathrm{ss}})]| = |\text{Tr}[\hat{O}(t) (\hat{\rho}_{\text{ini}} - \hat{\rho}_{\mathrm{ss}})]|,
\end{align}
where \(\hat{O}(t) = (\mathcal{U}^\dag)^t(\hat{O})\) is the time-evolved observable in the Heisenberg picture.
The relaxation time \( \tau_{\text{rel}} \) can be defined as the time at which \(\Delta O(t)\) drops below a certain threshold.
Here, we assume that the observable satisfies \( \text{Tr}[\hat{O} \hat{\rho}_{\mathrm{ss}}] = 0 \), as is the case for \( \hat{S}_x \) in the kicked collective spin model.
To proceed, we introduce an inner product for operators \( \hat{A} \) and \( \hat{B} \) by \( \langle \hat{A}, \hat{B} \rangle_{\text{ss}} := \text{Tr}[\hat{A}^\dag \hat{B} \hat{\rho}_{\text{ss}}] \), where \( \hat{\rho}_{\text{ss}} \) is the steady state.
By using the Cauchy-Schwarz inequality, we obtain
\begin{align}
|\text{Tr}[\hat{O}(t) (\hat{\rho}_{\text{ini}} - \hat{\rho}_{\mathrm{ss}})]| &= |\langle \hat{O}(t), \hat{\rho}_{\text{ini}} \hat{\rho}_{\text{ss}}^{-1} - \hat{I} \rangle_{\text{ss}}| \notag \\
&\leq \| \hat{O}(t) \|_{\text{ss}} \| \hat{\rho}_{\text{ini}} \hat{\rho}_{\text{ss}}^{-1} - \hat{I} \|_{\text{ss}},
\end{align}
where \( \| \hat{A} \|_{\text{ss}}^2 = \langle \hat{A}, \hat{A} \rangle_{\text{ss}} \).

Let \( \hat{\pi}_\alpha \ (\alpha=0, 1, \ldots) \) be the eigenmodes of \( \mathcal{U}^\dag \) with corresponding eigenvalues \( \lambda_\alpha^* \), which are the complex conjugates of those of \( \mathcal{U} \).
Note that \( \hat{\pi}_0 = \hat{I} \), corresponding to the eigenvalue \( \lambda_0 = 1 \).
The eigenmode expansion of \(\hat{O}(t) = (\mathcal{U}^\dag)^t(\hat{O})\) is given by
\begin{equation}
\hat{O}(t) = \sum_{\alpha > 0} b_\alpha (\lambda_\alpha^*)^t \hat{\pi}_\alpha,
\label{observable_eigenmode_expansion}
\end{equation}
where \( b_\alpha \) are the expansion coefficients of \( \hat{O} \).
The term \( \hat{\pi}_0 = \hat{I} \) does not appear in this expansion due to the assumption \( \text{Tr}[\hat{O} \hat{\rho}_{\mathrm{ss}}] = 0 \).
At long times, the mode corresponding to the eigenvalue \(\lambda_1\) with the second largest modulus dominates, giving:
\begin{equation}
\| \hat{O}(t) \|_{\text{ss}}^2 \sim |\lambda_1|^{2t} \sim e^{-2t\Delta},
\end{equation}
where \( \Delta \) is the Liouvillian gap.
Next, we evaluate the second factor in the inequality:
\begin{align}
\| \hat{\rho}_{\text{ini}} \hat{\rho}_{\text{ss}}^{-1} - \hat{I} \|_{\text{ss}}^2 &= \text{Tr}[( \hat{\rho}_{\text{ss}}^{-1} \hat{\rho}_{\text{ini}} - \hat{I}) (\hat{\rho}_{\text{ini}} \hat{\rho}_{\text{ss}}^{-1} - \hat{I}) \hat{\rho}_{\text{ss}}] \notag \\
&= \text{Tr}[\hat{\rho}_{\text{ini}}^2 \hat{\rho}_{\text{ss}}^{-1}] - 1 \notag \\
&= e^{S_2(\hat{\rho}_{\mathrm{ini}}|\hat{\rho}_{\mathrm{ss}})} - 1.
\end{align}
Assuming \( S_2(\hat{\rho}_{\mathrm{ini}}|\hat{\rho}_{\mathrm{ss}}) \gg 1 \), we have
\begin{equation}
\Delta O(t) \lesssim  \| \hat{O} \|_{\text{ss}} e^{-t\Delta + S_2(\hat{\rho}_{\mathrm{ini}}|\hat{\rho}_{\mathrm{ss}})/2},
\end{equation}
which implies the inequality \eqref{relaxation_time_gap_Renyi}.

\begin{figure}
\centering
\includegraphics[width=8.6cm]{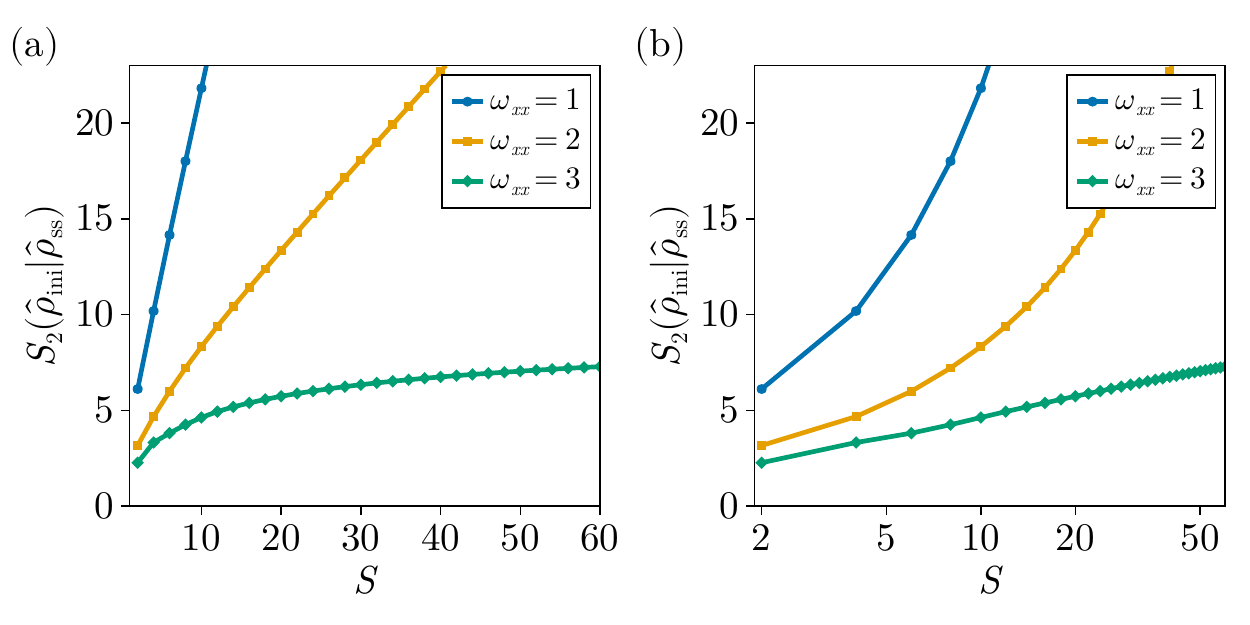}
\caption{Quantum R\'enyi 2-divergence \( S_2(\hat{\rho}_{\mathrm{ini}}|\hat{\rho}_{\mathrm{ss}}) \) as a function of \( S \) with \( \omega_{xx} = 1 \) (stable fixed point), \( \omega_{xx} = 2 \) (time crystal), and \( \omega_{xx} = 3 \) (time glass).
The other parameters are fixed at \( \omega_z = \pi/2 \), and \( \kappa = 1 \). 
The initial state is the coherent state with polar angle \( \theta = \pi/2 \) and azimuthal angle \( \phi = 0 \).
(b) The same plot as (a), but with the horizontal axis plotted on a logarithmic scale.
In the time glass phase, the trend \( S_2(\hat{\rho}_{\mathrm{ini}}|\hat{\rho}_{\mathrm{ss}}) \propto \log S\) is observed.}
\label{fig_collective_spin_initial_renyi}
\end{figure}

Figure \ref{fig_collective_spin_initial_renyi} shows \( S_2(\hat{\rho}_{\mathrm{ini}}|\hat{\rho}_{\mathrm{ss}}) \) as a function of \(S\) for three regimes of the kicked collective spin model: \( \omega_{xx}=1 \) (stable fixed point), \( \omega_{xx}=2 \) (time crystal), and \( \omega_{xx}=3 \) (time glass).
The initial state is the coherent state with polar angle \( \theta = \pi/2 \) and azimuthal angle \( \phi = 0 \).
For both the stable fixed point and the time crystal regimes, we observe
\begin{equation}
S_2(\hat{\rho}_{\mathrm{ini}}|\hat{\rho}_{\mathrm{ss}}) \propto S.
\label{renyi_S_time_crystal}
\end{equation}
The origin of this linear scaling is as follows.
In the fixed point phase, the initial state \( \hat{\rho}_{\mathrm{ini}} \) and the steady state \( \hat{\rho}_{\mathrm{ss}} \) are each localized wave packets on the Bloch sphere with angular width \( S^{-1/2} \) \cite{Dutta-25}.  
Since the two packets are separated by a finite angle, their overlap decays as \( e^{-kS} \) with \( k>0 \).  
Writing \( \hat{\rho}_{\mathrm{ss}} = \sum_{n} p_n \ket{n}\bra{n} \) in an orthonormal basis, we have  
\begin{equation}
\text{Tr}[\hat{\rho}_{\text{ini}}^2 \hat{\rho}_{\text{ss}}^{-1}] = \sum_n \frac{|\langle n | \theta, \phi \rangle|^2}{p_n}.
\end{equation}
The dominant contribution arises from basis states \( \ket{n} \) whose overlap with the initial coherent state is \( O(1) \); the corresponding weights \( p_n \) are exponentially small in \( S \) because those packets \( \ket{n} \) lie far from the support of \( \hat{\rho}_{\mathrm{ss}} \).  
Consequently the trace grows as \( e^{kS} \), and taking the logarithm yields Eq.~\eqref{renyi_S_time_crystal}.
In the time crystal phase, the steady state is an equal mixture of two symmetry-broken packets located symmetrically around the south pole.  
Each packet again has width \( S^{-1/2} \), so the same geometric argument leads to the scaling Eq.~\eqref{renyi_S_time_crystal}.  

In the time glass phase, Fig.~\ref{fig_collective_spin_initial_renyi} shows that the quantum R\'enyi 2-divergence exhibits logarithmic dependence:
\begin{equation}
S_2(\hat{\rho}_{\mathrm{ini}}|\hat{\rho}_{\mathrm{ss}}) \propto \log S.
\label{renyi_S_time_glass}
\end{equation}
This behavior reflects the delocalized nature of the steady state in the time glass phase.  
Unlike the stable fixed point or time crystal phases, where the steady state is concentrated in a narrow angular region, the steady state in the time glass phase is broadly extended over the Hilbert space.
For sufficiently strong kicks \( \omega_{xx} \), the steady state is well approximated by the maximally mixed state:
\begin{equation}
\hat{\rho}_{\mathrm{ss}} \simeq \frac{\hat{I}}{D}, \quad D = 2S + 1,
\end{equation}
where \( D \) is the dimension of the symmetric Hilbert space.  
Given that the initial state is the pure coherent state defined in Eq.~\eqref{rho_ini_coherent}, the R\'enyi 2-divergence becomes \(\text{Tr} [\hat{\rho}_{\mathrm{ini}}^{2} \hat{\rho}_{\mathrm{ss}}^{-1}] = D \text{Tr} [\hat{\rho}_{\mathrm{ini}}] = D\), which implies the logarithmic dependence in Eq.~\eqref{renyi_S_time_glass}.

The upper bound given by Eq.~\eqref{relaxation_time_gap_Renyi}, together with the logarithmic scaling of the R\'enyi divergence in Eq.~\eqref{renyi_S_time_glass}, is consistent with the logarithmic divergence of the relaxation time described by Eq.~\eqref{tau_rel_S}.  
Furthermore, Eqs.~\eqref{tau_rel_S}, \eqref{relaxation_time_gap_Renyi}, and \eqref{renyi_S_time_glass} together imply a relationship between the Liouvillian gap \( \Delta \) and the Kolmogorov-Sinai entropy \( h_{\mathrm{KS}} \) of the classical dynamics:
\begin{equation}
\Delta \sim h_{\mathrm{KS}}.
\end{equation}
This relationship is consistent with our main result given by Eq.~\eqref{gap_mixing_rate} because the mixing rate \( g \) is identical to the sum of positive Lyapunov exponents in classical dynamical systems with small noise \cite{Ruelle-86}. 

It is important to note that Eq.~\eqref{relaxation_time_gap_Renyi} does not always provide a tight upper bound on the relaxation time.
To clarify the situation, we summarize below the scaling behavior of the Liouvillian gap \( \Delta \), the relaxation time \( \tau_{\mathrm{rel}} \), and the R\'enyi-2 divergence \(S_2(\hat{\rho}_{\mathrm{ini}}|\hat{\rho}_{\mathrm{ss}})\) for the three relevant phases:
\begin{itemize}
\item {\it Disordered phase:} $\Delta \sim O(1)$, $\tau_{\text{rel}} \sim O(1)$, but $S_2 \sim O(S)$. 
The bound $S_2/\Delta \sim O(S)$ is non-tight.
\item {\it Time crystal phase:} $\Delta \sim e^{-c S}$, $\tau_{\text{rel}} \sim e^{c S}$, but $S_2 \sim O(S)$. 
The bound $S_2/\Delta \sim S e^{c S}$ is non-tight.
\item {\it Time glass phase:} $\Delta \sim O(1)$, $\tau_{\text{rel}} \sim \log S$, and $S_2 \sim \log S$. 
The bound $S_2/\Delta \sim \log S$ is tight.
\end{itemize}

We now explain why the bound becomes loose in the disordered and time crystal phases. 
In these phases, both the initial coherent state and the steady state form localized wavepackets whose widths scale as \( S^{-1/2} \).
Relaxation is therefore dominated by the \emph{motion of the packet center} toward the attracting point, rather than by an expansion of its width.
Because the Liouvillian gap faithfully captures the rate of this center motion, one finds $\tau_{\mathrm{rel}} \sim \Delta^{-1}$ in these phases.
The linear growth of the R\'enyi-2 divergence, $S_2 \propto S$, in these phases arises from the exponentially small overlap between two narrow packets whose centers are separated in phase space.
However, \( S_2 \) alone cannot distinguish whether the difference between two states arises from the \emph{shape} (width) of the packets or merely from their \emph{positions}.
As a result, \( S_2 \) overestimates the effective distance relevant for relaxation, leading to a loose upper bound in Eq.~\eqref{relaxation_time_gap_Renyi}.

In contrast, in the time glass phase, the steady state becomes nearly extended over the entire phase space.  
Relaxation is then governed not by the motion of a wavepacket center but by the \emph{broadening} of its width.
In this situation, the R\'enyi-2 divergence captures the physically relevant distance between the two states, and the bound in Eq.~\eqref{relaxation_time_gap_Renyi} becomes tight.
This explains why Eq.~\eqref{relaxation_time_gap_Renyi} yields the correct scaling \( \tau_{\mathrm{rel}} \propto \log S \) in the time glass phase.

\section{Discussion}
\label{sec:discussion}

The primary limitation of the present study is that our comprehensive analysis of spectral features and relaxation dynamics is restricted to fully-connected mean-field models, namely the kicked collective spin and the kicked spin chain with all-to-all coupling, both of which retain permutation symmetry among the spins. 
Such mean-field models offer significant analytical advantages: (i) they possess well-defined semiclassical limits, where the dynamics of the order parameter are governed by a closed, finite set of deterministic nonlinear equations, and (ii) by restricting the analysis to the permutation-invariant sector of the Hilbert space, the effective dimension of the Liouvillian is drastically reduced, enabling the study of spectral properties via exact diagonalization for moderately large system sizes. 
These analytical conveniences are lost in systems featuring short-range interactions where the permutation symmetry is broken.

The absence of permutation symmetry introduces subtle physical problems. 
For the kicked collective spin, the symmetric subspace (with the maximal total spin $S=N/2$) occupies only a polynomially small fraction of the entire exponentially large Hilbert space ($2^N$). 
Thus, even weak perturbations that break the permutation symmetry could potentially cause the system to scramble into the exponentially large asymmetric subspaces ($S < N/2$), thereby destroying the time glass phase. 
Furthermore, the absence of symmetry leads to the non-closure of the BBGKY hierarchy. 
For short-range interacting systems, the dynamics of macroscopic observables require an infinite hierarchy of coupled equations involving multi-body correlations. 
Consequently, the time evolution of the order parameter cannot be captured by a finite set of deterministic equations, and fluctuations in higher-order correlations pose a genuine danger of washing out the synchronized, coherent dynamics of the macroscopic order parameter.

It is important to note that our results for the kicked spin chain with a nonzero interaction exponent $\alpha>0$ suggest that the time glass phase survives despite the breakdown of strict permutation symmetry (see Fig.~\ref{fig_spin_chain_phase_diagram}).
This observed robustness implies that the synchronization mechanism driven by spin interactions is strong enough to overcome the entropic force pulling the system into the exponentially large asymmetric subspaces.
However, these preliminary calculations employ the Quantum Trajectory Gutzwiller Approximation, which inherently neglects non-local inter-site quantum correlations (entanglement). 
Therefore, verifying the stability of the time glass phase against permutation symmetry breaking in a fully quantum mechanical regime remains a paramount open question. 
Addressing this challenge requires deploying more sophisticated numerical techniques capable of capturing quantum correlations beyond the mean-field level, such as cluster mean-field theories or tensor network approaches.

In the remainder of this section, we discuss several open problems and promising directions for future research. 
For time crystals, the established spatial long-range order grants the system rigidity, ensuring the oscillation frequency is robust against small perturbations. 
A similar degree of robustness is naturally anticipated for the chaotic dynamics characteristic of the time glass phase. 
In isolated quantum many-body systems, the sensitivity of dynamics to local perturbations is quantified by the out-of-time-ordered correlator (OTOC), defined as:
\begin{equation}
C_{ij}(t) = \langle [\hat{S}_i(t), \hat{S}_j]^\dag [\hat{S}_i(t), \hat{S}_j] \rangle,
\end{equation}
where $\hat{S}_i(t)$ is the spin operator at site $i$ evolved in time $t$. 
The OTOC $C_{ij}(t)$ fundamentally characterizes how a local perturbation applied at site $i$ spreads and affects the observable at site $j$ after time $t$. 
In typical chaotic phases, the OTOC exhibits an initial exponential growth followed by a ballistic light-cone structure, indicating rapid information scrambling across the lattice \cite{Roberts-16, Nahum-17, Keyserlingk-18, Nahum-18, Khemani-18-1, Rakovszky-18, Khemani-18-2}. 
Several recent theoretical proposals have successfully extended the notion of OTOC to open quantum systems to quantify the scrambling processes associated with dissipative quantum chaos \cite{Zhang-19, Touil-21, Zanardi-21, Schuster-23}. 
An intriguing issue is how the time glass phase is distinguished from conventional many-body chaos via the OTOC. 
We hypothesize that, for the time glass, the persistent spatial long-range order ensures the phase's rigidity against local perturbations, leading to a strong suppression and smearing of the spatial spreading of the correlation $C_{ij}(t)$, which would manifest as an anomalous light-cone structure.

In this study, we have focused on periodically driven systems.
However, time-crystalline order can also emerge in undriven dissipative systems governed by a time-independent Liouvillian, a phenomenon known as the continuous time crystal.
This naturally raises the question of whether a time glass phase can also arise in systems with a time-independent Liouvillian.
Indeed, several previous studies report chaotic behavior of macroscopic observables in such autonomous settings \cite{Solanki-24-2, Yang-25}.
In this case, the Liouvillian spectrum lies in the left half of the complex plane with non-positive real parts, and the Liouvillian gap is defined as the smallest absolute value of the real part among the nonzero eigenvalues.
It is known that, in continuous time crystals, the Liouvillian gap closes algebraically with increasing system size \cite{Iemini-18, Cabot-24}, rather than exponentially.
Our results suggest that, even in this autonomous setting, the time glass phase may be characterized by a finite Liouvillian gap, which corresponds to the mixing rate of the classical dynamics in the thermodynamic limit.
Verifying this conjecture might pose significant challenges.  
Autonomous chaotic dynamics typically require more degrees of freedom than their periodically driven counterparts, and the exact diagonalization of such systems becomes numerically intractable as system size increases.

Throughout this work we have analysed models possessing a discrete \( \mathbb{Z}_{2} \) symmetry.
An important open question is how our conclusions are altered in systems with a continuous symmetry, such as \( U(1) \).
In isolated systems, the spontaneous breaking of a continuous symmetry leads to the emergence of gapless excitations, Goldstone modes.
A similar statement holds for dissipative many-body systems that support static long-range order [see Fig.~\ref{fig_time_glass}(b)], for example exciton-polariton condensates, where phase fluctuations remain gapless even in the presence of loss and drive \cite{Wouters-06, Wouters-07, Szymanska-06}.
If the time glass phase associated with the spontaneous breaking of a continuous \( U(1) \) symmetry exhibits a finite Liouvillian gap, it raises a fundamental question regarding the fate of the Goldstone modes.  
Understanding whether and how these modes become gapped in such nonequilibrium settings is crucial for extending the theoretical framework of time glasses to systems with continuous symmetries.

Finally, we comment on the possibility of realizing a time glass phase in periodically driven isolated systems.  
In such systems, the situation is more subtle because periodic driving typically induces heating toward an infinite-temperature state, corresponding to the disordered phase illustrated in Fig.~\ref{fig_time_glass}(a).  
A well-known mechanism to avoid this runaway heating is many-body localization induced by quenched disorder, as employed in the construction of isolated DTCs.
At first glance, the coexistence of time glass behavior and many-body localization appears unlikely.  
This is because many-body localized systems exhibit Poissonian level statistics in the spectrum of the Floquet operator, rather than the Wigner-Dyson statistics associated with quantum chaos.  
Such Poissonian statistics are generally interpreted as the absence of quantum chaos at the microscopic level.
However, it is important to distinguish between microscopic and macroscopic chaos.
While random matrix theory is a hallmark of microscopic chaos, signaling level repulsion and complex dynamics at the level of individual degrees of freedom, it does not necessarily characterize the chaotic behavior of macroscopic observables.
It remains an intriguing open question whether a system that lacks microscopic chaos, in the sense of random matrix level statistics, can nevertheless exhibit macroscopic chaos in the form of chaotic dynamics of collective degrees of freedom.
Clarifying this point is essential for understanding the feasibility of time glass phases in isolated many-body systems.

\section{Conclusion}
\label{sec:conclusion}

In this study, we introduced the concept of the \emph{time glass}, a dynamical phase in periodically driven dissipative quantum many-body systems that serves as a minimal model for macroscopic chaos. 
The time glass is rigorously defined by the coexistence of two key features in the thermodynamic limit: (i) spatial long-range order arising from the spontaneous breaking of an internal symmetry, and (ii) indefinitely persistent, temporally chaotic oscillations of the corresponding order parameter.

Our primary contribution lies in the spectral characterization of this phase. 
Through numerical studies of kicked dissipative Ising models, we demonstrated that, unlike time crystals where the Liouvillian gap $\Delta$ vanishes exponentially with system size, the time glass phase maintains a finite Liouvillian gap ($\Delta_\infty > 0$) in the thermodynamic limit. 
Crucially, we established a direct correspondence between microscopic quantum spectral properties and macroscopic classical dynamics by showing that this finite gap coincides with the decay rate $g$ of the order-parameter autocorrelation function in the thermodynamic limit: $\Delta_\infty = g$.
This result challenges the conventional understanding that spontaneous symmetry breaking necessitates the closing of the spectral gap.

Furthermore, we resolved the apparent paradox posed by the coexistence of a finite gap (implying rapid asymptotic decay) and indefinitely sustained chaotic dynamics (implying long-lived transients). 
We demonstrated that the relaxation time $\tau_{\mathrm{rel}}$, which marks the duration of the classical-like chaotic transient, diverges logarithmically with system size: $\tau_{\mathrm{rel}} \propto \log N$. 
This divergence is accommodated by the logarithmic growth of the quantum R\'enyi 2-divergence ($S_2 \propto \log N$) between the localized initial state and the extended steady state, confirming that the transient stage is extended by the ``distance" between these states, rather than the intrinsic slowing down of asymptotic decay.

While the present work characterizes the time glass within mean-field (fully-connected) models, several fundamental open questions remain. 
Foremost among these is fully verifying the robustness of the phase against permutation-symmetry-breaking perturbations, such as short-range interactions, which require numerical methods beyond the mean-field approximation.
Additionally, future theoretical efforts should aim to connect the time glass characteristics with diagnostics like the OTOC and the spectral statistics in open systems. 
Ultimately, the time glass provides a unique platform for exploring the fundamental physics of synchronization and chaos in nonequilibrium quantum matter.

\begin{acknowledgments}
This work was supported by JSPS KAKENHI Grant Number JP22K13983.
\end{acknowledgments}

\section*{DATA AVAILABILITY}
The data and codes that support the findings of this article are openly available \cite{codes}.

\appendix

\section{Quantum-classical correspondence of autocorrelations}
\label{sec:quantum_classical_correspondence}

\begin{figure*}
\centering
\includegraphics[width=\textwidth]{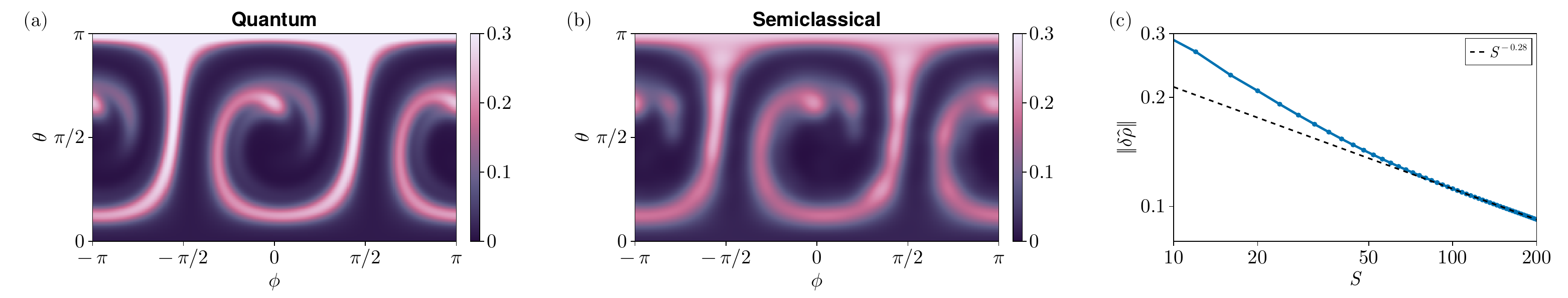}
\caption{Numerical test of the quantum-classical correspondence of the steady state, given in Eq.~\eqref{steady_state_quantum_classical_correspondence}.
(a) Husimi function of the quantum steady state $\hat{\rho}_{\mathrm{ss}}$.
(b) Husimi function of the semiclassical steady state $\hat{\rho}_{\mathrm{ss}}^{(\mathrm{cl})}$, which is given by Eq.~\eqref{semiclassical_steady_state}.
The total spin is $S=64$ in (a) and (b).
(c) Hilbert--Schmidt norm of the deviation $\delta \hat{\rho} = \hat{\rho}_{\mathrm{ss}} - \hat{\rho}_{\mathrm{ss}}^{(\mathrm{cl})}$ as a function of $S$.
The dashed line shows a power-law fit over the range $[100, 200]$, yielding $\| \delta \hat{\rho} \| \propto S^{-0.28}$.
The parameters are fixed at \( \omega_{xx} = 4 \), \( \omega_z = \pi/2 \), and \( \kappa = 1 \).}
\label{fig_steady_state_correspondence}
\end{figure*}

In this Appendix, we examine in detail the quantum-classical correspondence of autocorrelations, which is given by Eq.~\eqref{autocorrelation_quantum_classical_correspondence}.
For clarity and to make the discussion self-contained, we restate the setup in the context of the collective spin model.
Consider an observable $\hat{O}=f(\hat{\boldsymbol{S}}/S)$, where $\hat{\boldsymbol{S}}=(\hat{S}^x, \hat{S}^y, \hat{S}^z)$ and $f(\boldsymbol{n})$ is an arbitrary analytic function.
The associated autocorrelation function $C_O(t)$ $(t=0, 1, \ldots)$ is defined by
\begin{equation}
C_O(t) = \text{Tr}[\hat{O} \, \mathcal{U}^t (\hat{O} \hat{\rho}_{\mathrm{ss}})],
\end{equation}
where $\hat{\rho}_{\mathrm{ss}}$ is the stroboscopic steady state.
The classical dynamics of $\boldsymbol{m} = \langle \hat{\boldsymbol{S}}/S \rangle$ is described by Eqs.~\eqref{collective_spin_classical_equation_1} and \eqref{collective_spin_classical_equation_2}.
We assume that this classical dynamics is chaotic and ergodic.
The corresponding autocorrelation is defined as
\begin{equation}
C_O^{(\mathrm{cl})}(t) = \lim_{K \to \infty} \frac{1}{K} \sum_{k=1}^K f(\boldsymbol{m}(t+k)) f(\boldsymbol{m}(k)),
\end{equation}
where the average is taken along a single classical trajectory $\boldsymbol{m}(t)=(m^x(t), m^y(t), m^z(t))$.
Our conjecture for the quantum-classical correspondence can be written as
\begin{equation}
\lim_{S \to \infty} C_O(t) = C_O^{(\mathrm{cl})}(t),
\label{appendix_autocorrelation_quantum_classical_correspondence}
\end{equation}
for any fixed time $t$.
In the remainder of this Appendix, we establish this correspondence under a reasonable assumption on the structure of the steady state.

A key element in our argument is a theorem established by Carollo and Lesanovsky in Ref.~\cite{Carollo-24}, which guarantees the convergence of quantum expectation values to their classical counterparts.
Consider an initial state prepared as a coherent pure state, $\hat{\rho}(0) = \ket{\boldsymbol{n}_0} \bra{\boldsymbol{n}_0}$, where $\ket{\boldsymbol{n}}$ is the spin coherent state pointing along $\boldsymbol{n}=(\sin \theta \cos \phi, \sin \theta \sin \phi, \cos \theta)$, as defined in Eq.~\eqref{spin_coherent_state_def}.
The theorem proved by Carollo and Lesanovsky guarantees that, for any fixed time $t$, the quantum expectation value $\langle \hat{O}(t) \rangle=\text{Tr}[\hat{O} \hat{\rho}(t)]$ converges, in the limit $S \to \infty$, to the classical value $f(\boldsymbol{m}(t))$ obtained from the equations of motion, Eqs.~\eqref{collective_spin_classical_equation_1} and \eqref{collective_spin_classical_equation_2}, with initial condition $\boldsymbol{m}=\boldsymbol{n}_0$.
This theorem provides a rigorous foundation for the quantum-classical correspondence at the level of individual trajectories.

However, Carollo and Lesanovsky's theorem alone is not sufficient to establish Eq.~\eqref{appendix_autocorrelation_quantum_classical_correspondence}, because the theorem does not provide any information about the long-time behavior of $\hat{\rho}(t)$.
In particular, to relate the quantum autocorrelation to its classical counterpart, we must introduce an assumption concerning the connection between the quantum steady state $\hat{\rho}_{\mathrm{ss}}$ and the classical invariant distribution $\rho_{\mathrm{cl}}(\boldsymbol{n})$, which is defined by
\begin{equation}
\rho_{\mathrm{cl}}(\boldsymbol{n}) = \lim_{T \to \infty} \frac{1}{T} \sum_{t=1}^T \delta(\boldsymbol{n} - \boldsymbol{m}(t)),
\end{equation}
where $\boldsymbol{m}(t)=(m^x(t), m^y(t), m^z(t))$ is the classical trajectory.
To proceed, we introduce the following physically reasonable assumption regarding the structure of the steady state:
\begin{equation}
\lim_{S \to \infty} \left\| \: \hat{\rho}_{\mathrm{ss}} - \int d\boldsymbol{n} \rho_{\mathrm{cl}}(\boldsymbol{n}) \ket{\boldsymbol{n}} \bra{\boldsymbol{n}} \: \right\| = 0.
\label{steady_state_quantum_classical_correspondence}
\end{equation}
This assumption states that, in the semiclassical limit, the quantum steady state becomes indistinguishable from a classical mixture of coherent states weighted by the invariant distribution $\rho_{\mathrm{cl}}(\boldsymbol{n})$.
In the following, we denote the semiclassical approximation of the steady state as
\begin{equation}
\hat{\rho}_{\mathrm{ss}}^{(\mathrm{cl})} := \int d\boldsymbol{n} \rho_{\mathrm{cl}}(\boldsymbol{n}) \ket{\boldsymbol{n}} \bra{\boldsymbol{n}}.
\label{semiclassical_steady_state}
\end{equation}

With the assumption in Eq.~\eqref{steady_state_quantum_classical_correspondence}, the derivation of the quantum-classical correspondence of autocorrelations in Eq.~\eqref{appendix_autocorrelation_quantum_classical_correspondence} becomes straightforward.
For an observable of the form $\hat{O}=f(\hat{\boldsymbol{S}}/S)$, the action of $\hat{O}$ on a coherent state satisfies
\begin{equation}
\hat{O} \ket{\boldsymbol{n}} = f(\boldsymbol{n}) \ket{\boldsymbol{n}} + O(S^{-1/2}).
\end{equation}
Using this relation, the quantum autocorrelation can be expressed as
\begin{equation}
C_O(t) \simeq \int d\boldsymbol{n}_0 \rho_{\mathrm{cl}}(\boldsymbol{n}_0) f(\boldsymbol{n}_0) \text{Tr}[\hat{O} \, \mathcal{U}^t (\ket{\boldsymbol{n}_0} \bra{\boldsymbol{n}_0})].
\end{equation}
According to the theorem of Carollo and Lesanovsky, $\text{Tr}[\hat{O} \, \mathcal{U}^t (\ket{\boldsymbol{n}_0} \bra{\boldsymbol{n}_0})]$ converges, in the limit $S \to \infty$, to the classical value $f(\boldsymbol{m}(t; \boldsymbol{n}_0))$, where $\boldsymbol{m}(t; \boldsymbol{n}_0)$ is the classical trajectory that starts form $\boldsymbol{n}_0$.
Therefore, we have
\begin{equation}
\lim_{S \to \infty} C_O(t) = \int d\boldsymbol{n}_0 \rho_{\mathrm{cl}}(\boldsymbol{n}_0) f(\boldsymbol{n}_0) f(\boldsymbol{m}(t; \boldsymbol{n}_0)),
\end{equation}
which is exactly the classical autocorrelation $C_O^{(\mathrm{cl})}(t)$.

We remark that the convergence of $C_O(t)$ to $C_O^{(\mathrm{cl})}(t)$ can become extremely slow at large $t$.
The key reason is that the convergence of quantum expectation values to their corresponding classical trajectories is governed by the Ehrenfest timescale $\tau_{\text{cl}}$, beyond which the quantum and classical dynamics begin to deviate.
As discussed in Sec.~\ref{sec:relaxation_time}, this timescale grows only logarithmically with $S$ in chaotic systems, $\tau_{\text{cl}} \propto \log S$.
Since the convergence of the autocorrelations depends on this quantum-classical correspondence of trajectories, the same timescale $\tau_{\text{cl}}$ controls how long $C_O(t)$ remains close to $C_O^{(\mathrm{cl})}(t)$.
Therefore, due to the slow (logarithmic) growth of $\tau_{\text{cl}}$, the approach of $C_O(t)$ to its classical counterpart can be extremely slow for large $t$.

Finally, we numerically examine the validity of the quantum-classical correspondence of the steady state expressed in Eq.~\eqref{steady_state_quantum_classical_correspondence}.
Figures \ref{fig_steady_state_correspondence}(a) and (b) show the Husimi functions $Q(\theta, \phi)$, which is defined by Eq.~\eqref{Husimi_function_def}, for the quantum steady state $\hat{\rho}_{\mathrm{ss}}$ and the semiclassical steady state $\hat{\rho}_{\mathrm{ss}}^{(\mathrm{cl})}$, respectively.
The semiclassical steady state is calculated as
\begin{equation}
\hat{\rho}_{\mathrm{ss}}^{(\mathrm{cl})} = \lim_{T \to \infty} \frac{1}{T} \sum_{t=1}^T \ket{\boldsymbol{m}(t)} \bra{\boldsymbol{m}(t)},
\end{equation}
where $\boldsymbol{m}(t)$ is a classical trajectory.
The similarity between Fig.~\ref{fig_steady_state_correspondence}(a) and (b) indicates that $\hat{\rho}_{\mathrm{ss}}^{(\mathrm{cl})}$ provides a good approximation to the quantum steady state in the semiclassical regime.
Figure \ref{fig_steady_state_correspondence}(c) shows the Hilbert--Schmidt norm of the deviation $\delta \hat{\rho} = \hat{\rho}_{\mathrm{ss}} - \hat{\rho}_{\mathrm{ss}}^{(\mathrm{cl})}$ as a function of $S$.
We find a power-law behavior $\| \delta \hat{\rho} \| \propto S^{-0.28}$ at large $S$.
These results support the assumption stated in Eq.~\eqref{steady_state_quantum_classical_correspondence}.

\section{Linear stability analysis of a fixed point}
\label{sec:linear_stability_analysis}

\subsection{Kicked collective spin}

The classical dynamics of the kicked collective spin are governed by Eqs.~\eqref{collective_spin_classical_equation_1} and \eqref{collective_spin_classical_equation_2}.  
We analyze the linear stability of the down-spin state \( \boldsymbol{m} = (0, 0, -1) \).  
By noting that \( m^z = - \sqrt{1 - (m^x)^2 - (m^y)^2} \), we can eliminate \( m^z \) from Eq.~\eqref{collective_spin_classical_equation_1}.  
Keeping only terms linear in \( m^x \) and \( m^y \), we obtain the linearized equations  
\begin{equation}
\begin{split}
\frac{dm^x}{dt} &\simeq  - \kappa m^x - \omega_z m^y, \\
\frac{dm^y}{dt} &\simeq \omega_z m^x - \kappa m^y,
\end{split}
\label{collective_spin_classical_equation_1_linearized}
\end{equation}
where higher-order terms have been neglected. 
This system can be written in matrix form as 
\begin{equation}
\frac{d}{dt}
\begin{pmatrix}
m^x \\
m^y \\
\end{pmatrix}
= A 
\begin{pmatrix}
m^x \\
m^y \\
\end{pmatrix}
, \quad
A = 
\begin{pmatrix}
-\kappa & -\omega_z  \\
\omega_z & -\kappa \\
\end{pmatrix}.
\end{equation}
The linear time evolution of \( (m^x, m^y) \) is thus described by 
\begin{equation}
e^A = e^{-\kappa} 
\begin{pmatrix}
\cos \omega_z & -\sin \omega_z  \\
\sin \omega_z & \cos \omega_z \\
\end{pmatrix},
\end{equation}
which corresponds to a spiral motion around the origin, with angular frequency \( \omega_z \) and decay rate \( \kappa \).

Similarly, Eq.~\eqref{collective_spin_classical_equation_2} can be linearized as
\begin{equation}
\begin{split}
\frac{dm^x}{dt} &\simeq  0, \\
\frac{dm^y}{dt} &\simeq 2\omega_{xx} m^x,
\end{split}
\label{collective_spin_classical_equation_2_linearized}
\end{equation}
which corresponds to the evolution matrix  
\begin{equation}
B = 
\begin{pmatrix}
0 & 0 \\
2 \omega_{xx} & 0
\end{pmatrix}.
\end{equation}
The time evolution generated by \( B \) is given by the matrix exponential  
\begin{equation}
e^B = 
\begin{pmatrix}
1 & 0 \\
2 \omega_{xx} & 1
\end{pmatrix}.
\end{equation}

By combining the two processes, the linearized one-cycle evolution of \( (m^x, m^y) \) is given by
\begin{align}
U &= e^B e^A \notag \\
&= e^{-\kappa}
\begin{pmatrix}
\cos \omega_z & -\sin \omega_z  \\
2\omega_{xx}\cos \omega_z + \sin \omega_z & -2\omega_{xx}\sin \omega_z + \cos \omega_z \\
\end{pmatrix}.
\end{align}
The eigenvalues of \( U \) are  
\begin{align}
\lambda_U^{\pm} = e^{-\kappa} \bigg[ &-\omega_{xx}\sin \omega_z + \cos \omega_z \notag \\
&\pm \sqrt{(\omega_{xx}\sin \omega_z - \cos \omega_z)^2 - 1} \bigg].
\end{align}
The fixed point is stable when \( |\lambda_U^{\pm}| < 1 \) for both eigenvalues, and unstable when \( |\lambda_U^+| > 1 \) or \( |\lambda_U^-| > 1 \).
For \( \omega_z = \pi/2 \), the eigenvalues simplify to
\begin{equation}
\lambda_U^{\pm} = 
\left\{ 
\begin{alignedat}{2}
e^{-\kappa} \left(-\omega_{xx} \pm i \sqrt{1-\omega_{xx}^2} \right)  \quad (\omega_{xx}<1), \\
e^{-\kappa} \left(-\omega_{xx} \pm \sqrt{\omega_{xx}^2-1} \right) \quad (\omega_{xx} \geq 1).
\label{collective_spin_jacobian_eigenvalue_fixed_point}
\end{alignedat}
\right.
\end{equation}
For \( \omega_{xx} < 1 \), the modulus of the eigenvalues is \( |\lambda_U^{\pm}| = e^{-\kappa} \), so the fixed point is stable.  
When \( \omega_{xx} \geq 1 \), the larger modulus of the eigenvalues is \( |\lambda_U^{-}| = e^{-\kappa} (\omega_{xx} + \sqrt{\omega_{xx}^2 - 1}) \).  
The fixed point becomes unstable when this value reaches unity, at which point a limit cycle emerges.
The bifurcation point \( \omega_{xx}^c \) is thus given by
\begin{equation}
\omega_{xx}^c = \frac{e^{2\kappa} + 1}{2 e^{\kappa}}.
\end{equation}
For \( \kappa = 1 \), this yields \( \omega_{xx}^c \simeq 1.543 \), which is consistent with Fig.~\ref{fig_collective_spin_bifurcation}.

\subsection{Kicked spin chain with all-to-all coupling}

The mean-field dynamics of the kicked spin chain with all-to-all coupling are governed by Eqs.~\eqref{spin_chain_mean_field_dissipation} and \eqref{spin_chain_mean_field_kick}.
We analyze the linear stability of the down-spin state \( \boldsymbol{m} = (0, 0, -1) \). 
Defining the deviations $\delta m^x = m^x$, $\delta m^y = m^y$, and $\delta m^z = m^z+1$ from the fixed point, Eq.~\eqref{spin_chain_mean_field_dissipation} can be linearized as
\begin{equation}
\frac{d}{dt}
\begin{pmatrix}
\delta m^x \\
\delta m^y \\
\delta m^z \\
\end{pmatrix}
= A 
\begin{pmatrix}
\delta m^x \\
\delta m^y \\
\delta m^z \\
\end{pmatrix}
, \quad
A = 
\begin{pmatrix}
-\kappa/2 & -\omega_z & 0  \\
\omega_z & -\kappa/2  & 0 \\
0 & 0 & -\kappa
\end{pmatrix}.
\end{equation}
Similarly, Eq.~\eqref{spin_chain_mean_field_kick} can be linearized as
\begin{equation}
\frac{d}{dt}
\begin{pmatrix}
\delta m^x \\
\delta m^y \\
\delta m^z \\
\end{pmatrix}
= B 
\begin{pmatrix}
\delta m^x \\
\delta m^y \\
\delta m^z \\
\end{pmatrix}
, \quad
B = 
\begin{pmatrix}
0 & 0 & 0  \\
4J & 0  & 0 \\
0 & 0 & 0
\end{pmatrix}.
\end{equation}

By combining the two processes, the linearized one-cycle evolution of \( (\delta m^x, \delta m^y, \delta m^z) \) is given by
\begin{align}
U &= e^B e^A \notag \\
&= e^{-\kappa/2}
\begin{pmatrix}
\cos \omega_z & -\sin \omega_z & 0 \\
4J \cos \omega_z + \sin \omega_z & -4J \sin \omega_z + \cos \omega_z & 0 \\
0 & 0 & e^{-\kappa/2} \\
\end{pmatrix}.
\end{align}
The eigenvalues of \( U \) are  
\begin{align}
\lambda_U = e^{-\kappa/2} \bigg[ &-2J \sin \omega_z + \cos \omega_z \notag \\
&\pm \sqrt{(2J \sin \omega_z - \cos \omega_z)^2 - 1} \bigg], \: e^{-\kappa}.
\label{spin_chain_jacobian_eigenvalue_fixed_point}
\end{align}
For the case \( \omega_z = \pi/2 \), a similar argument to the collective spin case yields the bifurcation point  
\begin{equation}
J_c = \frac{e^{\kappa} + 1}{4 e^{\kappa/2}}.
\end{equation}
For \( \kappa = 1 \), we obtain \( J_c = 0.564 \), which is consistent with Fig.~\ref{fig_spin_chain_bifurcation}.

\section{Quantum trajectory method}
\label{sec:quantum_trajectory_method}

In this appendix, we present a detailed explanation of the quantum trajectory method \cite{Dalay-14}, which is applied to the kicked spin chain model.
We begin by considering the general quantum master equation given in Eq.~\eqref{master_equation_general}.  
To facilitate the stochastic unraveling of the dynamics, we introduce the effective non-Hermitian Hamiltonian,  
\begin{equation}
\hat{H}_{\mathrm{eff}} = \hat{H} - \frac{i}{2} \sum_k \hat{L}_k^\dag \hat{L}_k,
\end{equation}
where we omit the explicit time dependence of \( \hat{H} \) and \( \hat{L}_k \) for simplicity.

In this procedure, a single quantum trajectory \( \ket{\psi(t)} \) is computed as follows:  
\begin{enumerate}
\item Set the initial state \( \ket{\psi(0)} \).

\item Sample a uniform random number \( r \in [0, 1] \).

\item Solve the Schr\"odinger equation for the effective Hamiltonian with the initial condition \( \ket{\tilde{\psi}(0)} = \ket{\psi(0)} \):
\begin{equation}
i \partial_t \ket{\tilde{\psi}(t)} = \hat{H}_{\mathrm{eff}} \ket{\tilde{\psi}(t)}.
\end{equation}
Note that the norm \( \braket{\tilde{\psi}(t)| \tilde{\psi}(t)} \) decreases monotonically during this evolution.  
Let \( t_1 \) be the time at which the norm becomes equal to the sampled value \( r \), i.e., \( \braket{\tilde{\psi}(t_1)|\tilde{\psi}(t_1)} = r \).

\item For \( 0 \leq t \leq t_1 \), the normalized quantum trajectory \( \ket{\psi(t)} \) is obtained by
\begin{equation}
\ket{\psi(t)} = \frac{\ket{\tilde{\psi}(t)}}{\sqrt{\braket{\tilde{\psi}(t)| \tilde{\psi}(t)}}}.
\end{equation}

\item At time \( t_1 \), a quantum jump occurs.  
The index \( k \) of the jump operator \( \hat{L}_k \) is chosen with probability
\begin{equation}
p_k = \frac{\bra{\psi(t_1)} \hat{L}_k^\dag \hat{L}_k \ket{\psi(t_1)}}{\sum_j \bra{\psi(t_1)} \hat{L}_j^\dag \hat{L}_j \ket{\psi(t_1)}}.
\end{equation}
The state immediately after the jump is given by
\begin{equation}
\ket{\psi(t_1^+)} = \frac{\hat{L}_k \ket{\psi(t_1^-)}}{\sqrt{\bra{\psi(t_1^-)} \hat{L}_k^\dag \hat{L}_k \ket{\psi(t_1^-)}}}.
\end{equation}
Here, \( \ket{\psi(t_1^-)} \) and \( \ket{\psi(t_1^+)} \) denote the states just before and just after the quantum jump, respectively.

\item Repeat steps 2-5.
\end{enumerate}
The expectation value \( \langle \hat{A} \rangle_t \) of an observable \( \hat{A} \) at time \( t \) is given by
\begin{equation}
\langle \hat{A} \rangle_t = \overline{\bra{\psi(t)} \hat{A} \ket{\psi(t)}},
\end{equation}
where \( \overline{X} \) denotes the average over quantum trajectories.

In the actual computation, we assume that the quantum trajectory \( \ket{\psi(t)} \) can be approximated as a tensor product over individual sites, as given in Eq.~\eqref{phi_product_approximation}.
This approximation implies that entanglement between different sites is neglected.
We now discuss how to compute the time evolution of each \( \ket{\phi_i(t)} \) in Eq.~\eqref{phi_product_approximation}.
In the dissipative process during the first half of the cycle, each spin independently undergoes precession around the \( z \)-axis while relaxing toward the ground state.
During this stage, each \( \ket{\phi_i(t)} \) can be computed in the same way as in the single-spin case.
In the second half of the cycle, the unitary time evolution arising from spin interactions is described by the following mean-field Hamiltonian:
\begin{equation}
\hat{H}_1^{(i)} = \frac{2J}{C_{N,\alpha}} \sum_{j \ (\neq i)} \frac{1}{(r_{ij})^\alpha} \langle \hat{\sigma}_j^x \rangle \, \hat{\sigma}_i^x,
\end{equation}
where \( \langle \hat{\sigma}_j^x \rangle = \bra{\phi_j(t)} \hat{\sigma}^x \ket{\phi_j(t)} \) is the expectation value of the \( x \)-component of spin \( j \).
The time evolution of each \( \ket{\phi_i(t)} \) is then governed by
\begin{equation}
i \partial_t \ket{\phi_i(t)} = \hat{H}_1^{(i)} \ket{\phi_i(t)}.
\end{equation}

The quantity of interest in this study is the autocorrelation function of the magnetization.  
A general two-time correlation function \( C_{AB}(t, \tau) = \langle \hat{A}(t+\tau) \hat{B}(\tau) \rangle \) can be computed using the quantum trajectory method as follows \cite{Dalay-14}:
\begin{enumerate}
\item Starting from an initial condition, evolve the system up to time \( \tau \) to obtain the state \( \ket{\psi(\tau)} \).

\item Construct auxiliary states as
\begin{align}
\ket{\chi_{\pm}^R(0)} &= \frac{1}{\sqrt{\mu_{\pm}^R}} (\hat{I} \pm \hat{B}) \ket{\psi(\tau)}, \notag \\
\ket{\chi_{\pm}^I(0)} &= \frac{1}{\sqrt{\mu_{\pm}^I}} (\hat{I} \pm i \hat{B}) \ket{\psi(\tau)}.
\end{align}
Here, \( \mu_{\pm}^R \) and \( \mu_{\pm}^I \) are normalization constants chosen such that \( \braket{\chi_{\pm}^R(0)|\chi_{\pm}^R(0)} = \braket{\chi_{\pm}^I(0)|\chi_{\pm}^I(0)} = 1 \).

\item Evolve \( \ket{\chi_{\pm}^R(0)} \) and \( \ket{\chi_{\pm}^I(0)} \) for a time \( t \) to obtain the time-evolved states \( \ket{\chi_{\pm}^R(t)} \) and \( \ket{\chi_{\pm}^I(t)} \), respectively.

\item Compute the quantities
\begin{equation}
c_{\pm}^R(t) = \bra{\chi_{\pm}^R(t)} \hat{A} \ket{\chi_{\pm}^R(t)}, \:
c_{\pm}^I(t) = \bra{\chi_{\pm}^I(t)} \hat{A} \ket{\chi_{\pm}^I(t)}.
\end{equation}

\item The desired correlation function is then given by
\begin{equation}
C_{AB}(t, \tau) = \frac{1}{4} \left[\mu_+^R c_+^R(t) - \mu_-^R c_-^R(t) - i \mu_+^I c_+^I(t) + i \mu_-^I c_-^I(t) \right].
\end{equation}

\item Repeat steps 1-5 over multiple quantum trajectories to compute the ensemble average of \( C_{AB}(t, \tau) \).
\end{enumerate}

\section{Matrix elements in the symmetric sector of all-to-all coupled kicked spin chain}
\label{sec:symmetric_sector_spin_chain}

Let us consider the all-to-all coupling case of the kicked spin chain described by Eq.~\eqref{Hamiltonian_spin_chain_alpha_0}.
In this case, the system is invariant under permutations of the \( N \) spins.  
Therefore, by restricting to the subspace invariant under spin permutations, one can significantly reduce the dimensionality of the Liouvillian.  
In this appendix, we derive the matrix elements of the Liouvillian in the basis of permutation-symmetric states.

It is convenient to express the Liouvillians for the static part \( \mathcal{L}_0 \) and the kick \( \mathcal{L}_1 = -i[\hat{H}_1, \cdot] \) in vectorized form \cite{Prosen-12, Znidaric-15}:
\begin{align}
\mathcal{L}_0 =& \sum_{i = 1}^N \bigg[ - \frac{i \omega_z}{2} (\hat{\sigma}_{i, A}^z - \hat{\sigma}_{i, B}^z) \notag \\ 
&+ \kappa \bigg( \hat{\sigma}_{i, A}^- \hat{\sigma}_{i, B}^- - \frac{1}{2} \hat{\sigma}_{i, A}^+ \hat{\sigma}_{i, A}^- - \frac{1}{2} \hat{\sigma}_{i, B}^+ \hat{\sigma}_{i, B}^- \bigg) \bigg],
\end{align}
\begin{equation}
\mathcal{L}_1 = - \frac{iJ}{N} \sum_{i, j = 1}^N (\hat{\sigma}_{i, A}^x \hat{\sigma}_{j, A}^x - \hat{\sigma}_{i, B}^x \hat{\sigma}_{j, B}^x).
\end{equation}
Here, for each site \( i \), we introduce two spin-1/2 degrees of freedom, labeled A and B, which act on the ket and bra spaces, respectively.  
Operators acting on these spins are denoted as \( \hat{\sigma}_{i, A}^z \), \( \hat{\sigma}_{i, B}^z \), and so on.  
Note that the sum in \( \mathcal{L}_1 \) includes terms with \( i = j \); however, since \( (\hat{\sigma}_{i, A}^x)^2 = (\hat{\sigma}_{i, B}^x)^2 = 1 \), these terms do not affect the dynamics.
For each site, there are four basis states corresponding to the combined configurations of spins A and B:  
\( \ket{\downarrow \downarrow}, \ket{\downarrow \uparrow}, \ket{\uparrow \downarrow}, \ket{\uparrow \uparrow} \).  
We label these single-site basis states as  
\begin{equation}
\ket{0} = \ket{\downarrow \downarrow}, \quad \ket{1} = \ket{\downarrow \uparrow}, \quad \ket{2} = \ket{\uparrow \downarrow}, \quad \ket{3} = \ket{\uparrow \uparrow}.
\end{equation}
Accordingly, a basis state for the full \( N \)-site system can be denoted as \( \ket{3, 0, 1, \ldots, 2, 3} \), indicating the four-state configuration of each site.

To construct permutation symmetric basis states, one can superpose all basis states \( \ket{i_1, i_2, \ldots, i_N} \) obtained by permuting the site indices.  
Such a permutation symmetric basis can be labeled by the number of occurrences of the values \( 0 \), \( 1 \), \( 2 \), and \( 3 \).  
Let \( l \), \( m \), and \( n \) denote the number of sites in the state \( 0 \), \( 1 \), and \( 2 \), respectively.  
Then, the number of sites in state \( 3 \) is \( N - l - m - n \).  
We denote the corresponding permutation symmetric basis state as \( |l, m, n) \), which is defined by
\begin{equation}
|l, m, n) = \mathcal{N}_{l,m,n} \sum_{\#0 = l, \#1 = m, \#2 = n} \ket{i_1, i_2, \ldots, i_N},
\label{symmetric_basis}
\end{equation}
where the sum runs over all basis states \( \ket{i_1, i_2, \ldots, i_N} \) that contain \( l \) sites in state \( 0 \), \( m \) in state \( 1 \), and \( n \) in state \( 2 \).
The parameters \( l, m, n \) must satisfy the constraints \( l \geq 0, m \geq 0, n \geq 0 \), and \( N - l - m - n \geq 0 \).  
The total number of such integer triplets \( (l, m, n) \) is \( \binom{N+3}{3} \), which determines the dimension of the permutation symmetric sector.
The prefactor \( \mathcal{N}_{l,m,n} \) in Eq.~\eqref{symmetric_basis} ensures normalization \( (l, m, n | l, m, n) = 1 \), and is given by
\begin{equation}
\mathcal{N}_{l,m,n} = \left[ \frac{N!}{l! \, m! \, n! \, (N - l - m - n)!} \right]^{-1/2}.
\end{equation}

We now compute the matrix elements of the Liouvillian operators \( \mathcal{L}_0 \) and \( \mathcal{L}_1 \) with respect to the newly defined orthonormal basis \( \{ |l, m, n) \} \).  
To this end, we first list the action of relevant operators on the one-site basis states:  
\begin{equation}
\hat{\sigma}_A^z \ket{0} = - \ket{0}, \: \hat{\sigma}_A^z \ket{1} = - \ket{1}, \: \hat{\sigma}_A^z \ket{2} = \ket{2}, \: \hat{\sigma}_A^z \ket{3} = \ket{3}.
\end{equation}
\begin{equation}
\hat{\sigma}_B^z \ket{0} = - \ket{0}, \: \hat{\sigma}_B^z \ket{1} = \ket{1}, \: \hat{\sigma}_B^z \ket{2} = -\ket{2}, \: \hat{\sigma}_B^z \ket{3} = \ket{3}.
\end{equation}
\begin{equation}
\hat{\sigma}_A^x \ket{0} = \ket{2}, \: \hat{\sigma}_A^x \ket{1} = \ket{3}, \: \hat{\sigma}_A^x \ket{2} = \ket{0}, \: \hat{\sigma}_A^x \ket{3} = \ket{1}.
\end{equation}
\begin{equation}
\hat{\sigma}_B^x \ket{0} = \ket{1}, \: \hat{\sigma}_B^x \ket{1} = \ket{0}, \: \hat{\sigma}_B^x \ket{2} = \ket{3}, \: \hat{\sigma}_B^x \ket{3} = \ket{2}.
\end{equation}
\begin{equation}
\hat{\sigma}_A^- \hat{\sigma}_B^- \ket{0} = \hat{\sigma}_A^- \hat{\sigma}_B^- \ket{1} = \hat{\sigma}_A^- \hat{\sigma}_B^- \ket{2} = 0, \: \hat{\sigma}_A^- \hat{\sigma}_B^- \ket{3} = \ket{0}.
\end{equation}
\begin{equation}
\hat{\sigma}_A^+ \hat{\sigma}_A^- \ket{0} = \hat{\sigma}_A^+ \hat{\sigma}_A^- \ket{1} = 0, \: \hat{\sigma}_A^+ \hat{\sigma}_A^- \ket{2} = \ket{2}, \: \hat{\sigma}_A^+ \hat{\sigma}_A^- \ket{3} = \ket{3}.
\end{equation}
\begin{equation}
\hat{\sigma}_B^+ \hat{\sigma}_B^- \ket{0} = \hat{\sigma}_B^+ \hat{\sigma}_B^- \ket{2} = 0, \: \hat{\sigma}_B^+ \hat{\sigma}_B^- \ket{1} = \ket{1}, \: \hat{\sigma}_B^+ \hat{\sigma}_B^- \ket{3} = \ket{3}.
\end{equation}

We first compute the matrix elements of diagonal operators:
\begin{equation}
\sum_{i = 1}^N \hat{\sigma}_{i, A}^z |l, m, n) = (N - 2l - 2m) |l, m, n),
\end{equation}
\begin{equation}
\sum_{i = 1}^N \hat{\sigma}_{i, B}^z |l, m, n) = (N - 2l - 2n) |l, m, n),
\end{equation}
\begin{equation}
\sum_{i = 1}^N \hat{\sigma}_{i, A}^+ \hat{\sigma}_{i, A}^- |l, m, n) = (N - l - m) |l, m, n),
\end{equation}
\begin{equation}
\sum_{i = 1}^N \hat{\sigma}_{i, B}^+ \hat{\sigma}_{i, B}^- |l, m, n) = (N - l - n) |l, m, n).
\end{equation}

Next, we compute the matrix elements of the off-diagonal operators:
\begin{align}
&\sum_{i = 1}^N \hat{\sigma}_{i, A}^x |l, m, n) \nonumber \\
=& \sqrt{l (n+1)} (1-\delta_{l, 0}) |l-1, m, n+1) \nonumber \\
&+ \sqrt{(l+1) n} (1-\delta_{n, 0}) |l+1, m, n-1) \nonumber \\
&+ \sqrt{m (N-l-m-n+1)} (1-\delta_{m, 0}) |l, m-1, n) \nonumber \\
&+ \sqrt{(m+1) (N-l-m-n)} (1-\delta_{l+m+n, N}) |l, m+1, n).
\end{align}
\begin{align}
&\sum_{i = 1}^N \hat{\sigma}_{i, B}^x |l, m, n) \nonumber \\ 
=& \sqrt{l (m+1)} (1-\delta_{l, 0}) |l-1, m+1, n)  \nonumber \\
&+ \sqrt{(l+1) m} (1-\delta_{m, 0}) |l+1, m-1, n) \nonumber \\
&+ \sqrt{n (N-l-m-n+1)} (1-\delta_{n, 0}) |l, m, n-1) \nonumber \\
&+ \sqrt{(n+1) (N-l-m-n)} (1-\delta_{l+m+n, N}) |l, m, n+1),
\end{align}
\begin{align}
&\sum_{i = 1}^N \hat{\sigma}_{i, A}^- \hat{\sigma}_{i, B}^- |l, m, n) \nonumber \\
=& \sqrt{(l+1) (N-l-m-n)} (1-\delta_{l+m+n, N}) |l+1, m, n).
\end{align}
By combining these expressions, we obtain all necessary matrix elements for constructing the Liouvillians \( \mathcal{L}_0 \) and \( \mathcal{L}_1 \) in the permutation-symmetric basis.

\section{Liouvillian gap of a classical stochastic system in the noiseless limit}
\label{appendix:Liouvillian_gap_of_stochastic_system}

\begin{figure}
\centering
\includegraphics[width=8.6cm]{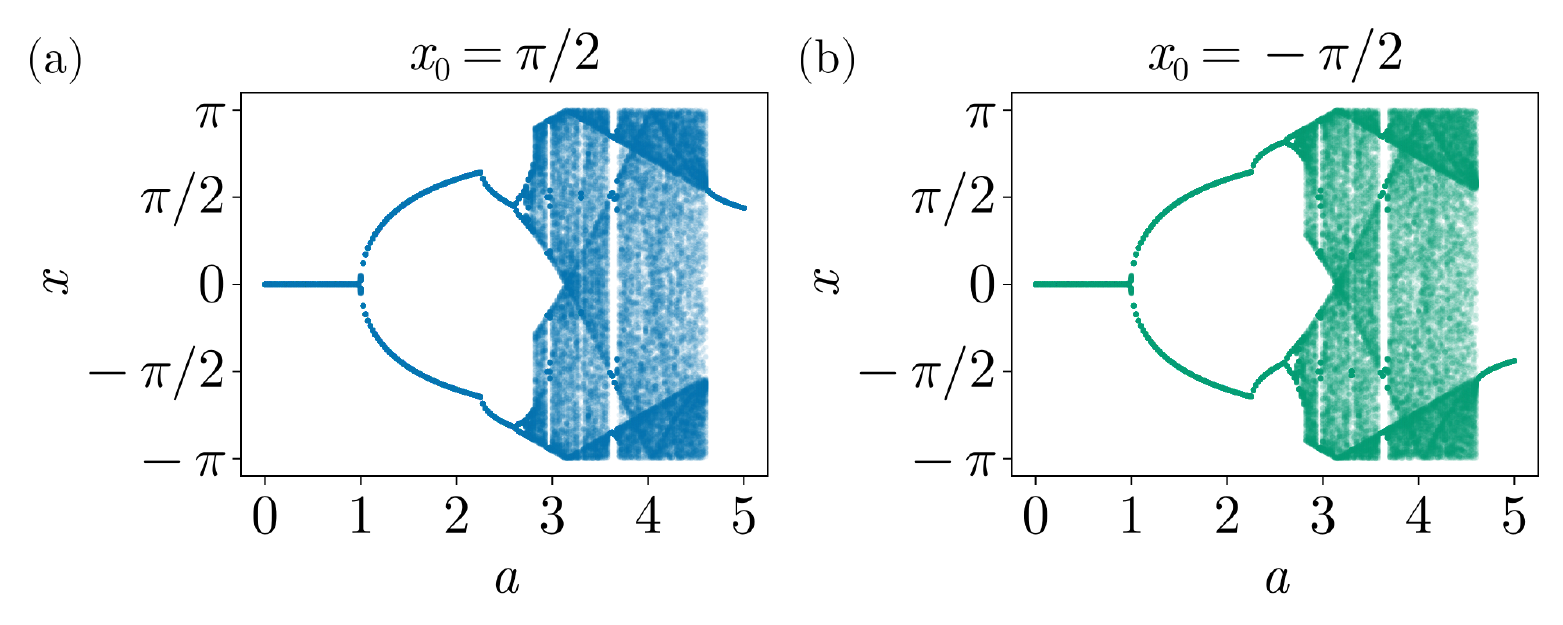}
\caption{Bifurcation diagrams of the deterministic sine map with \( \sigma = 0 \).  
Panels (a) and (b) show the trajectories starting from initial conditions \( x_0 = \pi/2 \) and \( x_0 = -\pi/2 \), respectively.  
The initial transients over the first 100 steps are discarded.  
The dependence of the bifurcation diagrams on the initial conditions indicates spontaneous symmetry breaking.}
\label{fig_sin_map_bifurcation}
\end{figure}

In this appendix, we investigate the behavior of the Liouvillian gap in a classical stochastic system as the noise intensity approaches zero.
It is known that dissipative quantum systems can be approximated by effective classical stochastic systems under suitable conditions.
In such cases, the large system-size limit of the quantum system often corresponds to the noiseless limit of the associated stochastic model.  
Therefore, understanding the behavior of the Liouvillian gap in classical stochastic systems sheds light on the spectral properties of dissipative quantum systems.

We follow the general framework presented in Sec.~\ref{sec:Liouvillian_gap_of_stochastic_system}.  
The time evolution of a one-dimensional dynamical system subjected to noise is described by Eq.~\eqref{stochastic_dynamical_system_general} with a map \( f(x) \).
When considering \( f(x) \) defined on a bounded interval, it is convenient for numerical calculations to impose periodic boundary conditions.  
We consider a \( 2\pi \)-periodic map \( f(x) \) defined on the interval \( [-\pi, \pi) \).
Inspired by the analogy to \( \mathbb{Z}_2 \) symmetry in spin models, we require that the map \( f(x) \) be symmetric under coordinate inversion \( x \to -x \), that is, it satisfies \( f(-x) = -f(x) \).
Furthermore, motivated by the discrete time crystal, we impose the additional condition \( f'(0) < 0 \), so that period-2 oscillations can occur.
As the simplest map satisfying these conditions, we consider  
\begin{equation}
f(x) = -a \sin x \mod 2\pi \quad (a > 0).
\end{equation}
Taking into account the periodicity, we assume that the sequence \( x_t \; (t = 0, 1, \ldots) \) lies within the interval \( [-\pi, \pi) \).

Considering the periodic boundary conditions, the propagator given in Eq.~\eqref{stochastic_dynamics_propagator} is modified as follows:  
\begin{equation}
k(x|y) = C \exp \left( \frac{\cos[x - f(y)] - 1}{\sigma^2} \right),
\label{stochastic_dynamics_propagator_periodic}
\end{equation}
where \( C \) is the normalization constant.  
Note that \( k(x|y) \) has periodicity \( 2\pi \), and it reduces to Eq.~\eqref{stochastic_dynamics_propagator} in the small-\( \sigma \) limit.  

In numerical calculations of the Liouvillian eigenvalues, it is necessary to discretize the integral in Eq.~\eqref{stochastic_dynamics_Liouvillian}.
To this end, we divide the interval \( [-\pi, \pi) \) into \( N \) segments and introduce a mesh defined by \( x_i = 2\pi i/N - \pi \; (i = 1, \ldots, N) \).
With this discretization, the integral in Eq.~\eqref{stochastic_dynamics_Liouvillian} is approximated as
\begin{equation}
\int_{-\pi}^{\pi} k(x|y) \, p(y) \, dy \simeq \sum_{i=1}^N k(x|x_i) \, p(x_i) \, \Delta x,
\end{equation}  
where \( \Delta x = 2\pi / N \).
Accordingly, we define an \( N \times N \) matrix \( A \) by  
\begin{equation}
A_{ij} = k(x_i|x_j).
\end{equation}
Here, the normalization constant \( C \) in Eq.~\eqref{stochastic_dynamics_propagator_periodic} is chosen such that \( \sum_{i=1}^N A_{ij} = 1 \) holds.  
The Liouvillian eigenvalues are then obtained by diagonalizing the matrix \( A \).
The Liouvillian gap is defined by \( \Delta = -\log |\lambda_1| \), where \( \lambda_1 \) is the eigenvalue with the second largest modulus.

\begin{figure}
\centering
\includegraphics[width=8.6cm]{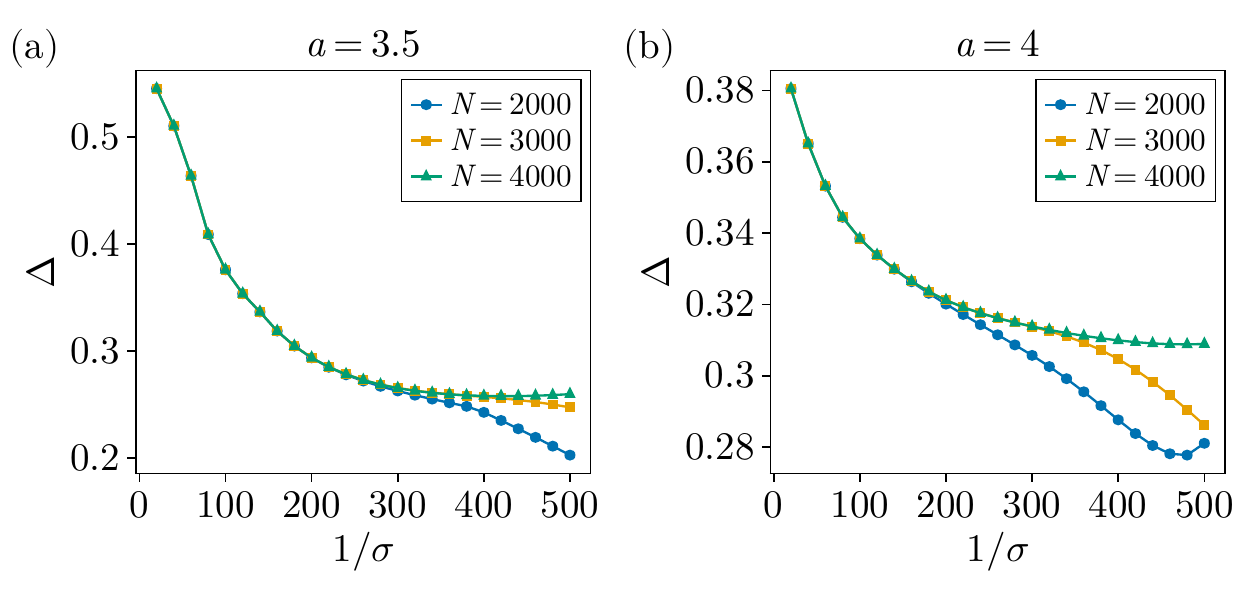}
\caption{Liouvillian gap \( \Delta \) as a function of the inverse noise intensity \( 1/\sigma \) for different discretization numbers \( N \).  
Panels (a) and (b) show \( \Delta \) for \( a = 3.5 \) and \( a = 4 \), respectively, both of which correspond to the global chaos regime.  
The gap converges to a nonzero value in the noiseless limit \( \sigma \to 0 \).}
\label{fig_sin_map_gap}
\end{figure}

Firstly, let us consider the deterministic dynamics of the sine map with \( \sigma = 0 \).  
Figure~\ref{fig_sin_map_bifurcation} shows the bifurcation diagrams obtained from different initial conditions, \( x_0 = \pi/2 \) and \( x_0 = -\pi/2 \).
The dependence of the bifurcation diagrams on the initial conditions indicates spontaneous symmetry breaking.
For \( a < 1 \), the point \( x = 0 \) corresponds to a stable fixed point.  
At \( a = 1 \), this fixed point becomes unstable, leading to the emergence of a limit cycle.  
Around \( a = 2.7 \), the system enters a chaotic regime.  
At \( a = \pi \), two chaotic regions merge, resulting in global chaos.  
For \( a > 4.6 \), chaos disappears, and a stable fixed point reappears.

We now consider the behavior of the Liouvillian gap \( \Delta \) in the noiseless limit \( \sigma \to 0 \), when the corresponding deterministic dynamics exhibit global chaos.   
Note that the discretization number \( N \) must be taken to infinity before taking the noiseless limit.  
Figure~\ref{fig_sin_map_gap} shows the Liouvillian gap \( \Delta \) as a function of the inverse noise intensity \( 1/\sigma \) for different values of the discretization number \( N \).
Panels (a) and (b) show \( \Delta \) for \( a = 3.5 \) and \( a = 4 \), respectively. 
For a fixed \( \sigma \), the convergence of \( \Delta \) is evident as \( N \) increases.  
We can observe that \( \lim_{\sigma \to 0} \lim_{N \to \infty} \Delta \) remains nonzero.  
This result is consistent with the observation that the time glass phase possesses a finite gap.

\begin{figure}
\centering
\includegraphics[width=8.6cm]{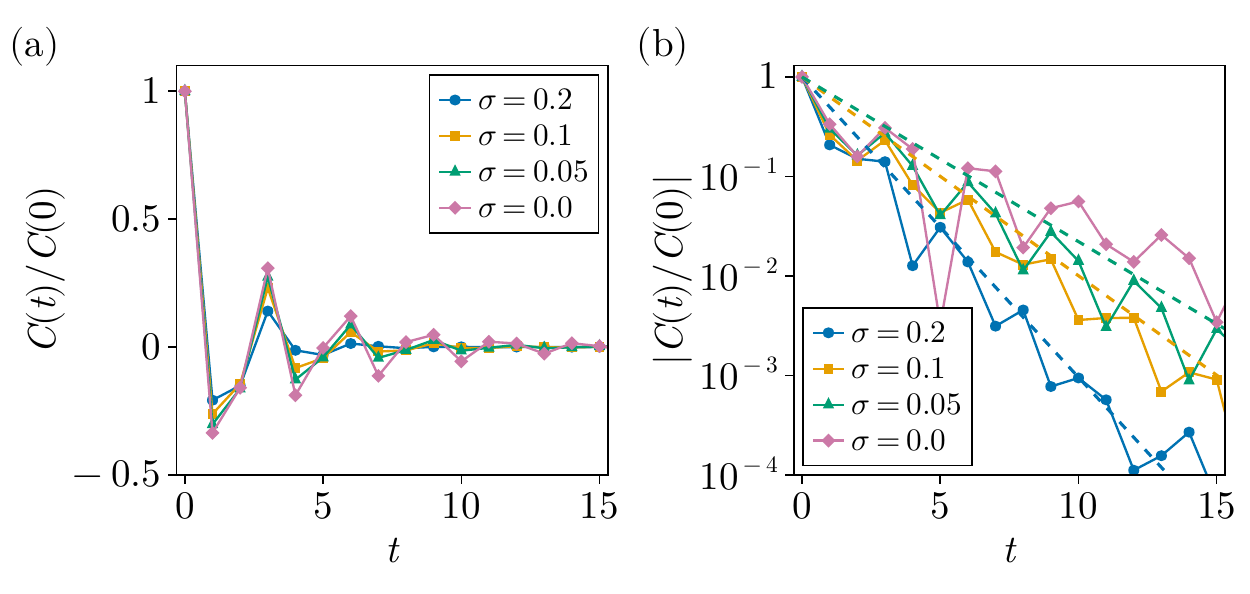}
\caption{Normalized autocorrelation \( C(t) / C(0) \) for \( a = 4 \) with different noise intensities \( \sigma = 0.2 \), \( 0.1 \), \( 0.05 \), and \( 0 \) (deterministic).  
(b) The absolute value \( |C(t) / C(0)| \) is shown in a semi-logarithmic plot.  
The dashed lines represent \( e^{-\Delta t} \), where \( \Delta \) is the Liouvillian gap obtained from the diagonalization of the matrix \( A_{ij} \) with the corresponding parameters indicated by the same colors.  
We observe that, for nonzero \( \sigma \), the Liouvillian gap coincides with the decay rate of \( C(t) \).}
\label{fig_sin_map_correlation}
\end{figure}

Next, we consider the relationship between the Liouvillian gap and the decay rate of the autocorrelation function.
The autocorrelation \( C(t) \) is defined by
\begin{equation}
C(t) = \lim_{T \to \infty} \frac{1}{T} \sum_{\tau=1}^{T} (x_{t+\tau} - \bar{x}) (x_\tau - \bar{x}),
\end{equation}
where \( \bar{x} \) denotes the time average of \( x_t \).
Figure~\ref{fig_sin_map_correlation} shows the normalized autocorrelation \( C(t)/C(0) \) for \( a = 4 \) at different noise intensities \( \sigma = 0.2 \), \( 0.1 \), \( 0.05 \), and \( 0 \) (deterministic).
In Fig.~\ref{fig_sin_map_correlation}(b), the dashed lines correspond to \( e^{-\Delta t} \), where \( \Delta \) is the Liouvillian gap obtained by diagonalizing the matrix \( A_{ij} \) using the corresponding parameters.
We observe that, for nonzero \( \sigma \), the Liouvillian gap precisely matches the decay rate of \( C(t) \).
This result strongly supports the conclusion that, in the noiseless limit \( \sigma \to 0 \), the Liouvillian gap converges to the decay rate of \( C(t) \) associated with the deterministic dynamics.

\section{Clarifying the discrepancy with Ref.~\cite{Carollo-24}}
\label{appendix:clarifying_the_discrepancy}

\begin{figure}
\centering
\includegraphics[width=8.6cm]{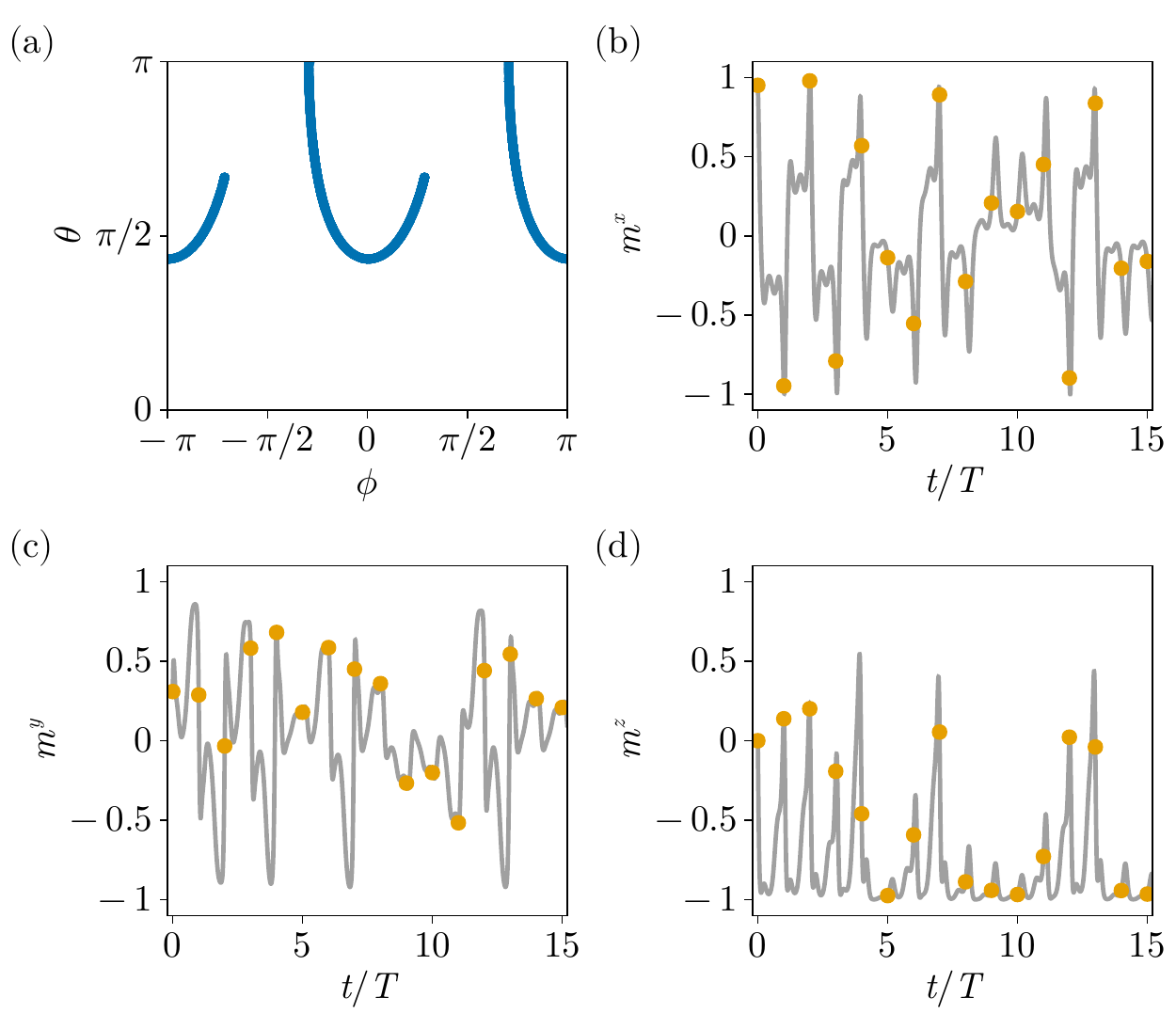}
\caption{Classical dynamics of the Carollo model. 
(a) Poincar\'e map plotted in the plane of the polar angle $\theta$ and azimuthal angle $\phi$.
(b)-(d) Time evolution of the spin components $m^x$, $m^y$, and $m^z$.
The markers indicate stroboscopic points sampled at every period $t = nT$ $(n = 0, 1, \ldots)$. 
The initial state is given by $\theta = 0.5\pi$ and $\phi = 0.1\pi$.
The string-like attractor and the oscillatory behavior of $m_x$ signify a small maximum Lyapunov exponent or, equivalently, a small mixing rate.
The parameters are $g=\gamma=h_0=h_1=1$ and $T=2\pi$.}
\label{fig_Carollo_model_classical_dynamics}
\end{figure}

\begin{figure*}
\centering
\includegraphics[width=\textwidth]{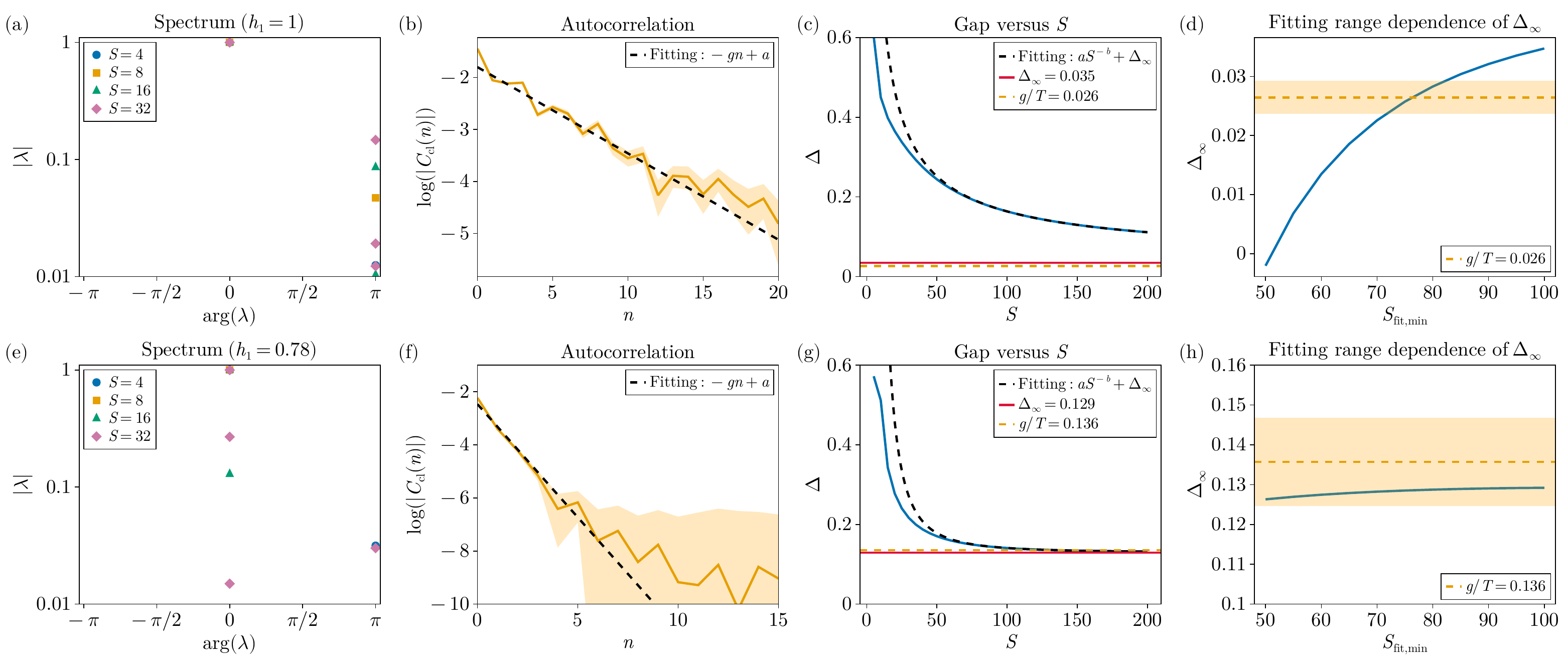}
\caption{Relationship between the Liouvillian gap and the classical mixing rate.
The driving amplitude is set to \( h_1 = 1 \) in panels (a)-(d) and \( h_1 = 0.78 \) in panels (e)-(h), both corresponding to the time glass phase.
Panels (a) and (e) display the eigenvalues \( \lambda_\alpha \) of the Floquet map, plotted in terms of their modulus \( |\lambda_\alpha| \) and argument \( \mathrm{arg}(\lambda_\alpha) \), for \( S = 4 \), \( 8 \), \( 16 \), and \( 32 \).
Panels (b) and (f) show the logarithm of the autocorrelation \( C_{\mathrm{cl}}(n) \) for the classical stroboscopic dynamics, while the shaded areas represent the standard deviation.
The dashed lines indicate linear fits of the form \( -gn + a \) over the ranges \( [1, 10] \) in (b) and \( [0, 6] \) in (f).
Panels (c) and (g) present the Liouvillian gap \( \Delta \) as a function of \( S \).
The dashed black curves show fits to \( a S^{-b} + \Delta_\infty \) within the range \( [100, 200] \).
The horizontal solid lines indicate the thermodynamic-limit value \( \Delta_\infty \) obtained from the fitting, and the horizontal dashed lines denote the classical mixing rate \( g / T \).
Panels (d) and (h) display the dependence of the fitted value \( \Delta_\infty \) on the fitting range, where the range is defined as \( [S_{\mathrm{fit,min}}, S_{\mathrm{fit,min}} + 100] \).
The horizontal dashed lines again represent the classical mixing rate \( g / T \), and the shaded areas indicate the standard error estimated from the fitting of \( C_{\mathrm{cl}}(n) \).
The parameters are $g=\gamma=h_0=1$ and $T=2\pi$.}
\label{fig_Carollo_model_gap}
\end{figure*}

In this appendix, we address the apparent discrepancy between our results and those reported in Ref.~\cite{Carollo-24} by F. Carollo and I. Lesanovsky.
In Ref.~\cite{Carollo-24}, the authors studied a periodically driven dissipative Ising model and found that, within a certain parameter regime, its classical (mean-field) dynamics exhibit chaotic oscillations, which correspond to the time-glass behavior according to our definition.
They also analyzed the Liouvillian gap \( \Delta \) of the corresponding Floquet map and concluded that, in the time glass phase, the gap closes following a power-law scaling \( \Delta \propto N^{-1} \), where \( N \) denotes the number of spins.
This observation appears to contradict our finding that the Liouvillian gap remains finite in the thermodynamic limit.
In the following, we perform a detailed investigation of the behavior of the Liouvillian gap for the model studied in Ref.~\cite{Carollo-24}, which is referred to as the Carollo model.

The collective spin operators $\hat{S}^x$, $\hat{S}^y$, and $\hat{S}^z$ are defined in Eq.~\eqref{def_collective_spin_operator}.
The time-periodic Hamiltonian is given by
\begin{equation}
\hat{H}(t) = - \frac{2g}{S} (\hat{S}^x)^2 + 2[h_0 + h_1 \sin(\chi t)] S^z,
\end{equation}
where $S=N/2$ is the total spin, and $g$, $h_0$, and $h_1$ are tunable parameters.
(Here, we denote $\Delta_0$ and $\Delta_1$ in Ref.~\cite{Carollo-24} as $h_0$ and $h_1$, respectively, to avoid confusion with the Liouvillian gap $\Delta$.)
The driving period is $T=2\pi/\chi$.
The quantum master equation governing the system's dissipative dynamics is written as
\begin{equation}
\partial_t \hat{\rho} = - i [\hat{H}(t), \hat{\rho}] + \frac{\gamma}{S} \left( \hat{S}^- \hat{\rho} \hat{S}^+ - \frac{1}{2} \{ \hat{S}^+ \hat{S}^-, \hat{\rho} \} \right),
\end{equation}
where \( \gamma \) represents the dissipation strength.
In the following, the parameters are fixed to $g=\gamma=h_0=1$ and $T=2\pi$.
We analyze the spectrum of the Floquet map, which is obtained by integrating this master equation over one driving period, from \( t = 0 \) to \( t = T \).

We first analyze the classical (mean-field) dynamics of this model.
In the large-spin limit \( S \to \infty \), the normalized spin components \( m^\nu = \langle \hat{S}^\nu \rangle / S \) (\( \nu = x, y, z \)) obey the following equations of motion:
\begin{equation}
\begin{split}
\frac{dm^x}{dt} &= -2[h_0 + h_1 \sin(\chi t)] m^y + \gamma m^x m^z, \\
\frac{dm^y}{dt} &= 4g m^x m^z + 2[h_0 + h_1 \sin(\chi t)] m^x + \gamma m^y m^z, \\
\frac{dm^z}{dt} &= -4g m^x m^y - \gamma [(m^x)^2 + (m^y)^2].
\label{Carollo_classical_equation}
\end{split}
\end{equation}
Figure \ref{fig_Carollo_model_classical_dynamics} presents the classical dynamics in the time glass regime for $h_1=1$.
Figure \ref{fig_Carollo_model_classical_dynamics}(a) shows the Poincar\'e map plotted in the plane of the polar angle $\theta$ and azimuthal angle $\phi$.
The attractor exhibits a string-like structure and consists of two disconnected branches.
The corresponding Lyapunov exponents are estimated as \( \lambda^+ = 0.12 \) and \( \lambda^- = -1.7 \).
The resulting Lyapunov dimension, \( 1 + \lambda^+ / |\lambda^-| = 1.07 \), being close to unity, explains the quasi-one-dimensional form of the attractor.
The stroboscopic evolution of \( m^x \), shown in Fig.~\ref{fig_Carollo_model_classical_dynamics}(b), reveals nearly periodic oscillations where the sign of \( m^x \) alternates between positive and negative, reflecting the two disconnected branches of the attractor.
This near-oscillatory behavior, together with the small positive Lyapunov exponent \( \lambda^+ \), indicates a weakly mixing regime characterized by slow decorrelation.

Figure \ref{fig_Carollo_model_gap} examines the relationship between the Liouvillian gap and the classical mixing rate.
The driving amplitude is set to \( h_1 = 1 \) in panels (a)-(d) and \( h_1 = 0.78 \) in panels (e)-(h), both corresponding to the time glass phase.
Panels (a) and (e) show the eigenvalues \( \lambda_\alpha \) of the Floquet map for various values of \( S \).
For \( h_1 = 1 \) [see panel~(a)], the leading eigenvalue \( \lambda_1 \) with the second largest modulus lies on the negative real axis, reflecting oscillatory behavior with an effective period of two.
In contrast, for \( h_1 = 0.78 \) [see panel~(e)], \( \lambda_1 \) lies on the positive real axis, indicating that these two cases belong to distinct subclasses of the time glass phase.
In both cases, the modulus of \( \lambda_1 \) increases with \( S \); however, this behavior does not necessarily imply the closing of the Liouvillian gap.

Let \( (m^x(n), m^y(n), m^z(n)) \) (\( n = 0, 1, \ldots \)) denote the stroboscopic solution of Eq.~\eqref{Carollo_classical_equation} evaluated at each driving period \( t = nT \).
The autocorrelation function of the classical dynamics is defined as
\begin{equation}
C_{\mathrm{cl}}(n) = \lim_{K \to \infty} \frac{1}{K} \sum_{k=1}^K m^y(n+k) m^y(k),
\end{equation}
where we focus on the component \( m^y(n) \) because it exhibits a clear exponential decay.
Figures \ref{fig_Carollo_model_gap}(b) and (f) display the logarithm of \( C_{\mathrm{cl}}(n) \) for \( h_1 = 1 \) and \( h_1 = 0.78 \), respectively.
To extract the exponential decay behavior \( C_{\mathrm{cl}}(n) \propto e^{-gn} \) characterized by the mixing rate \( g \),
we fit the data to the functional form \( -gn + a \), as indicated by the dashed lines.

Figures \ref{fig_Carollo_model_gap}(c) and (g) present the Liouvillian gap \( \Delta \) as a function of the spin size \( S \) for \( h_1 = 1 \) and \( h_1 = 0.78 \), respectively.
In the definition of \( \Delta \), we divide \( -\log |\lambda_1| \) by the driving period \( T = 2\pi \).
To estimate the thermodynamic-limit value \( \Delta_\infty \), the data are fitted to the functional form \( a S^{-b} + \Delta_\infty \), as shown by the dashed curves.
The horizontal solid lines indicate the extrapolated values of \( \Delta_\infty \), while the horizontal dashed lines represent the classical mixing rate \( g / T \) obtained from the fitting of \( C_{\mathrm{cl}}(n) \).
For \( h_1 = 0.78 \) [see panel~(g)], \( \Delta \) converges well to a finite \( \Delta_\infty \) within the accessible range of \( S \), and the resulting \( \Delta_\infty \) agrees with the classical mixing rate.
For \( h_1 = 1 \) [see panel~(c)], which corresponds to the parameter regime studied in Ref.~\cite{Carollo-24}, the classical mixing rate is very small and the convergence of \( \Delta \) with \( S \) is significantly slower.
In Ref.~\cite{Carollo-24}, the Liouvillian gap was computed up to \( S = 50 \) (\( N = 100 \)) and reported to vanish as \( \Delta \propto S^{-1} \).
However, our analysis suggests that this apparent closing of the gap may arise from finite-size effects, reflecting the slow convergence of \( \Delta \) within the limited system-size range.
Figures \ref{fig_Carollo_model_gap}(d) and (h) show how the fitted value of \( \Delta_\infty \) depends on the fitting range, defined as \( [S_{\mathrm{fit,min}}, S_{\mathrm{fit,min}} + 100] \).
For \( h_1 = 1 \) [see panel~(d)], the observation that the fitted \( \Delta_\infty \) increases as the fitting range shifts toward larger \( S \) may support the interpretation that \( \Delta \) converges to a small but finite \( \Delta_\infty \) in the thermodynamic limit.

In summary, the available data for the model studied in Ref.~\cite{Carollo-24} are not inconsistent with our interpretation, which assumes the presence of a finite Liouvillian gap in the time glass phase.
To draw a definitive conclusion, further investigations involving larger system sizes would be necessary.
It is worth emphasizing, however, that the possible existence of a finite gap in the time glass phase does not affect the validity of the main result of Ref.~\cite{Carollo-24}, where the authors rigorously proved the convergence of the quantum dynamics to its mean-field counterpart in the thermodynamic limit.
As discussed in Sec.~\ref{sec:relaxation_time}, the timescale over which the quantum evolution faithfully follows the corresponding classical dynamics can diverge even when the Liouvillian gap remains finite.
Therefore, the existence of a finite gap and the validity of the mean-field description are logically independent issues.


\end{document}